\documentclass[12pt,superscriptaddress]{article}

\usepackage[utf8]{inputenc}
\usepackage{charter}
\usepackage{fullpage}
\usepackage{hyperref}
\usepackage{graphicx}
\usepackage{lipsum}
\usepackage[sort&compress,numbers]{natbib}
\usepackage{hyperref}
\hypersetup{hidelinks}
\usepackage{amsmath}
\usepackage{amssymb, xspace}
\usepackage[total={6.5in,9in},top=1in,headsep=0.1in,headheight=1in]{geometry}
\usepackage{siunitx} % For proper unit typesetting
\usepackage{aas_macros}
\usepackage{multirow}
\usepackage[normalem]{ulem}
\usepackage{microtype}

\newcommand\snowmass{
\begin{center}
  \rule[-0.2in]{\hsize}{0.01in}\\
  \rule{\hsize}{0.01in}\\
  \vskip 0.1in
  Submitted to the Proceedings of the US Community Study\\ 
  on the Future of Particle Physics (Snowmass 2021)\\
  \rule{\hsize}{0.01in}\\
  \rule[+0.2in]{\hsize}{0.01in}\\[-2em]
\end{center}
}
\def\fnl{{f_{\rm NL}}}
\def\fnlloc{{f^{\rm loc}_{\rm NL}}}
\def\fnleq{{f^{\rm eq}_{\rm NL}}}
\def\fnlorth{{f^{\rm ortho}_{\rm NL}}}
\def\beq{\begin{equation}}
\def\eeq{\end{equation}}
\def\Neff{{N_{\rm eff}}}
\def\bea{\begin{align}}
\def\eea{\end{align}}

\def\gw#1{gravitational wave#1 (GW#1)\gdef\gw{GW}}

\usepackage[firstpage=true]{background}
\backgroundsetup{contents={\parbox{6.5in}{\snowmass}}, scale=1,placement=top,opacity=1,color=black,position={3.25in,1.2in}}

\usepackage{fancyhdr}
\fancypagestyle{plain}{%
  \fancyhf{}%
  \fancyhead[C]{}
  \fancyfoot[C]{\thepage}
}

\fancypagestyle{empty}{%
  \fancyhf{}%
  \fancyhead[C]{{\it Snowmass2021 Cosmic Frontier Topical Group Report}}
  \fancyfoot[C]{\thepage}
}
\title{Report of the Topical Group on Cosmic Frontier 5 \\
Dark Energy and Cosmic Acceleration: \\
Cosmic Dawn and Before for Snowmass 2021}
\date{}

\usepackage{authblk}

\author[1]{C.L.~Chang} 
\author[2]{L.~Newburgh}
\author[3]{D.~Shoemaker}
\author[4]{\\ \vspace{.7cm}S.~Ballmer}
\author[5]{D.~Green} 
\author[6]{R.~Hlozek}
\author[7]{K.~Huffenberger}
\author[8]{K.~Karkare}
\author[9]{A.~Liu}
\author[10]{V.~Mandic}
\author[11]{J.~McMahon}
\author[12]{B.~Wallisch}

\affil[1]{High Energy Physics Division, Argonne National Laboratory}
\affil[2]{Department of Physics, Yale University}
\affil[3]{Center for Gravitational Physics, Department of Physics, University of Texas at Austin}
\affil[4]{Department of Physics, Syracuse University}
\affil[5]{Department of Physics, University of California at San Diego}
\affil[6]{Dunlap Institute, Department of Astronomy and Astrophysics, University of Toronto}
\affil[7]{Department of Physics, Florida State University}
\affil[8]{Kavli Institute for Cosmological Physics, University of Chicago}
\affil[9]{Department of Physics, McGill University}
\affil[10]{School of Physics and Astronomy, University of Minnesota}
\affil[11]{Department of Astronomy and Astrophysics, Enrico Fermi Institute, University of Chicago}
\affil[12]{Department of Physics, University of California San Diego  and School of Natural Sciences, Institute for Advanced Study, Princeton}
\begin{document}

\maketitle

%%%%%%%%%%%%%%%%%%%%%%%%%%%%%%%%%%%%%%%%%%%%%
%%%%%%%%%%% TEXT START  %%%%%%%%%%%%%%%%%%%%%
%%%%%%%%%%%%%%%%%%%%%%%%%%%%%%%%%%%%%%%%%%%%%

%%%%%%%%%%%%%%%%%%%%%%%%%%%%%%%%%%%%%%%%%%%%%
%%%%%%%%%%% INPUTS %%%%%%%%%%%%%%%%%%%%%%%%%%
%%%%%%%%%%%%%%%%%%%%%%%%%%%%%%%%%%%%%%%%%%%%%
\begin{abstract}
  This report summarizes the envisioned research activities as gathered from the Snowmass 2021 CF5 working group concerning Dark Energy and Cosmic Acceleration: Cosmic Dawn and Before. The scientific goals are to study inflation and to search for new physics through precision measurements of relic radiation from the early universe. The envisioned research activities for this decade (2025–35) are constructing and operating major facilities and developing critical enabling capabilities. The major facilities for this decade are the CMB-S4 project, a new Stage-V spectroscopic survey facility, and existing gravitational wave observatories. Enabling capabilities include aligning and investing in theory, computation and model building, and investing in new technologies needed for early universe studies in the following decade (2035+).     
  %This report summarizes the envisioned research activities as gathered from the Snowmass 2021 CF5 working group concerning dark energy and cosmic acceleration.  The scientific goals are  inflation and the discovery of new physics and the research activities are major facilities and enabling capabilities.  Enabling capabilities includes aligning and investing in theory, computation and model building with the facilities and investing in new technologies needed for the discovery era of the 2030+.     
\end{abstract}

\begin{center}
  \normalsize\bfseries\MakeUppercase{Executive Summary}
\end{center}

The early universe is a unique and powerful tool for fundamental science. From the validation of ``Big Bang cosmology'' to precision measurements of our cosmological model, studies of the early universe have transformed our understanding of high energy physics. This report summarizes the major themes arising from activities of the Snowmass CF5 working group. The envisioned timeframe is 2025--35 with an eye towards 2035--50. 

The scientific goals fall broadly into two categories.
\begin{itemize}
    \item The first category is the topic of inflation where the goals are to discover/constrain the amplitude of inflationary gravitational waves ($r$), make precision measurements of the shape of the primordial power spectrum and its Gaussian/non-Gaussian statistics, and to test for deviations from the scale invariant spectrum of inflationary gravitational waves.
    \item The second category is the discovery of new physics via precision measurements of relic radiation. In the Standard Model, the only relic radiation apart from photons is the Cosmic Neutrino Background, which has a precisely predicted energy density. Thus, measuring any signal that differs from the predicted CNB would be an unambiguous discovery of new physics.
\end{itemize}

The envisioned research activities for 2025--35 fall into two directions: major facilities and enabling capabilities. Major facilities for this decade will drive transformational impact through searches for $r$, primordial features and statistics, and searches for the stochastic background of primordial gravitational waves. These facilities include:
\begin{itemize}
    \item constructing and operating the CMB-S4 experiment, 
    \item operating and upgrading existing gravitational wave observatories (LIGO), and 
    \item developing, constructing and operating a Stage-V spectroscopic facility 
\end{itemize}
Research into enabling capabilities includes:
\begin{itemize}
    \item Research into theory with a program that is aligned with the major facilities described above and encompasses a continuum of research including: theoretical model building, predicting and calculating new observable phenomena, modeling and simulating astrophysical and cosmological signals, and building analysis pipelines.
    \item Investing in new technologies to provide the needed technical foundation to execute the next major facilities in 2035+. These technologies include developing new CMB detectors and instrumentation; developing new technologies for future gravitational wave observatories (e.g.\ CBE); and developing technologies for long-wave intensity mapping surveys including 21-cm and mm-wave. This technology development will include fielding smaller-scale instruments to provide a staged approach to developing the needed technical maturity for executing a major survey in the next decade.
\end{itemize}
The research program described in this report is ambitious, which reflects the excitement and discovery potential of early universe studies.

\section{Introduction}
\label{sec:intro}

Fourteen billion years ago, the Universe began, generating particles and planting the seeds that would later develop into the galaxies and large-scale structure we measure today. During the first fraction of a second, the Universe conducted the most extreme high-energy physics experiment ever.  That experiment provides us with a unique window on two important areas of interest: inflation and particle relics from the Hot Big Bang. 

\textit{Inflation --- } The leading paradigm to describe the first moments is inflation, characterized by rapid, accelerated expansion at energies potentially as high as the scale of grand unification. The violent expansion generates gravitational waves and imprints specific features into the primordial density field.  This report presents observational targets for three important signatures
of inflation: primordial gravitational waves, primordial non-Gaussianity, and primordial
features. If we detect these gravitational waves, we will have indirectly observed quantum fluctuations in the spacetime metric and thus the quantum nature of gravity. 
We will have learned about high-energy physics more generally, for example, by constraining
axion physics and moduli, the fields that control the shapes and sizes of the internal
manifolds in string theory. In additional, primordial features and non-Gaussianity reveal the dynamics, particle content and interactions that govern the inflationary epoch. Combining theoretical advances, new analysis techniques, and tremendous increases in raw sensitivity, upcoming and planned surveys offer the potential for dramatic discoveries about the nature of cosmic acceleration
in the very early universe, and will probe physics on the smallest scales and at the highest energies.

\textit{Relic Radiation --- } 
Many well-motivated extensions of the Standard Model (SM) predict the existence of yet-unknown relic radiation (e.g.\
light species, gravitational waves, axions). The hunt is more
than a blind search. There are many models that aim to explain the
physics of the dark sector, address the strong CP problem, solve the hierarchy problem, and
account for short baseline neutrino anomalies share a feature. Many of these models contain new light degrees of freedom that upcoming cosmological observations 
can detect or severely constrain. In addition, phase transitions associated with symmetry breaking at a variety of energy scales may generate gravitational waves measurable by next-generation observatories.  

\begin{figure}
    \centering
    \includegraphics[width=6in]{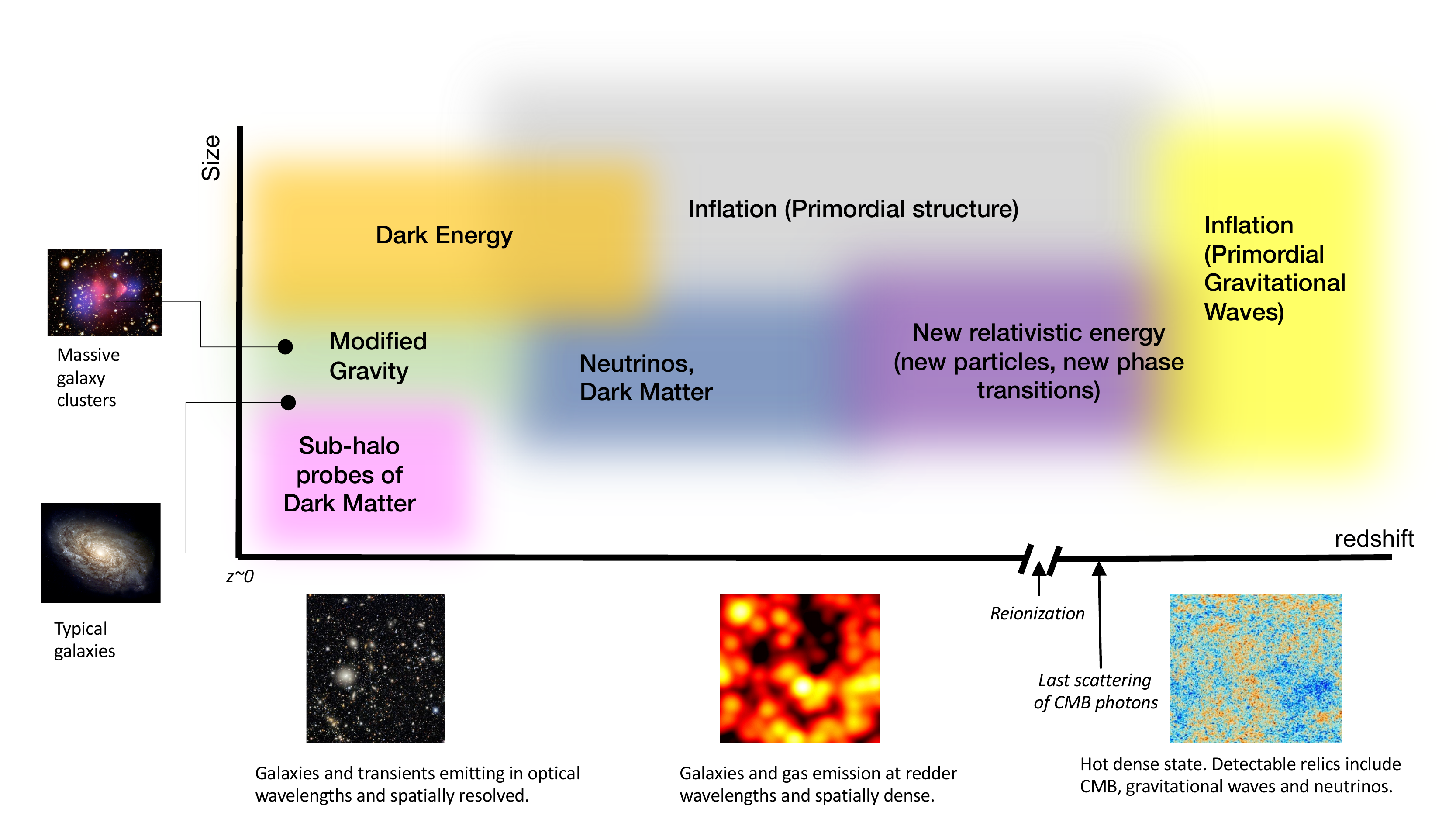}
    \caption{The nature of gravity, Dark Matter, inflation, and Dark Energy can be explored using cosmological surveys spanning different ranges in redshift (x-axis) and spatial scale (y-axis). Physics associated with the early universe (e.g. inflation and relics) can be explored with high redshift techniques (e.g. CMB, GWO) and large volume galaxy surveys.}
    \label{fig:CF5cartoon}
\end{figure}

Measurements of inflation and light relics follow three broad classes of complementary observations: Cosmic Microwave Background (CMB) surveys, Large-Scale Structure (LSS) surveys, and measurements by Gravitational Wave Observatories (GWOs).
For inflationary gravitational waves in the nearer term, CMB measurements  can detect the signature in the B-mode polarization power spectrum.  In the longer term, GWOs can probe the waves directly and constrain or measure the gravitational-wave amplitude (a probe of inflation's energy scale) and shape of the spectrum (a probe of the particular inflationary model).   
Interactions during inflation leave evidence in the non-Gaussianity of primordial fluctuations. The CMB records the two-dimensional projection of the fluctuations, while three-dimensional LSS surveys  will likely deliver large improvements in the constraining power, in particular with optical galaxy surveys and line intensity mapping (in 21\,cm or other lines). 
For early Universe, light-relic particles outside the Standard Model, CMB and LSS measurements provide indirect probes via the shape of the power spectra.  GWOs can directly probe a gravitational-wave component generated during phase transitions.

In the near term, we would target large surveys that are technically ready and have an established track-record of success: CMB (CMB-S4), LSS (DESI, LSST, WFMOS), and GWOs (LIGO/Virgo). This generation of experiments will make definitive measurements of or constraints on the amplitude of gravitational waves from inflation, and place interesting constraints on the abundance of light relics.  
In parallel with this large survey program is an R\&D program advancing new technologies to further these goals beyond 2035, in particular in new measurements of LSS via line intensity mapping and next generation GWOs.  
In concert and aligned with this experimental program is a research program across theory, computing, and analysis.

\section{Inflation}
\label{sec:pGW}

\subsection{Introduction}
The current leading scenario for the origin of structure in our Universe is cosmic inflation, a period of accelerated expansion prior to the hot big bang, as discussed in the dedicated Snowmass 2021 White Papers~\cite{Snowmass2021:Inflation,Snowmass2021:TheoryCosmo}, also see \cite{Inflation-NonG,Inflation-PowerS}. Quantum fluctuations during inflation were blown up to large scales, and manifest as density perturbations in the hot particle plasma that followed -- these perturbations are visible in the primordial power spectrum. Eventually, the density perturbations developed into the structure (galaxies, clusters of galaxies, the cosmic web) in the Universe today. If gravitational waves from inflation are measured in the CMB B-mode spectrum at the level achievable by CMB-S4, inflation would have occurred near the energy scale associated with grand unified theories, thus a detection would provide evidence for new physics at energy scales far beyond the reach of any terrestrial collider experiment.

Inflation is defined by two fundamental properties. First, it is a period of of nearly exponential expansion, such that the expansion rate $H(t) = \dot a/a$ is nearly constant, $|\dot H| \ll H^2$. Second, inflation includes a physical degree of freedom that behaves like a clock, effectively telling the universe when to end inflation.  Like any clock, it is subject to errors from quantum fluctuations giving rise to the density fluctuations.

\begin{figure}[t]
\centering
\includegraphics[width=3.5in]{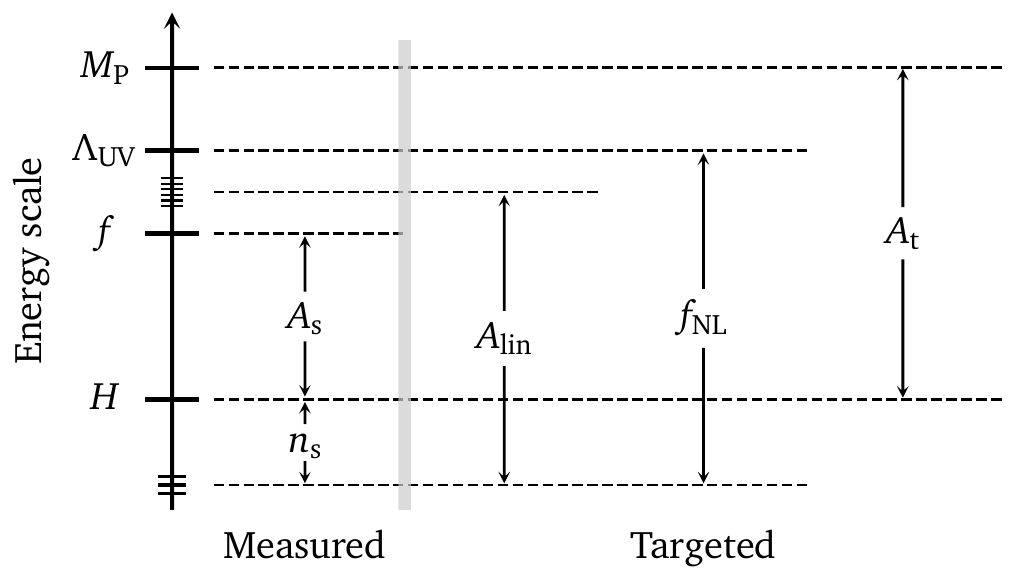}
\caption{
Sketch connecting the known energy scales relevant to inflationary cosmology to observables that have been measured and will be targeted in the next decade. The energy scales in descending order are the Planck scale, $M_p$, the scale beyond which the scalar fluctuations become strongly coupled, $\Lambda_{\textrm{UV}}$, the scale which controls the size of the scalar primordial fluctuations, $f$ , and the Hubble scale, $H$, during inflation. Additional scales that are determined by observations, but are more model dependent, are also included. The amplitude of the scalar power spectrum of initial fluctuations, $A_s$, and its spectral tilt, $n_s$, have already been measured. In addition, we indicate primordial features, primordial non-Gaussianity and primordial tensor modes by the amplitude of linear oscillatory features, $A_{\textrm{lin}}$ (as a proxy for more general features), the relative bispectrum amplitude, $f_{\textrm{NL}}$, and the tensor amplitude, $A_t$, respectively. A detection of these three prime observables, which are targeted in the next decade, will be sensitive to higher energy scales. Reproduced from~\cite{Snowmass2021:Inflation}. }
\label{fig:inflation}
\end{figure}

The physics of inflation is characterized by energy scales that are relevant for various physical processes, illustrated in Figure~\ref{fig:inflation}. In practice, what we observe are dimensionless ratios involving the Hubble scale during inflation, $H$, and these other physical scales.  Measurements to date determine the scale of the background evolution $f$, which in the case of conventional slow roll inflation is given by the speed of the background scalar field, $f^2 = |\dot \phi|$.  The spectral tilt, $n_s$, encodes the time evolution of inflationary parameters such as $\epsilon\equiv-\dot H / H^2 \ll 1$.  

Future probes of inflation target three parameters, 
$A_{\rm{t}}$, $f_{\rm NL}$ and $A_{\rm lin}$.  The first parameter, $A_{\rm{t}}$, is the amplitude of primordial gravitational waves and is typically fixed in terms of $H$ and the Planck scale, $M_{\rm pl}$,
\beq
A_t = \frac{1}{2\pi^2} \frac{4 H(t_\star)^2}{M_{\rm pl}^2}
\eeq
where $t_\star$ is a reference time (which is usually translated to a pivot scale $k_\star$ by $k_\star=a(t_\star) H(t_\star)$). This parameter is often expressed in terms of the tensor to scalar ratio, $r \equiv A_t/A_s$. 

In conventional slow-roll, $f^4 = - 2 M_{\rm pl}^2 \dot H= \dot \phi^2$, $3 H^2 M_{\rm pl}^2  \approx V(\phi)$, and $r = - 16 \dot H/H^2= 16 \epsilon $.

The second parameter, $\fnl$, is the amplitude of primordial non-Gaussianity associated with a particular three-point function, or bispectrum. Non-Gaussianity is a direct reflection of the mechanism of inflation, particle content, and interactions during the inflationary era that can be inferred from the amplitude of the particular non-Gaussian `shapes' in the bispectrum. Deviations from non-Gaussianity come in many shapes, and the amplitudes of three typical shapes are given by $\fnlloc$ (local shape), $\fnleq$ (equilateral shape), and $\fnlorth$ (orthogonal shape), described in more detail below.

The third parameter, $A_{\rm lin}$, is a characteristic scale of features in the inflationary power spectrum (and/or higher-point correlators). This is a signal of the breaking of scale invariance, which indicates that there is a characteristic time-scale associated with inflation beyond $f$. The precise form of the features encodes the physics responsible.

\subsection{Observable: Amplitude of Primordial Gravitational Waves, \texorpdfstring{ $r \equiv \frac{A_{\rm t}}{A_{\rm s}}$}{r} }
\label{sec:inflation_r}

During inflation, quantum fluctuations were imprinted on all spatial scales in the Universe. These fluctuations seeded the density perturbations that developed into all the structure in the Universe today. While 
there are still viable alternative models for the early history of the Universe, 
the simplest models of inflation are exceptionally successful in describing the data. 

Tantalizingly, the observed scale dependence of the amplitude of density perturbations has quantitative implications for the amplitude of primordial gravitational waves, commonly parameterized by $r$, the ratio of fluctuation power in gravitational waves to that in density perturbations. All inflation models that naturally explain the observed deviation from scale invariance and that also have a characteristic scale equal
to or larger than the Planck scale 
predict $r \gtrsim 0.001$.  
The observed departure from scale invariance is a potentially important clue that strongly motivates exploring down to $r = 10^{-3}$. 

For these simple models of inflation, the tensor to scalar ratio can be related to the energy scale of inflation: 

\beq
V^{1/4} = 1.04\times10^{16}~\mathrm{GeV}~\left( \frac{r}{0.01} \right)^{1/4}
\eeq

This highlights that for a tensor-to-scalar ratio within reach of CMB observations, inflation would have occurred near the energy
scale associated with grand unified theories, thus a detection would provide evidence for new physics at energy densities far beyond the reach of any terrestrial experiment. In addition, this is the power spectrum associated with quantum fluctuations in
the metric. A detection of this signal would therefore provide evidence for the quantum nature of gravity.

%%%%%%%%%%%%
\subsubsection{Cosmic Microwave Background}
\begin{figure}[h!]
\centering
\includegraphics[width=6in]{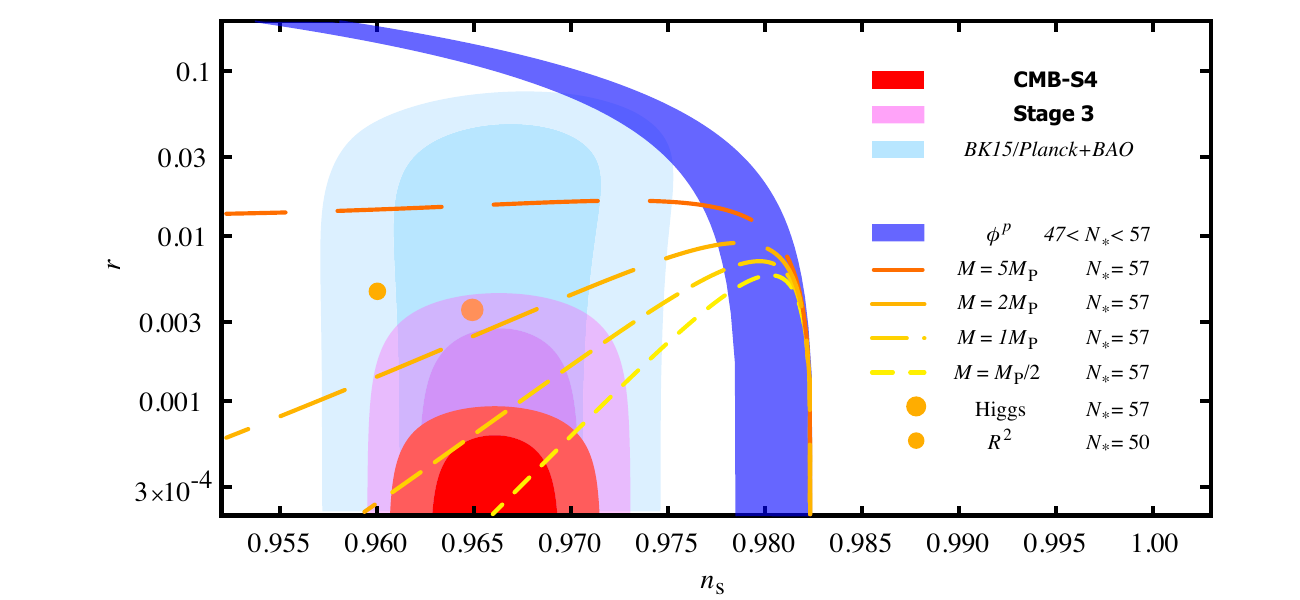}
\caption{Forecasts and constraints on primordial gravitational waves from inflation, via measurements of the CMB B-mode polarization signal. The current best constraints from a combination of the BICEP2/Keck Array experiments and Planck is shown in light blue. Projected ``Stage-3'' constraints from the South Pole Observatory and Simons Observatory are shown in purple. The red region shows projected constraints for CMB-S4. The orange circles correspond to the Starobinsky model and Higgs inflation. The lines show classes of models that naturally explain the observed value of $n_s$ with various ``characteristic scales,'' $M$. The Planck scale plays an important role because the gravitational scale and the characteristic scale share a common origin. The number of e-folds, $N_{*}$, chosen for the figure corresponds to nearly instantaneous reheating, which leads to the smallest values for $r$ for a given model. Reproduced from~\cite{2019arXiv190704473A}.}
\label{fig:Bmodes}
\end{figure}

Primordial gravitational waves (PWGs) from inflation leave an imprint in the temperature and polarization anisotropies. In particular, PGWs generate divergence-free (parity-odd) B-mode polarization and are the only source of B-modes at recombination at
linear order. As such, B modes of the CMB provide a unique window to PGWs \cite{1997PhRvL..78.2058K,1997PhRvL..78.2054S}, typically targeting scales of $\theta\sim1^{\circ}$ from the ground. Given the significance of the implications of a detection of PWGs,
many current- and next-generation experiments are designed to go after this `B-mode signature' in the CMB \cite{2021PhRvL.127o1301A, 2020PhRvD.101l2003S,2011ApJS..192...18K,2012ApJ...760..145Q,2016SPIE.9914E..1KH,2018JCAP...09..005K,2020A&A...641A...6P,2020ApJ...897...55P,2021arXiv210313334S}. In the last $\sim$10 years, the uncertainty on $r$ has tightened by about two
orders of magnitude~\cite{2021PhRvL.127o1301A}. Looking forward, the search for CMB B modes will continue to advance through new experiments across complementary ground-based, sub-orbital, and satellite platforms~\cite{Chang:2022tzj}. 
For ground-based facilities, 
upcoming experiments such
as the Simons Observatory~\cite{2019JCAP...02..056A} and South Pole Observatories~\cite{2018SPIE10708E..07H,2022ApJS..258...42S} are projected to cross an important threshold: $r<0.01$, which is associated with monomial
models and a super-Planckian excursion in field space that would provide strong evidence for
the existence of an approximate shift symmetry in quantum gravity. However, to reach $r<0.001$, which is associated with the simplest models of inflation that naturally predict the
observed value of the scalar spectral index $n_{s}$ and have a characteristic scale that exceeds the
Planck scale requires an experiment at the scale of CMB-S4, as shown in Figure~\ref{fig:Bmodes}.

%%%%%%%%%%%%
\subsubsection{Gravitational Wave Observatories}
\begin{figure}[h!]
\begin{center}
\includegraphics[width=6in]{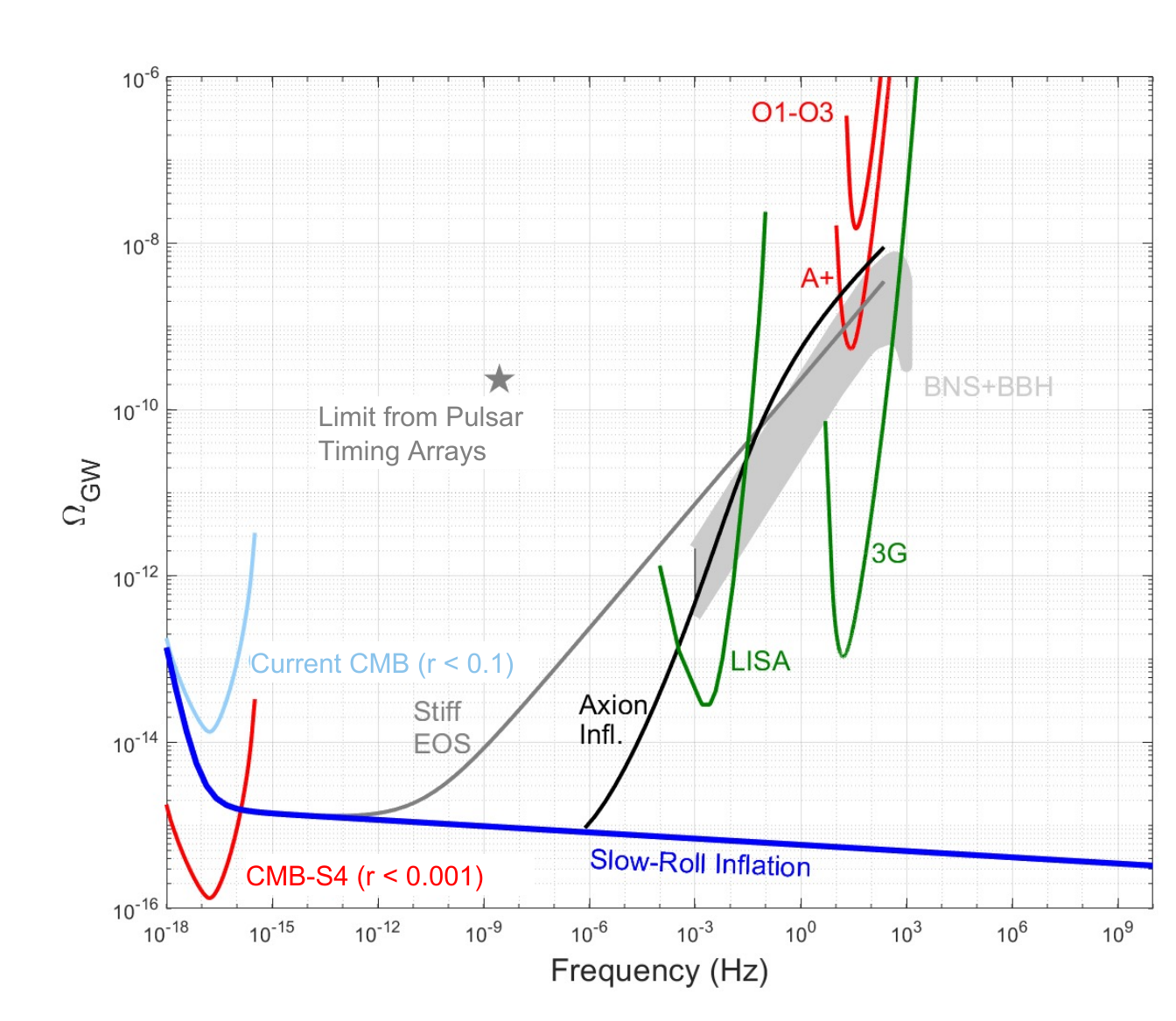}
\end{center}
\caption{Landscape of gravitational wave cosmology. Experimental results include: O1-O3 LIGO-Virgo upper limits~\cite{PhysRevD.104.022004}, CMB limits~\cite{PhysRevX.6.011035}, and Parkes pulsar timing limit~\cite{PhysRevX.6.011035}, as well as projected sensitivities of the third generation (3G) terrestrial GW detectors~\cite{EinsteinTelescope,CosmicExplorer}, LISA~\cite{LISA} and CMB-S4. Theoretical models include examples of slow-roll inflation~\cite{turner}, Axion Inflation~\cite{peloso_parviol}, hypothetical stiff equation of state in the early universe~\cite{boylebuonanno}, and foregrounds due to binary black hole/neutron stars~\cite{PhysRevD.104.022004}.}
\label{fig:landscape_inflation}
\end{figure}

The direct detection of gravitational waves through CMB polarization motivates
the opportunity for detecting these PGWs from inflation directly using GWOs. Although the standard inflationary paradigm generally predicts PGW spectra that are nearly undetectable by GWOs, the existence of new physical processes during and/or after inflation could lead to significantly stronger high frequency signal that may be detectable by upcoming experiments (see Fig.~\ref{fig:landscape_inflation}). For example, scenarios in which the inflaton couples to a gauge field or the existence of a new pre-radiation dominated epoch with equation of state $1/3 \leq w \leq 1$ can lead to strongly blue-tilted PGW spectra and the existence of a new matter-dominated era with $w=0$ or phase transitions during the inflationary epoch can lead to kinks and oscillatory features in the PGW spectrum.

The combination of CMB and GWO provides a complementary suite of measurements of inflationary PGWs and their spectrum. Although the favored focus of these observations is inflationary physics, we note that the combination of measurements are also probes of alternatives to the inflationary hypothesis. For example, the pre-Big Bang~\cite{Gasperini:1992em} model would generate a red PGW spectrum with additional power at higher frequencies within the design sensitivity of aLIGO/Virgo and/or LISA~\cite{Gasperini:2016gre}; the ekpyrotic model~\cite{Khoury:2001wf} predicts a very blue PGW specturm with negligible gravitational waves on cosmological scales; the string gas cosmology~\cite{Brandenberger:1988aj,Battefeld:2014uga} predicts a slightly blue spectrum (compared to the slightly red spectrum generated by canonical inflation); and the matter bounce scenario~\cite{Finelli:2001sr,Brandenberger:2012zb} would generate large gravitational waves, such that $r\sim1$, which is a prediction in tension with current CMB measurements.

%%%%%%%%%%%%
\subsection{Observable: Interactions during Inflation, \texorpdfstring{$f_{\rm NL}$}{fNL}   }
\label{sec:inflation_fnl}
%%%%%%%%%%%%

\begin{figure}[ht!]
\centering
\includegraphics[width=4.5in]{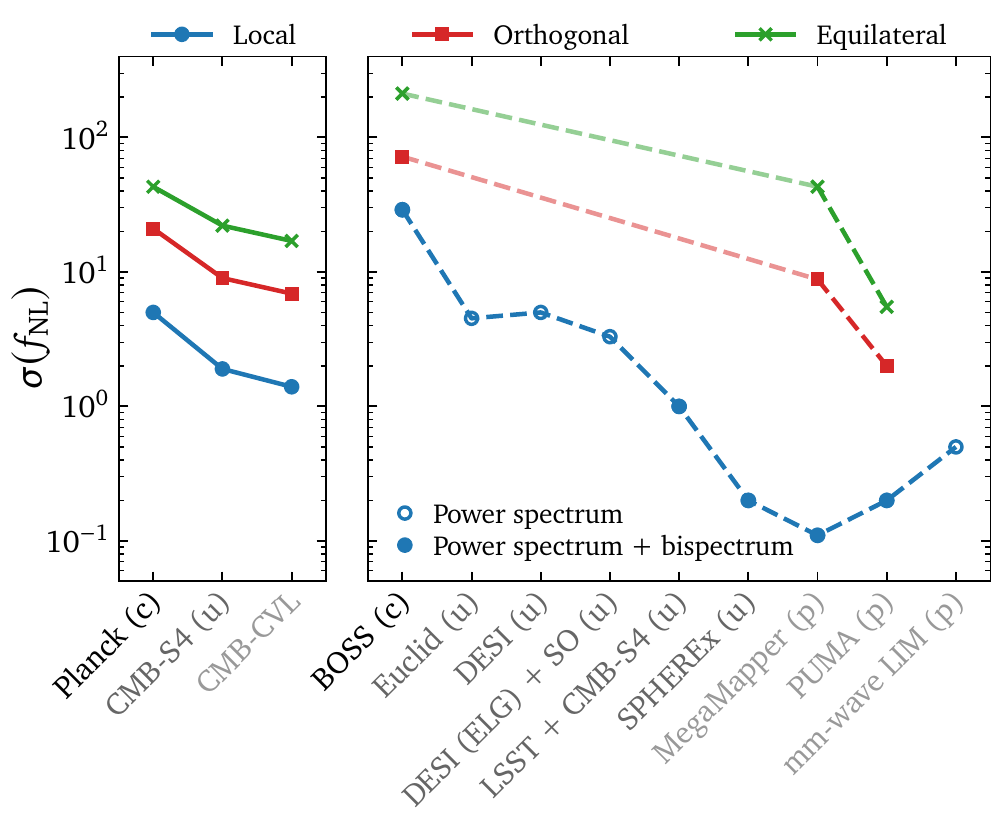}
\caption{A comparison of constraints on three types of primoridal non-Gaussianity from a small subset of completed (‘c’), upcoming (‘u’) and proposed (‘p’) experiments. We also forecast a cosmic-variance-limited (CVL) CMB experiment up to $\ell^{\mathrm{T}}_{\mathrm{max}}$ = 3000 and $\ell^{\mathrm{P}}_{\mathrm{max}}$= 5000, but note that these limits could be further improved by the use of delensing or the inclusion of Rayleigh-scattering anisotropies. The numbers for the (e)BOSS, Euclid, DESI and mm-LIM assume only power spectrum information. The DESI+SO forecast includes the cross-correlation between the ELG sample of DESI and SO SZ maps. The forecast for LSST+CMB-S4 considers the power spectra and the bispectra, including the cross-correlations between galaxy and lensing maps to remove sample variance. All other LSS probes are forecasted including the bispectrum. There are two important caveats to the results shown here: (i) scale-dependent bias measurements hinge on the ability to measure the largest scales at high precision and most
of the underlying forecasts contain only a limited assessment of the impact of observational systematics; (ii) while much theoretical progress has been made in recent years, there remains a large degree of uncertainty over several aspects of these forecasts and, therefore, the achievable constraints may be better or worse as these issues are resolved. Reproduced from~\cite{Snowmass2021:Inflation}. }
\label{fig:pNG}
\end{figure}

Quantum fluctuations of the inflaton or, equivalently, the scalar mode of the metric would change the amount of inflation that occurred in different parts of the universe, and thus correspond to a change in the physical energy densities from place to place.  After reheating, these density perturbations grow and evolve as they re-enter the cosmic horizon.  This gives rise to a primordial spectrum of fluctuations, $\zeta(\vec k)$ where where $\vec k$ is a comoving wave-number, or the comoving scale of the fluctuations. In the simplest single-field inflationary models, these fluctuations only self-interact gravitationally, and deviations from Gaussianity (primordial non-Gaussianity, or PNG), would be evidence for additional interactions present in the early Universe. Measuring PNG will thus allow us to answer fundamental questions, such as: How did the background evolve and is it consistent with slow-roll inflation scenarios? Are there additional scalar fields at play during inflation? If so, how did they evolve and interact? Were there heavy fields ($\sim$ Hubble scale) present during inflation?

There are various measures of non-Gaussianity. We focus on the scalar three-point correlation function, the bispectrum. It has been the most studied and analyzed observable in the
literature, because it often is the dominant non-Gaussian signature in weakly coupled models
of inflation. For translational, rotational and scale-invariant perturbations, the bispectrum is:

\beq
\langle \zeta(\vec k_1)\zeta(\vec k_2)\zeta(\vec k_3) \rangle \propto f_{\rm{NL}}^{\rm{type}} \delta^{3}(\vec k_1+\vec k_2+\vec k_3)\frac{S^{\rm{type}}(k_{1},k_{2},k_{3})}{k_{1}k_{2}k_{3}}  \ 
\eeq

\noindent Here, $f_{\rm{NL}}$ parameterizes the size of PNG and the dimensionless shape function $S^{\rm{type}}$ controls the
overall size of PNG as a function of the triangle formed by the momenta. The shape dependence
encodes information about the specific dynamical mechanism that generated the non-Gaussian
signal, thus serving as a discriminator between various inflationary models.
Studies of various inflationary models have demonstrated several broad classes of scale invariant PNG with large, potentially detectable $f_{\rm{NL}}$. We list them below with emphasis on the physics that they probe:

\begin{itemize}
\item `Local' non-Gaussianity and multi-field inflation ($\fnlloc$): This is also defined by a large signal in the squeezed limit ($k_3 \to 0$). The single-field consistency conditions forbid any contribution in this region in the absence of extra particles; the squeezed limit is therefore an excellent probe of single- versus multi-field inflation. Non-Gaussian correlators of this type are particularly sensitive to the mass and spin of additional fields in addition to being a probe of light fields.
\item Orthogonal non-Gaussianity ($\fnlorth$): This is the flattened limit ($k_1 = 2 k_2 =2 k_3$). An enhanced signal in the flattened limit is associated with excited states (with respect to the Bunch-Davis vacuum). 
\item `Equilateral' non-Gaussianity and single-field inflation ($\fnleq$, where $k_1 \sim k_2 \sim k_3$):  Both $\fnleq$ and $\fnlorth$ get large contributions from this region. This is typical of self-interactions of the inflaton and is often used as a test of canonical single-field slow-roll inflation which predicts $\fnleq,\fnlorth<1$. Interactions between the inflaton and heavy fields also contributes significantly in this region. A detection of equilateral non-Gaussianity without a large $\fnlorth$ is a signal of the quantum origin of structure.
\end{itemize}

These three PNG bispectrum shape estimators provide a mechanism to use non-Gaussianity to characterize inflation. Because there are many more shapes that may test for non-Gaussian signatures, this bispectrum-based framing is a significant under-estimate of the opportunity from PNG measurements to probe inflation.

To measure PNG, we note that these quantum fluctuations would change the amount of inflation that occurred in different parts of the universe, and thus correspond to a change in the physical energy densities from place to place.  After reheating, these density perturbations grow and evolve as they re-enter the cosmic horizon.  As a result, these fluctuations would eventually appear as temperature anisotropies in the CMB and also dictate where structure would preferentially form, leading to a connection between the primordial fluctuations and structure formation. The different components of the energy density and their fundamental interactions shape this subsequent evolution, giving rise (at linear order) to:
\beq
\delta_{i}(\vec k,z) = T_i(k,z) \zeta(\vec k)
\eeq
where $\vec k$ is again comoving wave-number, $k= |\vec k|$, $\zeta(\vec k)$ is the scalar metric fluctuation, $\delta_{i}(\vec k,z)$ is the density contrast of species $i$ and $T_i(k,z)$ is its transfer function. As a result, any level of non-Gaussianity in the statistics of the primordial fluctuations ($\zeta(\vec k)$) will be transferred to the maps of the cosmic microwave background and large-scale structure. 

The best constraints on non-Gaussianity have come from the CMB, as shown in Figure~\ref{fig:pNG}, and CMB-S4 can improve on the $\fnl$ parameters by a factor of a few before reaching a fundamental floor based on the number of modes available in the two-dimensional sky area available to the experiment.   
Because large scale structure surveys have access to three-dimensional volumes of modes, in principle they can significantly improve upon the CMB constraints, in particular for $\fnleq$ and $\fnlorth$. As a result, LSS surveys including 21cm intensity mapping (e.g. PUMA) and optical galaxy surveys (e.g. MegaMapper) expect to significantly improve upon the CMB. The power of these surveys are shown in Figure~\ref{fig:pNG}: the CMB essentially `ends' on the left-hand side, and future LSS surveys project dramatic improvements. Although they can access far more modes than are available in the CMB, these surveys observe the primordial power spectrum through a more complicated matter transfer function $T_{m}(k,z)$ that includes non-linear local gravitational effects. Non-linear effects become worse at lower redshift because there has been more time for structure to evolve based on its local gravitational environment, and at smaller scales where the local environment operates more efficiently, thus the limitations of LSS surveys at redshifts $z<6$ will ultimately be related to the modeling of small-scale non-linearities. LSS surveys at higher redshifts, where the signal is more pristine ($z>10$, `Dark Ages'), may be uniquely possible using 21\,cm intensity mapping techniques but with more complex instrument requirements (such as deploying on the lunar surface to avoid human-generated radio frequency interference). 

Local non-Gaussianity ($\fnlloc$) and other shapes with a contribution in the squeezed limit are made easier to observe through their non-local effect on the formation of halos, namely scale-dependent bias. The halos essentially form in proportion to the Newtonian potential in a way that causes large changes to the power spectrum at small $k$ (large distances), well away from the nonlinear regime. This effect can even be measured without cosmic variance, either from using multiple populations of halos or by cross correlating with the matter density inferred from gravitational lensing (e.g. CMB lensing). Forecasts show an order of magnitude or more improvement in $\fnlloc$ is realistic for a number of surveys and could be enhanced by cross-correlations with the CMB-S4 lensing maps.

%%%%%%%%%%%%
\subsection{Observable: Inflationary Potential, \texorpdfstring{ $A_{\rm{lin}}$}{Alin} }
\label{sec:inflation_alin}
%%%%%%%%%%%%
\begin{figure}[ht!]
\centering
\includegraphics[width=4.5in]{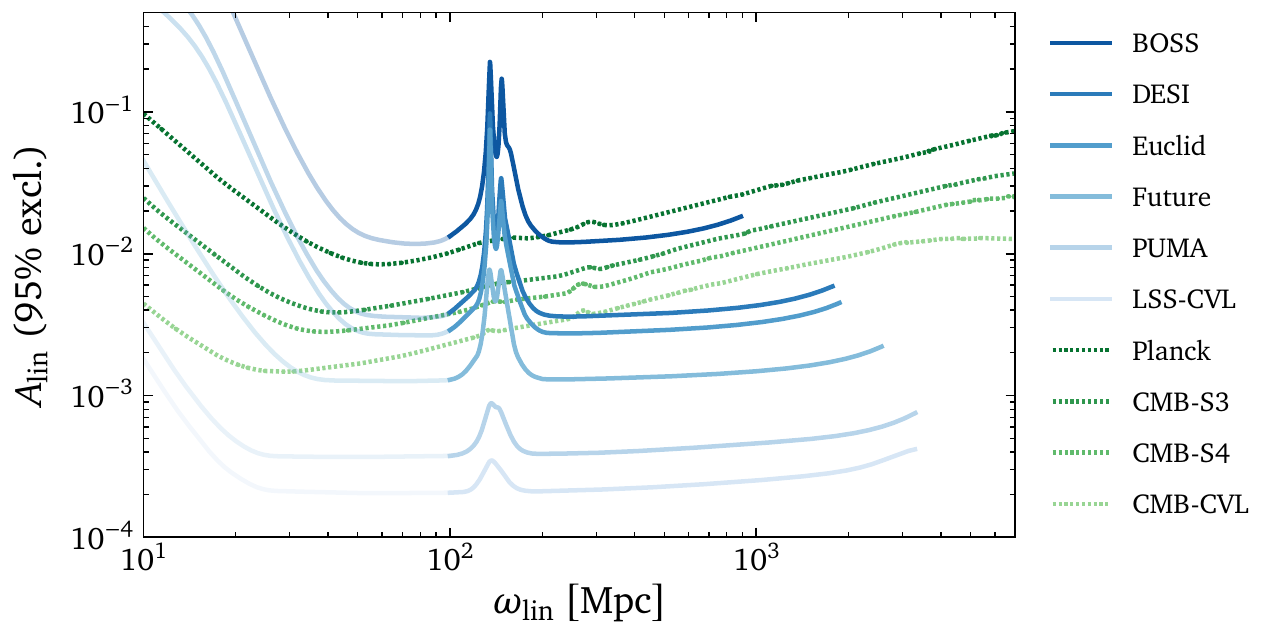}
\caption{Forecasted sensitivity for the ‘feature spectrometer’ of linear features. The potential reach of various CMB (dashed) and LSS (solid) experiments to constrain the feature amplitude $A_{\mathrm{lin}}$ at a confidence level of 95\% (under the assumption the true amplitude is zero) is presented as a function of their frequency $\omega_{\mathrm{lin}}$. The positive semi-definite nature of $A_{\mathrm{lin}}$ is taken into account in the displayed estimates. The underlying experimental specifications for planned surveys are similar to those of BOSS, DESI, Euclid, Planck, a CMB-S3-like experiment, and CMB-S4, respectively. To illustrate the potential future reach, we also show the expected
sensitivity of a future LSS survey with $10^{8}$ objects up to redshift $z_{\mathrm{max}}$ = 3 over half the sky using
a maximum wavenumber $k_{\mathrm{max}}$ = 0.5$h \mathrm{Mpc}^{-1}$, which yields a slightly less sensitivity than the proposed galaxy survey MegaMapper and the proposed 21 cm experiment PUMA. In addition, we include cosmic-variance-limited (CVL) observations of LSS
up to $z_{\mathrm{max}}$ = 6, over half the sky with $k_{\mathrm{max}}$ = 0.75$h \mathrm{Mpc}^{-1}$, and of the CMB up to $\ell^{T}_{\mathrm{max}}$ = 3000
and $\ell^{P}_{\mathrm{max}}$= 5000 over 75\% of the sky. The LSS forecasts with $\omega_{\mathrm{lin}} \leq$100 Mpc should be treated
cautiously since these low frequencies are more sensitive to the details of signal modeling. We also note that a reconstruction efficiency of 50\% was assumed which should be surpassed in
the future thanks to further theoretical developments. Overall, LSS surveys have the potential to improve over the CMB by more than an order of magnitude, while the CMB will always dominate the reach in feature frequency. Reproduced from~\cite{Snowmass2021:Inflation}, see references therein.}
\label{fig:pf}
\end{figure}

A conventional picture of a slow-rolling scalar field on a flat potential is consistent with observations. In this context, the measurement of $n_s \simeq 0.97$ is consistent with a mild time dependence of the inflationary background due to the slope of the potential.  However, observations do not forbid more dramatic deviations from scale invariance that can arise from sharp or oscillatory features in the potential, particle production events, and more.

Like deviations from Gaussianity, departures from scale invariance can come in many forms.  Observations of the CMB and LSS certainly forbid power laws for the power spectrum that are dramatically different from $k^{-3}$ at cosmological scales.  That still leaves three much less constrained possibilities: (i) oscillatory features in the power spectrum (ii) power-law changes to the power spectrum on small scales and/or (iii) scale-dependent non-Gaussian correlators.

Oscillatory features are a particularly well-motivated target for several reasons.  First, they arise naturally within a variety of microscopic models of inflation.  The flatness of the potential required for successful inflation can be broken to a discrete symmetry by non-perturbative corrections.  These periodic corrections to the potential give rise to logarithmically-spaced oscillations in the power spectrum.  Alternatively, particle production can give rise to linear oscillations.  Phenomenologically, these features can also evade current constraints without hiding the signal in modes that will be difficult to measure.  In fact, these signals are visible even on nonlinear scales, thus allowing for strong constraints from current and future galaxy surveys.

Although there is no specific theoretical target for these kinds of features, future observations can make dramatic improvements in sensitivity that present significant discovery potential.  The forecasts are shown in Figure~\ref{fig:pf} for the amplitude of linear oscilations, $A_{\rm lin}$, in the matter or CMB power spectrum.  We see that next generation large scale structure experiments like PUMA or MegaMapper could improve on current constraints by up to a factor of 100.  Furthermore, unlike other LSS signals from inflation, current analyses have already reached CMB sensitivity and produced constraints on $A_{\rm lin}$ at the level of these forecasts.

\section{Relics of the Hot Big Bang} 
\label{sec:RE}
In addition to the direct inflationary signatures described in the section above, the early universe presents another avenue for exploring new fundamental physics through the measurement of relic radiation. In the simplest narrative of the early universe, the only relic radiation (apart from CMB photons) is the Cosmic Neutrino Background. Because the Standard Model precisely predicts this neutrino energy density, measurements of relic radiation from the early universe have tremendous potential--- measuring any departure from the predicted neutrino background would be a clear sign of new physics. This is discussed in the Snomwass 2021 White Paper~\cite{Dvorkin:2022jyg} and references therein.

The landscape for new physics is broad ranging from around $\mathcal{O}(1)$ MeV, which is constrained by Big Bang Nucleosynthesis (BBN), up to at least $\mathcal{O}(10^{16})$ GeV. For example, Standard Model extensions with new light degrees of freedom can lead to thermal relic radiation. A stochastic background of gravitational waves and its associated spectrum will carry the imprint of the pre-BBN era, phase transitions in the early universe, and of any particle production associated with pre-heating/reheating. Similarly, axion-like particles generate additional relativistic degrees of freedom and potentially carry the imprint of the inflationary epoch. Measuring and constraining relic radiation from the early universe presents a unique tool for exploring high energy physics complementing collider-based approaches, which are sensitive to different energies and scales.

\begin{figure}[!t]
\begin{center}
\includegraphics[width=6in]{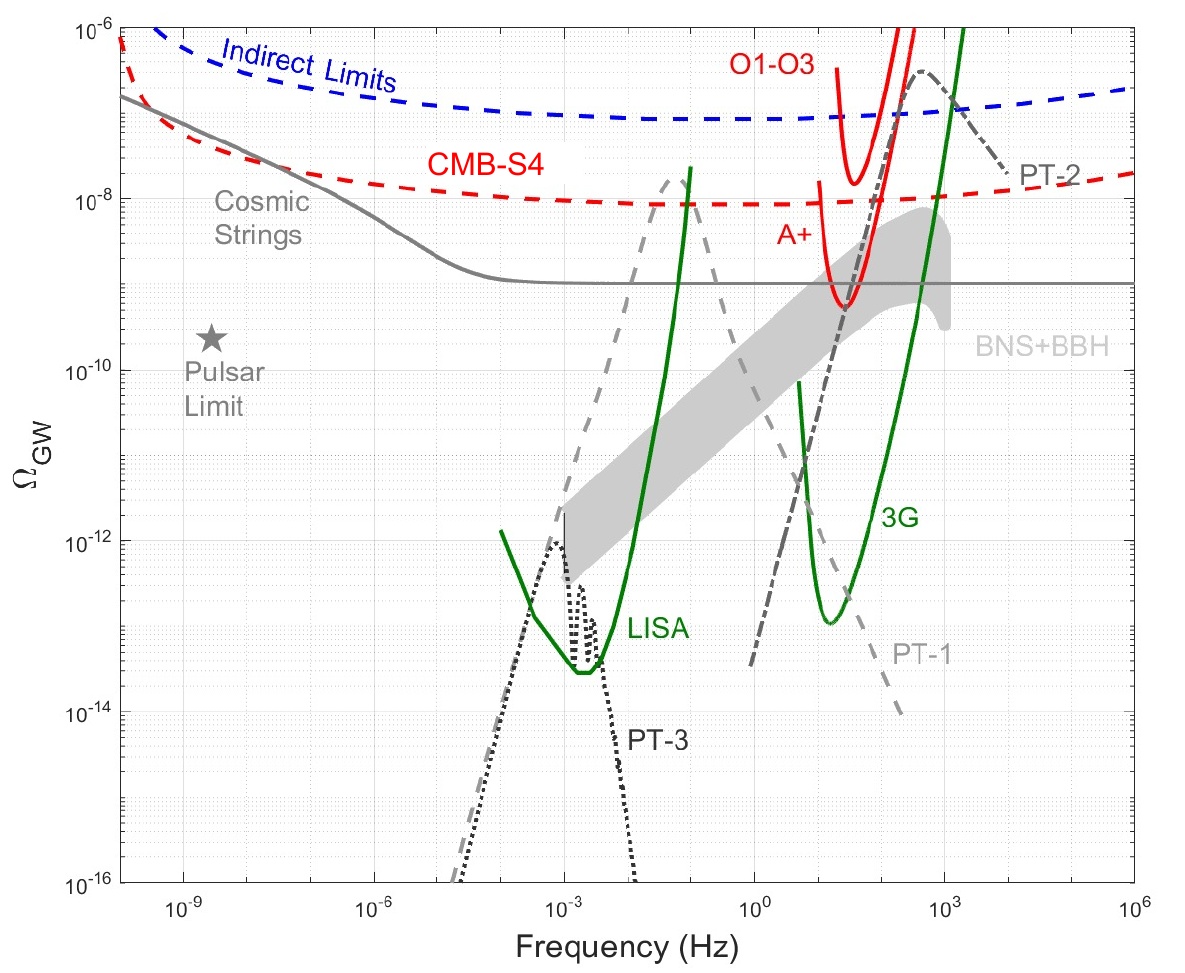}
\end{center}
\caption{Same as Fig. \ref{fig:landscape_inflation}, but focused on models of phase transitions and topological defects. Theoretical models include examples of first-order phase transitions (PT-1~\cite{Ellis:2019oqb}, PT-2~\cite{DelleRose:2019pgi}, and PT-3~\cite{An:2022}), cosmic strings~\cite{Siemens:2006yp}, and foregrounds due to binary black hole/neutron stars~\cite{PhysRevD.104.022004}. Also included is the indirect limits from big bang nucleosynthesis~\cite{PhysRevX.6.011035} and corresponding projections from CMB-S4. }
\label{fig:landscape_PT}
\end{figure}

\subsection{Light Relics}

After reheating, all the components of the Standard Model were in equilibrium, likely at temperatures well above the weak scale, $T \gg 1$ TeV.  At these enormous temperatures, any number of additional (beyond the Standard Model) particles could also have been in equilibrium with the Standard Model and would have been produced with number densities similar to that of the photons (or any other relativistic particle). 

Particles and/or dark sectors that decouple from the Standard Model while they are relativistic will carry a large amount of entropy.  For sufficiently heavy particle, $m \gg 1$ MeV, it is possible for these particles to decay back to the Standard Model while leaving little/no observational signature.  In contrast, for $m < 1$ MeV, there will be a measurable impact on the amount of radiation in the universe.  This is particularly straight-forward for light particles, $m \ll 1$ eV.  Because light particle remain relativistic through recombination and contribute to the total amount of radiation, during radiation domination, in the same way as a neutrino.  

\begin{figure}[h!]
\centering
\includegraphics[width=4.5in]{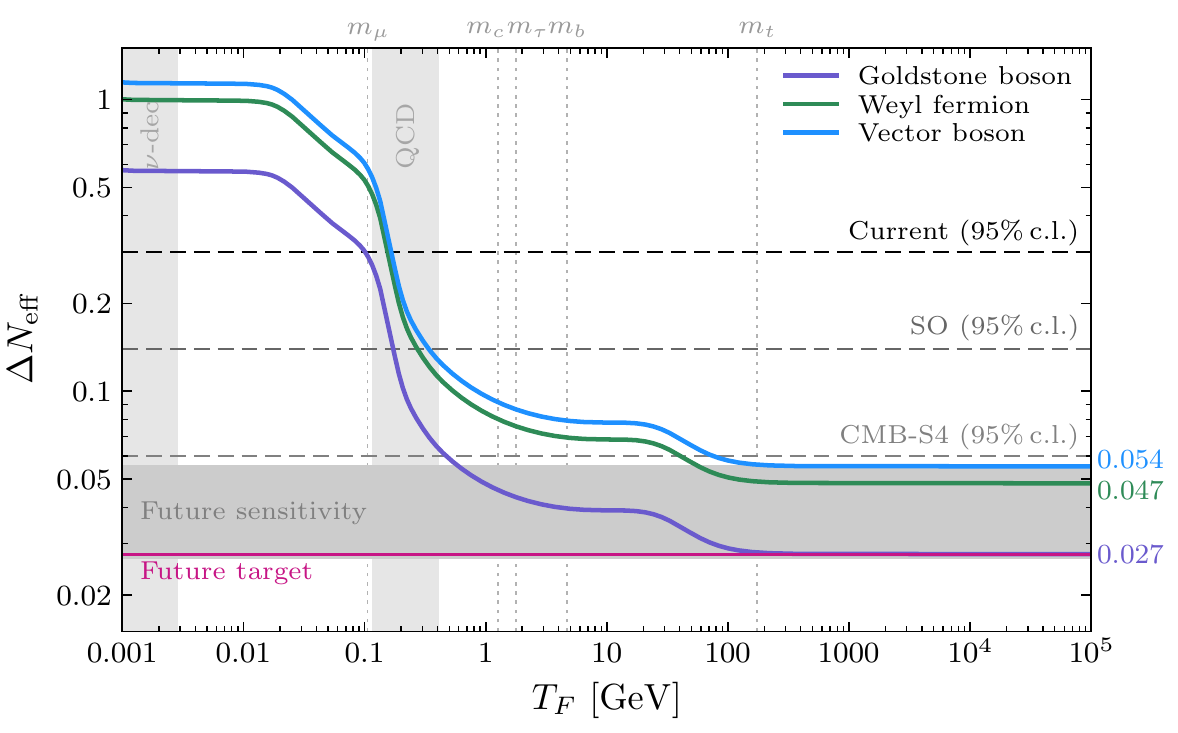}
\caption{Contributions to $\Delta \Neff$  light particle that decouple from the Standard Model at temperature $T_{\rm F}$, reproduced from~\cite{Dvorkin:2022jyg}. }
\label{fig:neff}
\end{figure}

It is conventional to define the total radiation density during this epoch as
\beq
\rho_{r}=\rho_{\gamma}\left(1+\frac{7}{8}\left(\frac{4}{11}\right)^{4 / 3} \Neff \right)
\eeq
such that $\Neff \approx 3$ in the Standard Model (reflecting the 3 species of neutrinos). In detail, the energy density of neutrinos in the Standard model is equivalent to $\Neff = 3.045$.  Additional light particles add to this energy density so that $\Neff =3.045 +\Delta \Neff $, with $\Delta \Neff > 0$.  Because there is no way to ride these sectors of their entropy after decoupling from the Standard Model, their contribution to $\Delta\Neff$ is determined by the number of degrees of freedom of the additional particle(s) and the entropy of the Standard Model at the temperature.  These universal results are shown in Figure~\ref{fig:neff}.

Current observations constrain $\Delta \Neff < 0.3$ (95\%), which probes individual particles decoupling during or after the QCD phase transition ($T_F \approx 100 MeV$). The next generation of cosmic surveys is poised to reach very exciting targets in $\Delta \Neff$.  CMB-S4 is expected to limit $\Delta \Neff < 0.06$ (95 \%) which would be sensitive to new particles with spin decoupling at $T_\approx 100$ GeV and real scalars at 1 GeV, just prior to the QCD phase transition.  The later is particularly important for axion-like particles coupling to heavy fermions, where CMB-S4 would be the most sensitive experimental or observational probe by orders of magnitude. Building off CMB-S4, more furturistic surveys like PUMA, LIM, Megamapper or high CMB-HD could reach the ambitious goal of excluding $\Delta \Neff = 0.027$ at 95\%, which would be sensitive to any particle that was in thermal equilibrium which the Standard Model at any time after reheating.

More generally, cosmic surveys are sensitive to a wide range of well motivated targets through potential contributions to $\Neff$ at recombination and/or suppression of small scale clustering at late times. Dark sectors of many varieties are motivated by solutions to the hierarchy problem, the Strong CP problem and the cosmological constant problem, and of course as models of dark matter.  These dark sectors can include both very light ($m \ll eV$) and light-but-massive $(m\approx$ eV) relics.  In some parameter regimes, these models correspond to changes to $\Neff$ and $\sum m_\nu$.  Somewhat heavier relics produce changes to the matter power spectrum that are similar to massive neutrinos but are distinguishable in principle.  

\subsection{Phase Transitions}
First order phase transitions (FOPTs) in the early Universe produce gravitational waves and offer a unique way of probing particle physics models at energy scales otherwise inaccessible. Such phase transitions could occur at nearly any time during or after inflation, and the GW spectrum, with examples shown in Fig.~\ref{fig:landscape_PT}, is sensitive to the shape of the effective potential, which depends on the symmetry breaking pattern and the particle content of the theory. This provides access to regions of parameter
space unexplored so far in various extensions of the SM. Moreover, thermal phase transitions and weak transitions source GWs with different spectral shapes (see e.g.~\cite{Hindmarsh:2017gnf,Hindmarsh:2019phv,Niksa:2018ofa,Caprini:2009yp,RoperPol:2019wvy,Jinno:2020eqg}), as illustrated in Fig. \ref{fig:landscape_PT}, allowing the possibility of reconstructing the conditions during and after the FOPT.

GWs from a strong FOPT have a plethora of motivations in the early universe. For instance, new states at the electroweak scale can catalyze a strongly first order electroweak phase transition \cite{Ramsey-Musolf:2019lsf,Profumo:2007wc,Delaunay:2007wb,Huang:2016cjm,Chala:2018ari,Croon:2020cgk,Grojean:2006bp,Alves:2018jsw,Alves:2020bpi,Vaskonen:2016yiu,Dorsch:2016nrg,Chao:2017vrq,Wang:2019pet,Demidov:2017lzf,Ahriche:2018rao,Huang:2017rzf,Mohamadnejad:2019vzg,Baldes:2018nel,Huang:2018aja,Ellis:2019flb, Alves:2018oct, Alves:2019igs,Cline:2021iff,Chao:2021xqv,Liu:2021mhn,Zhang:2021alu,Cai:2022bcf} and large lepton asymmetries or different  quark masses can make the QCD transition strong \cite{Schwarz:2009ii,Middeldorf-Wygas:2020glx,Caprini:2010xv,vonHarling:2017yew}. Beyond this, a strong transition can occur in multistep phase transitions\footnote{See Refs. \cite{Weinberg:1974hy,Land:1992sm,Patel:2012pi,Patel:2013zla,Blinov:2015sna} for the viability of a multistep phase transition} \cite{Niemi:2018asa,Croon:2018new,Morais:2018uou,Morais:2019fnm,Angelescu:2018dkk,TripletGW2022}, B-L breaking \cite{Jinno:2016knw,Chao:2017ilw,Brdar:2018num,Okada:2018xdh,Marzo:2018nov,Bian:2019szo,Hasegawa:2019amx,Ellis:2019oqb,Okada:2020vvb} (or B/L breaking \cite{Fornal:2020esl}), flavour physics \cite{Greljo:2019xan,Fornal:2020ngq}, axions \cite{Dev:2019njv,VonHarling:2019rgb,DelleRose:2019pgi}, GUT symmetry breaking chains \cite{Hashino:2018zsi,Huang:2017laj,Croon:2018kqn,Brdar:2019fur,Huang:2020bbe}, supersymmetry breaking \cite{Fornal:2021ovz,Craig:2020jfv,Apreda:2001us,Bian:2017wfv}, hidden sector involving scalars \cite{Schwaller:2015tja,Baldes:2018emh,Breitbach:2018ddu,Croon:2018erz,Hall:2019ank,Baldes:2017rcu,Croon:2019rqu,Hall:2019rld,Hall:2019ank,Chao:2020adk,Dent:2022bcd}, neutrino mass models \cite{Li:2020eun,DiBari:2021dri,Zhou:2022mlz} and confinement \cite{Helmboldt:2019pan,Aoki:2019mlt,Helmboldt:2019pan,Croon:2019ugf,Croon:2019iuh,Garcia-Bellido:2021zgu,Huang:2020crf,Halverson:2020xpg,Kang:2021epo}.

Of particular interests are phase transitions associated with physics that can is also explored by current and upcoming collider experiments. A high priority topic is electroweak symmetry breaking (EWSB), which in the Standard Model {with a 125 GeV Higgs boson}, occurs via a smooth cross-over rather than a FOPT~\cite{Kajantie:1996mn}. However, there are compelling theoretical arguments to expect new physics coupled to the SM Higgs not far from the TeV energy scale \cite{Ramsey-Musolf:2019lsf} (for example, light supersymmetric particles, such as stops or additional scalars in non minimal SUSY extensions, coupled to the SM Higgs). These new physical processes could alter the nature of the {EWSB transition} possibly making it a first order transition. The existence of such a transition is a necessary ingredient for electroweak baryogenesis~\cite{Morrissey:2012db,Barrow:2022gsu,Asadi:2022njl} and could provide a source for observable gravitational radiation. 

Another potential FOTP arises from spontaneous R-symmetry breaking \cite{Nelson:1993nf} in viable SUSY models. A study~\cite{Craig:2020jfv} has investigated the conditions where this transition can be first order, leading to GWs, and demonstrated that the resulting GW spectrum covers the frequency range accessible to current and future GW detectors. Moreover, once the SUSY breaking mediation scheme is specified, the peak of the GW spectrum is correlated with the typical scale of the SM superpartners, and a visible GW signal would imply superpartners within reach of future colliders.

\subsection{Topological Defects}

Phase transitions associated with symmetry breaking generically result in topological defects~\cite{Jeannerot:2003qv} that could have different forms~\cite{Kibble:1976sj,Vilenkin:2000jqa}. Three types of topological defects have been shown to produce SGWB: domain walls \cite{Vilenkin:1981zs,Sakellariadou:1990ne,Sakellariadou:1991sd,Gleiser:1998na,Hiramatsu:2010yz,Dunsky:2021tih}, textures \cite{Fenu:2009qf} and cosmic strings \cite{Vachaspati:1984gt,Blanco-Pillado:2017oxo,Blanco-Pillado:2017rnf,Ringeval:2017eww,Vilenkin:1981bx,Hogan:1984is,Siemens:2006yp,DePies:2007bm,Olmez:2010bi,Vachaspati:2015cma}. In all cases the amplitude of the GW signal grows with the symmetry breaking scale, implying that topological defects are effective probes of high energy physics. Furthermore, topological defects can also arise in superstring theories \cite{Sarangi:2002yt,Jones:2002cv,polchinski}, implying
that GW experiments can provide a novel and powerful way to test string theory~\cite{2006AAS...209.7413H,LIGOScientific:2017ikf}.

Among these, cosmic strings have been studied the most extensively, either as global strings (e.g. in axion dark matter models where a U(1) is broken to a vacuum with a discrete symmetry \cite{Chang:2019mza}) or as local strings in symmetry breaking chains that result from SO(10) breaking to the SM \cite{Dunsky:2021tih}. 
Considering all possible spontaneous symmetry breaking patterns from the GUT down to the SM gauge group, it was shown \cite{Jeannerot:2003qv} that cosmic string formation is unavoidable. 

Here we briefly highlight the SGWB production by local cosmic strings (for a more extensive discussion, see~\cite{GW_EarlyUniv_WP} and references therein).
Local strings with no internal structure can be described by the Nambu-Goto action, and are expected to quickly reach the scaling regime \cite{Kibble:1976sj}. The predicted SGWB spectrum is then defined by the dimensionless power spectrum for a string loop of a given length and by the number density of loops, both of which are studied in theoretical and numerical modeling~\cite{Blanco-Pillado:2017oxo,DV2,Ringeval:2017eww,Martins:1996jp,Martins:2000cs,Blanco-Pillado:2013qja,Blanco-Pillado:2011egf,Ringeval:2005kr,Lorenz:2010sm,Auclair:2019zoz}. The resulting SGWB spectrum is roughly constant over many decades of frequency, assuming standard cosmological history~\cite{Auclair:2019wcv}. %, as illustrated in Fig. \ref{fig:LHC_LISA_complementarity}. 
Detection of the SGWB due to cosmic strings could therefore be used to test for any departures from a standard cosmological picture \cite{Cui:2017ufi,Cui:2018rwi, Gouttenoire:2019kij}: probe new equations of state of the early universe, probe new particle species, and probe (pre-)inflationary universe~\cite{GW_EarlyUniv_WP}. Indeed, searches for cosmic string SGWB have already been conducted, placing upper limits on the string tension $G \mu \lesssim 9.6 \times 10^{-9}$ by LIGO-Virgo \cite{LIGOScientific:2021nrg}, and $G \mu \lesssim 10^{-10}$ by pulsar timing arrays \cite{Ellis:2020ena,Blanco-Pillado:2021ygr}. 
NANOGrav has reported a possible hint of cosmic strings \cite{NANOGrav:2020bcs}, although their observation may also be of instrumental origin.
Future experiments covering a wide frequency range will further improve the sensitivity to GW signals from cosmic strings, including Einstein Telescope, Cosmic Explorer, AEDGE, DECIGO, BBO, $\mu$Ares and Theia \cite{Punturo:2010zz,Yagi:2011wg,AEDGE:2019nxb,Hild:2010id,Sesana:2019vho,Theia:2017xtk}.
Cosmic string tension in the range of $G \mu \approx 10^{-16}- 10^{-15}$ or bigger could be detectable by LISA, with the galactic foreground affecting this limit more than the astrophysical background \cite{Boileau:2021gbr,Auclair:2019wcv}.

\subsection{Axion Like Particles}
Axions or axion-like particles (ALPs) are pseudo Nambu-Goldstone bosons that result from spontaneous symmetry breaking. Originally introduced to solve the strong CP problem, axions solve the hierarchy problem, inflation naturalness and naturally arise in string theory as modulus fields from dimensional compactification (see \cite{2016PhR...643....1M} for a recent review of axions in cosmology). Axions are typically light, and impact cosmology both at early and late times. Late-time effects of $\simeq 10^{-22}$ eV ALPs include the suppression of clustering of dark matter and galaxies, while $m_a < 10^{-27}$ eV ALPs generate late-time acceleration and change the amplitude of the Integrated Sachs Wolf (ISW) plateau in the CMB \cite{1992PhLB..289...67H,2006PhLB..642..192A,Hlo_ek_2018}. In the early universe, ALPs behave as relativistic species and can be mapped onto potential deviations from the Standard Model value of $N_\mathrm{eff}$, with deviations $\Delta N_\mathrm{eff}$ expected to be within the measurement sensitivity of experiments like CMB Stage IV \cite{cmbs4_sciencebook,2022arXiv220314923J}.

If the axion symmetry-breaking occurs during inflation, ALPs would source isocurvature perturbations with an amplitude set by the energy scale of inflation, which also sets the amplitude of the tensor spectrum from GW \cite{2017PhRvD..96f1301C,2013MNRAS.434.1619C}. Upcoming polarization CMB measurements will constrain $H_I$ while complementary constraints from the temperature spectrum constrain the contribution of ALPs to the total cosmic energy budget. For ALPs in the $10^{-25} \, {\rm eV} < m_a < 10^{-24} \, {\rm eV}$ mass range,  current data allow a roughly $10\%$ contribution of ALPs to the total dark matter budget, with around a 1\% contribution to isocurvature and tensors. This will significantly improve in the coming decade with increases in sensitivity to CMB polarization. Alternatively, if the ULA U(1) symmetry is broken \textit{after the end} of inflation, a white noise power spectrum of isocurvature would be produced. For these models, the expected sensitivity of experiments like CMB-S4 is to axions with masses as `high' as $10^{-17}$ eV. 

A background of oscillating ALPS with the standard $\frac{g_{a\gamma}}{4}aF\bar{F}$ coupling to photons leads to the rotation of linear polarization by ALP dark matter \cite{1998PhRvL..81.3067C,2008PhRvD..78j3516L,2009PhRvL.103e1302P,2009PhRvD..79f3002F,carosi_2013,2016PhR...643....1M,2018MNRAS.476.3063H}. 
It is the parity breaking associated with this coupling of a non-stationary background field to electromagnetism that generates this `birefringence' for the propagation of opposite-helicity photons.

If the source of the cosmic birefringence is \textit{spatially varying}, then the polarization rotation will be anisotropic across the sky \cite{Fedderke_2019,Namikawa_2020}. Many models of birefringence predict such anisotropies in the signal, or produce both an anisotropic and an isotropic birefringence signal. 
A measurement of anisotropic birefringence signal will strongly constrain these models\cite{2013PhRvD..87d7303G}.
While current bounds on the anisotropic CMB birefringence signal limit the axion-photon coupling to $g_{a \gamma} < 4.0 \times 10^{-2} / H_I$ \cite{Namikawa_2020}, future CMB observations should tighten limits by a few orders of magnitude. 

 An added effect is the washing out of polarization at the last scattering surface due to the early-time oscillations of the axion field. This would lower the polarized fraction measured through the TE and EE cross power spectra compared
to the standard prediction \cite{Fedderke_2019}. 

\subsection{Observation}
Relic radiation can be measured in one of two ways: direct interaction between the radiation on a detector, and indirectly through influencing other cosmological observables. The recent detection of GWs presents the tantalizing prospect of directly measuring relic GWs from the early universe. Current and upcoming GWOs could not only measure the energy density of relic GWs, but also the spectral density and perhaps even the spatial distribution. Gravitational and non-gravitational (e.g. axions, thermal relics, neutrinos) can also be measured through its impact on various cosmological observables. Relic radiation influences cosmological observables through its contribution to the scale of matter-radiation equality, which then changes the short wavelength modes of the matter power spectrum, and the phase of oscillations in the primordial photon-baryon fluid. Both of these signals can be observed via smaller-scale measurements of CMB and LSS. Though the search for new physics is often framed as searching for additional relativistic energy ($\Delta N_\mathrm{eff} > 0$), we note that it is permissible for new physics to result in ($\Delta N_\mathrm{eff} < 0$), which would carry significant implications for our understanding of neutrinos and our thermal history.

\section{Facilities}
\label{sec:obs}

\begin{figure}[t!]
    \centering
    \includegraphics[width=6in]{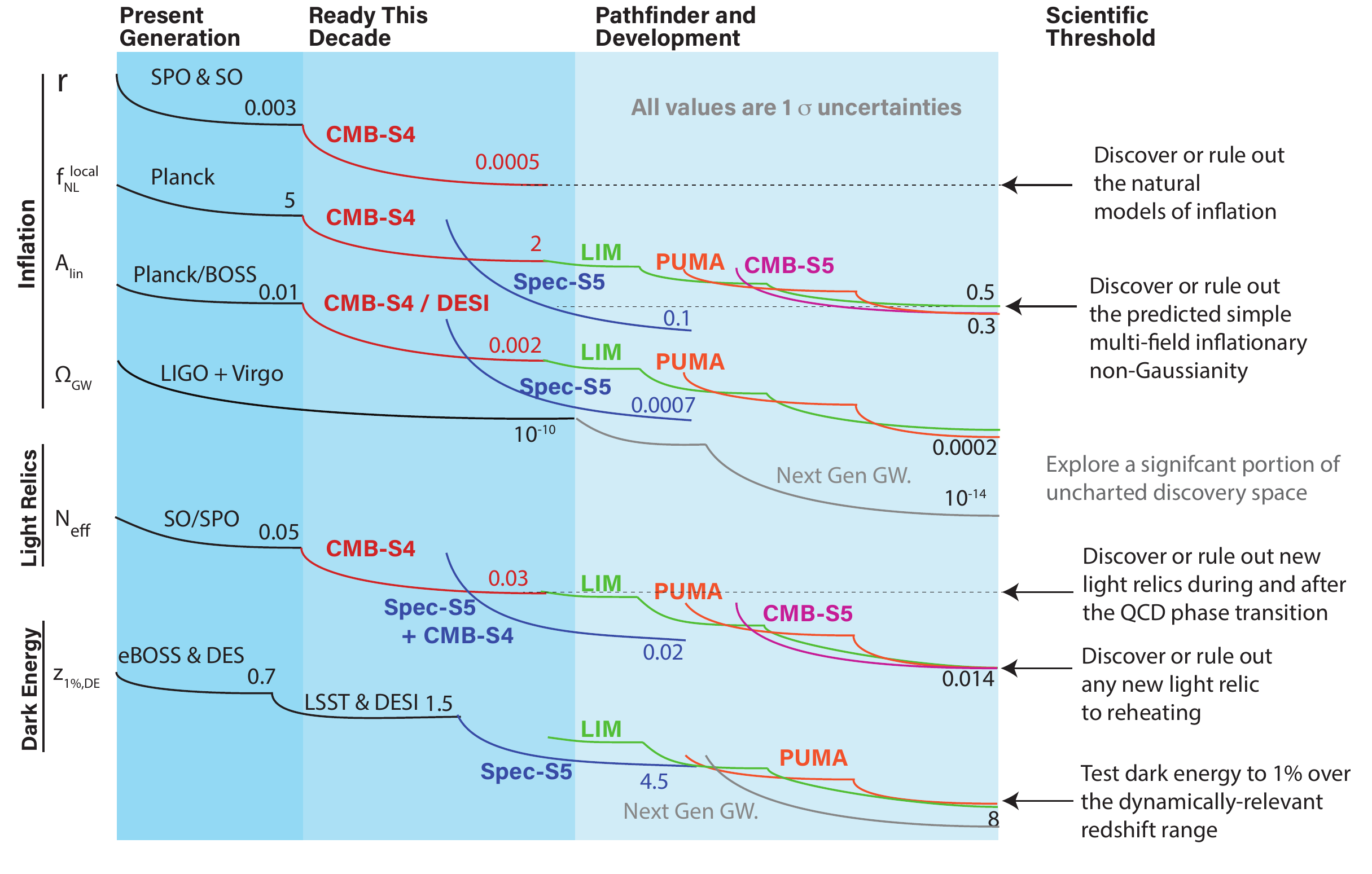}
    \caption{Illustration of  the growing scientific reach of cosmological facilities broken
into experiments that are technically ready to begin operation in this decade (2025-2035)
and more ambitious experiments requiring staged R\&D to realize facilities in the next
decade (2035-50).  This figure showcases six key cosmological observables and the associated scientific thresholds that future cosmic microwave background, spectroscopic, and line intensity mapping surveys will be able to achieve: primordial gravitational  waves as parameterized by the tensor-to-scalar ratio~$r$; local primordial non-Gaussianity as measured by~$\fnlloc$; features in the primordial spectra parameterized by their relative linear amplitude~$A_\mathrm{lin}$; the stochastic gravitational wave background as constrained through its energy density~$\Omega_\mathrm{GW}$; dark radiation as parameterized by the effective number of relativistic degrees of freedom~$N_{\rm eff}$; and dark energy as captured by the maximum redshift where its density is determined to better than 1\% of the total mass-energy density of the Universe, $z_{1\%,\mathrm{DE}}$.
Constraints were taken from the respective forecasting whitepapers; for the CMB-S5 we assumed CMB-HD \cite{2022arXiv220305728T}.}
    \label{fig:waterfall}
\end{figure}

All of the observable signals described above directly connect to new fundamental physics presenting an exciting opportunity for discovery. Robust detection and measurement of these signals is challenging requiring large facilities that employ sophisticated detector systems and technologies. Successful construction of these facilities requires that the selected technologies demonstrate an appropriate degree of technical readiness prior to implementation. This need for technical readiness shapes the experimental program for studies of the early universe (see Fig.~\ref{fig:waterfall}).

In the next ten years (2025-35), the early universe experimental program has two major activities. The first is the construction and operation of large facilities that are technically ready. These facilities include the CMB-S4 project, a new Wide-Field Multi-Object Spectrometer in the optical-infrared (Spec-S5), and upgrades to the operating Advanced LIGO, Advanced Virgo and KAGRA observatories. The second major activity is technology R\&D (including pathfinder experiments) focused on advancing new technologies to be implemented in large facilities in the following decade (2035-50). These technologies include new capabilities with line intensity mapping, new gravitational wave detection technologies, and new CMB instrumentation. In order to ensure sufficient technical maturity, this R\&D activity must extend beyond technology development in the lab and requires fielding smaller scale experiments using these novel techniques.  

\subsection{Large Facilities ready for construction and upgrade in 2025-35} 
The first major activity in 2025--35 is carrying out early universe measurements with large facilities that are ready for operation in this decade. These include the CMB-S4 experiment, a new Wide-Field Multi-Object Spectrometer (Spec-S5), and operations and upgrades to the currently running Gravitational Wave Observatories (GWOs). These facilities will target key scientific goals to realize substantial improvements in our understanding of new physics in the early universe.

\begin{figure}[h!]
    \centering
    \includegraphics[width=6.5in]{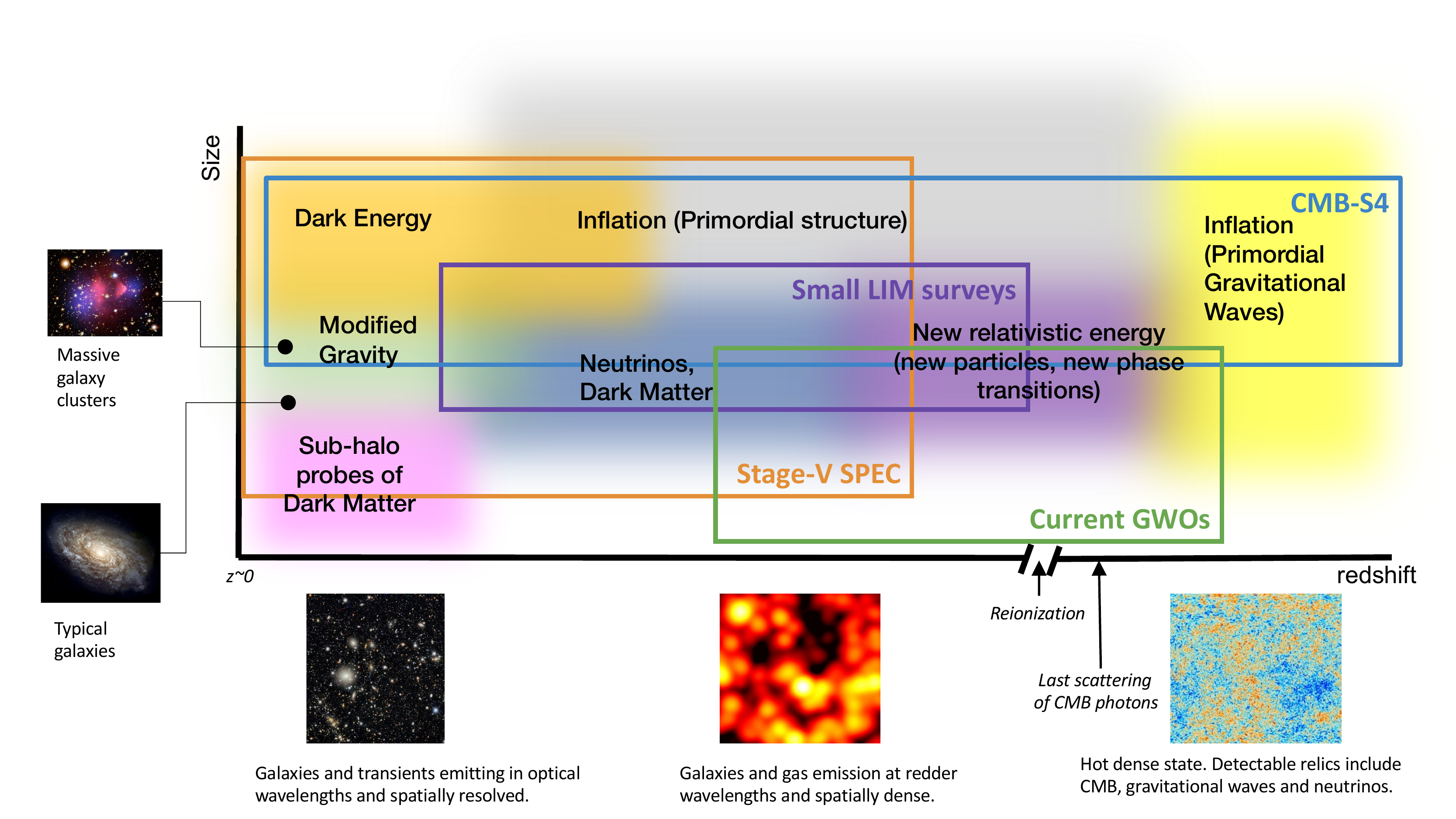}
    \caption{Landscape for Early Universe facilities operating during 2025--35 highlighting the overlapping and complementary approach to Early Universe science. The 2025--35 early universe science program includes three major facilities: CMB-S4, operating/upgrading existing GWOs, and a new Stage V spectroscopic facility (Spec-S5). The 2025--35 decade will also develop new technology critical for future surveys including fielding of small-scale instruments (e.g. LIM) and invest in key theoretical research to provide the needed tools to analyze the data.}
    \label{fig:facilities_thisdecade}
\end{figure}

\subsubsection{CMB-S4}
CMB-S4 \cite{2022arXiv220308024A} is a ``Stage-4'' cosmic microwave background project that plans to field multiple telescopes at the South Pole and in the Atacama desert, Chile. See the Snowmass 2021 White Paper~\cite{Chang:2022tzj} for a discussion on the broader experimental context.
\paragraph{Science goals.}
CMB-S4 has an enormously broad science case, including searches for primordial gravitational waves through the B-mode signal in the CMB
as predicted from inflation (detecting $r> 3\times 10^{-3}$ at $5\sigma$ or limiting $r\leq 10^{-3}$ at 95\% confidence if $r$ is very small) and for the imprint of relic particles including neutrinos (measuring $\sigma_{N_{\rm eff}}$ with uncertainty $\leq 0.06$ at 95\% confidence).  CMB-S4 will also offer unique insights into dark energy and tests of gravity on large scales, find large samples of high-redshift galaxy clusters, elucidate the role of baryonic feedback on galaxy formation and evolution, open a window onto the transient Universe at millimeter wavelengths, and explore objects in the outer Solar System, among other investigations.

\paragraph{Instrument description.} Current
CMB-S4 plans call for 500,000 polarization-sensitive bolometers that measure the sky at frequencies from 20--280 GHz.  The superconducting quantum interference device detectors will be read out using  time-domain multiplexed electronics and will be distributed between a set of telescopes at two site: two 6-meter cross-Dragone reflecting telescopes in Chile, a 5-meter three-mirror-astigmatic reflecting telescope at the South Pole, and eighteen 0.5-meter refracting telescopes at the South Pole, grouped as triplets on six mounts, with each mount sharing a cryogenic system.  

\paragraph{Technology status.}
The preliminary baseline design for CMB-S4 uses proven technology scaled up to much higher detector counts.  The already-established integrated project office has addressed the technical challenges of that scale-up with a detailed design and implementation plan that includes a full work-breakdown structure.  The project office also maintains a register of risks and a detailed cost estimation plan.  The preliminary, technology-limited project schedule contains nearly nine thousand milestones and catalogs the dependent relationships between them.  A dedicated group within the project oversees the production of prototype detectors and leads the development of a unified detector fabrication plan that covers multiple fabrication sites.    

\subsubsection{Stage-V spectroscopic facility (Spec-S5)} 

The upcoming Stage-V spectroscopic facility (Spec-S5) will employ highly-multiplexed spectroscopy on a large telescope to deliver a spectroscopic galaxy survey complementing and building upon the currently operating photometric galaxy surveys (e.g. LSST/VRO). 
\paragraph{Science goals.}

The overarching goal of the upcoming Stage-V spectroscopic facility is implementing large-format spectroscopy to capitalize on the large-area images currently coming online. With spectroscopy providing a fundamental and complementary observable, such a facility will lead to a broad suite of scientific results encompassing astrophysics to cosmology. Of relevance to the US High Energy Physics program are substantial improvements in astrophysical studies of Dark Matter and advancing the exploration of Dark Energy. Of particular overlap with early universe science is precision measurements of the primordial power spectrum and its statistics, which is enabled by the large volume survey of over 100~M galaxies out to high redshifts.

\paragraph{Instrument description.} Spec-S5 would consist of a 6-12~m Optical/IR telescope with a large ($>$5~deg) field-of-view. The instrument's massively multiplexed spectrometer consists of fiber-fed spectrographs multiplexed by $>$10,000 robotically positioned fibers.

\paragraph{Technology status.}
The spectroscopic survey technique is well developed with an established track record (e.g. eBOSS, DESI) making a Stave-V spectrograph an instrument with high technical maturity. The only technologies requiring development are the compact fiber positioners and low-noise CCDs at long wavelengths. Several concepts are already under development (MSE, MegaMapper, SpecTel) and recently, the Astro2020 Decadal Survey identified highly-multiplexed spectroscopy as a strategic priority and recommended that a major (MSRI-2 scale) investment could be made in a large, dedicated facility late this decade.

\subsubsection{Operating gravitational-wave observatories} 
The two Advanced LIGO detectors in the US and the Advanced Virgo detector in Italy were the first machines to directly observe gravitational-waves from binary merger events. Also operating, but in the early stages of commissioning, is the Japanese KAGRA detector.

\paragraph{Science goals.} Originally built for enabling the first detection of gravitational waves, the primary science goal of current gravitational-wave observatories is to study the population of compact mergers, to look for other potential signals from pulsars or other sources, to study any potential deviations from general relativity, and to constrain stochastic gravitational-wave background radiation, see Fig.~\ref{fig:landscape_inflation}.
GWs offer the unique possibility to probe the evolution of the universe within the first minute after the big bang, and the corresponding high-energy physics. (Roughly 1 minute after the big bang is when nucleosynthesis took place, of which we have observational evidence via the abundance of lightest nuclei.) During the first minute, the primordial plasma was opaque to both photons and neutrinos, so they cannot serve as messengers about this early epoch. 
Hence, GWs could tell us about inflation, possible additional phases of evolution (i.e. between inflation and radiation domination), phase transitions (multiple possibilities exist such as SUSY, QCD, electroweak and other transitions), and topological defects (cosmic strings, branes). Many of the proposed models (of inflation, particle physics etc) predict a stochastic GW background that could be directly measured by GWOs.

\paragraph{Instrument description.}
The Advanced LIGO and Advanced Virgo observatories are ground-based, L-shaped laser interferometers of 4~km and 3~km respective arm length. They are currently undergoing a significant upgrade to their readout and quantum noise reduction system, as well as the low frequency thermal noise. The upgrades are referred to as ``A+'' and ``Advanced Virgo+'' respectively. Their next one-year observation run (O4) is expected to start at the beginning of 2023, followed by a 2.5-year run (O5) between 2025 and 2028 at the full upgrade sensitivity--about twice the current sensitivity and corresponding to observing more than one binary merger per day.
The Japanese KAGRA observatory is also commissioning their detector, planning to join the O5 run.

\paragraph{Technology status.}
Planning on possible post-O5 run observatory upgrades has started. Possible scenarios include detector upgrades that allow increasing the low frequency sensitivity, while at the same time testing technology for the next-generation observatories Cosmic Explorer in the US and Einstein Telescope in Europe.

\subsection{R\&D for future facilities}
The second major activity in 2025-35 is technology R\&D to deliver key technologies for future large surveys after 2035. The science discussed above illustrates that there is significant discovery potential beyond the reach of the CMB, LSS and GW facilities described earlier, but going beyond those ambitious projects requires developing new technology and achieving a sufficient level of technical readiness for implementation in a future large survey instrument. With this objective, technology R\&D in 2025-35 will focus not only on advancing new instrumentation, but will require fielding new small-scale instruments to develop the needed experience with systems-level integration and systematics control.

\begin{figure}[h!]
    \centering
    \includegraphics[width=6.5in]{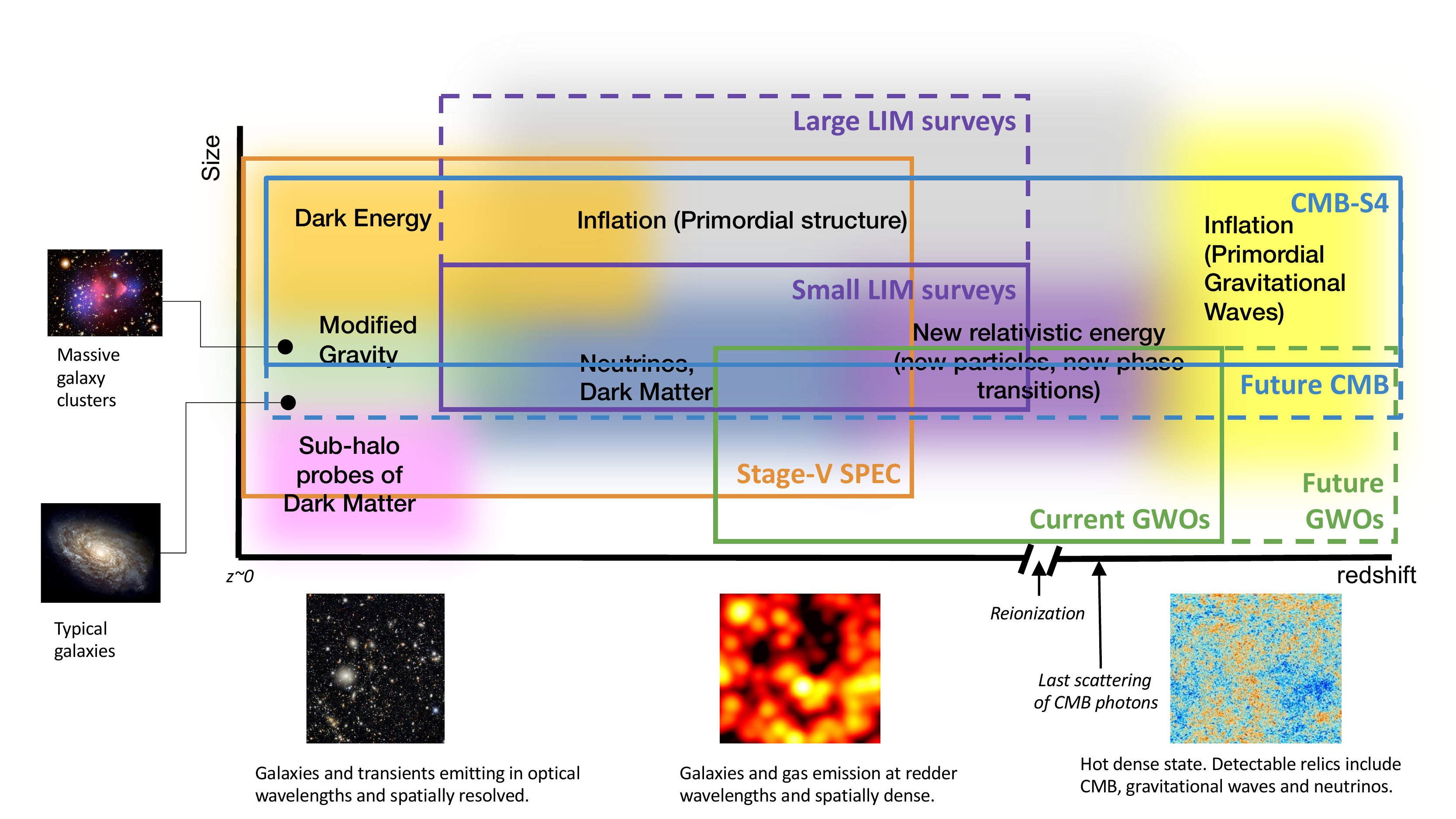}
    \caption{Potential landscape for early universe facilities post 2035 and the focus of technology R\&D in 2025--35. The large-scale nature of these facilities and novel survey approach require that the 2025-35 development include fielding instruments to sufficiently advance the readiness of the new technologies for a large scale experiment.}
    \label{fig:facilities_future}
\end{figure}

\subsubsection{R\&D for future 21~cm line intensity mapping}
\label{sec:21cmRD}
\paragraph{Science goals}
Neutral hydrogen in the Universe emits 21\,cm radiation across cosmic ages and hence forms a probe of structure starting from the earliest epoch after the CMB was formed (the discussion below draws liberally from ~\cite{PUMAAPC, puma_snowmass} and references therein). In particular:

\textit{Dark Ages -- }~$150 \gtrsim z \gtrsim 20$ -- The Dark Ages are prior to the formation of the first luminous sources are a particularly clean probe of a large number of modes and thus may be used to detect features in the primordial power spectrum, PNG, test statistical isotropy and homogeneity, enable measurements of the neutrino mass, constraints on the existence of warm dark matter, and exotic searches. Although the instrumentation is challenging, the Decadal panel on Astronomy and Astrophysics 2020 has identified Dark Ages cosmology as an especially attractive discovery area.
    
\textit{Cosmic Dawn and Reionization -- }~$20 \gtrsim z \gtrsim 5.5$ --  21\,cm emission traces the first luminous objects as they begin to form in this era. Measurements of velocity-induced acoustic oscillations may provide standard rulers at high redshifts and enable precision measurements of the Hubble expansion rate. 
\textit{High Redshift Galaxy Surveys -- }~$z\lesssim 5$ --  21\,cm observations at lower redshift form a measurement of large-scale structure through the redshift desert ($z\sim 1$--$3$) and beyond ($3 \lesssim z \lesssim 6$) where optical spectroscopy is challenging. These experiments target measurements of Dark Energy\citep{Knox2019HubbleHunter}, particularly in the context of proposed explanations that have non-trivial time evolution of dark energy at $z \gtrsim 1$; PNG; features in the power spectrum; and may constrain the sum of the neutrino masses to    otentially the number relativistic degrees of freedom and the sum of the neutrino masses $\lesssim 20\,{\rm meV}$ in combination with other probes.

\paragraph{Instrument description}

Detections of the $\mathcal{O}$(100\,mK) 21\,cm signal from high redshift large scale structure requires high sensitivity on few-degree spatial scales on the sky and good redshift resolution ($\delta z /z \sim 10^{-3}$). As a result the instrumentation is driven towards many (hundreds to thousands) radio detectors built as an interferometer for high sensitivity to a wide variety of spatial scales. Because frequency maps directly to redshift of 21\,cm emission, they also operate across a wide bandwidth for sensitivity to a wide redshift range, with many frequency channels commensurate with a spectroscopic survey. These arrays are physically large, with a digital correlator capable of processing many inputs and frequency channels. Future arrays designed for inflation science goals at redshifts $0.3 < z < 6$ will require a close-packed array of $>$10,000 dishes at least 6\,m in diameter~\cite{PUMAAPC} across $\sim$ km scales. These future arrays will require us to overcome a few challenges: 

\textit{Foreground removal -- } There is now $\sim$ 1 decade of experience using radio telescopes to measure cosmological neutral hydrogen in `intensity mapping' mode, with detections of large scale structure in combination with optical surveys~\cite{CHIMEresults,2022MNRAS.510.3495W,2021RAA....21...30L,2020MNRAS.498.5916T,2010Natur.466..463C,2013ApJ...763L..20M,2010PhRvD..81j3527M}, limits on $\Omega_{\rm{HI}}$~\cite{2013MNRAS.tmpL.125S}, limits on IGM heating at high redshift~\cite{2015ApJ...809...62P}, the 21\,cm power spectrum~\cite{2019ApJ...883..133K,2019ApJ...887..141L,2016ApJ...833..102B,2016MNRAS.460.4320E}, and a tension has appeared between results from different global experiments~\cite{EdgesDetection,SARAS}. Current results are limited by their ability to remove bright astrophysical foregrounds from galactic and extragalactic synchrotron emission\cite{aguirre2019roadmap}. Removing this foreground emission relies on differentiating the frequency dependence between the foregrounds and signal of interest~\cite{2017PASA...34...33P,pober_et_al2013b,seo_and_hirata2016,2012MNRAS.419.3491L} which requires good knowledge and control of the instrumental frequency response~\cite{Shaw:2014vy,Shaw:2013tb}.

\textit{Big Data from Large Arrays -- } The data sets resulting from these future arrays will need to be calibrated and processed in real-time to compress the data for transfer, storage, and analysis. 

\paragraph{Technology R\&D roadmap}
 Taken together, overcoming these challenges will require a dedicated and phased R\&D effort:
\begin{itemize}
    \item Current generation experiments (e.g. CHIME, HERA, MWA): These experiments are actively pursuing calibration and analysis techniques, data compression, and RFI removal for their science goals. Their results will be critical for informing the design of future arrays.  
    \item Near-term pathfinders (e.g. LuSEE night, CHORD, HIRAX): These experiments are working towards building a suite of simulations for instrument design. In particular, the individual telescopes comprising the array must be uniform to efficiently use current real-time gain stabilization algorithms which feeds back to manufacturing tolerances. Similarly, a flat bandpass response is necessary for foreground mitigation, and simulations are required to quantify the amount of bandpass variation allowed in the design. A uniform array allows efficient compression schemes for correlating the data and redundant array configurations provide the possibility of real-time gain stabilization with a sky model.  
    \item R\&D for future arrays: Instrument stability would be made more tractable if digitization could occur at or near the focus of the dish, however a high degree of shielding from the digitizer's radio frequency interference (RFI) would be necessary. LuSEE-Night is likely to act as the first demonstration of nearby digitization. Digitization near the dishes also requires excellent (better than 1\,ps) timing between the array elements, and so timing must be distributed across a large footprint with a mechanism for precise timing corrections. In addition, 21\,cm arrays find beam measurements challenging because they are designed to have broad beams with poor sensitivity to point sources, and are typically designed as transit telescopes (stationary, non-steerable) to reduce costs. New digital calibrators have been proposed for beam measurements from drones~\cite{2022arXiv220111806B} and are being explored by LuSEE-Night as well. 
\end{itemize}

\subsubsection{R\&D for future mm-wave LIM}

\paragraph{Science goals} Millimeter-wave line intensity mapping uses low angular resolution, spectroscopic observations of rest-frame far-IR atomic or molecular emission lines to trace large-scale structure \cite{karkare2022, kovetz2019}. Ground-based observations in the 80--310 GHz atmospheric window are sensitive to multiple CO rotational lines and the [CII] ionized carbon fine structure line originating from an extremely wide redshift range, $0 < z < 10$. A wide-field, high-sensitivity survey could therefore offer a valuable complement to galaxy surveys at $z \lesssim 2$ and a unique probe of higher redshifts. Such a survey would provide access to ultra-large scales and an unprecedented number of modes for testing primordial non-Gaussianity \cite{moradinezhaddizgah2019}, constrain the expansion history in the matter-dominated regime and a wide array of dark energy and modified gravity models \cite{karkare2018}, and limit neutrino masses beyond the capability of current LSS surveys \cite{moradinezhaddizgah2022}.

\paragraph{Instrument description} The observational requirements of mm-wave LIM---high-sensitivity, low-systematics measurements of faint, diffuse structure over large areas of sky---are largely met by contemporary CMB experiments operating in the Atacama Desert or the South Pole. Instead of broadband detectors, however, the LIM measurement requires moderate-resolution spectroscopy ($R \gtrsim 300$) to resolve fluctuations along the line of sight. We envision a straightforward replacement of current or future CMB experiments (ACT, SPT, SO, CMB-S4) with focal planes of high-density, on-chip mm-wave spectrometers \cite{karkare2022}. Millimeter-wave LIM surveys can be parametrized by \textit{spectrometer-hours}, and generally become competitive with galaxy surveys in the $\sim 10^7$ spectrometer-hour range.  Advances in technology (see below) should allow spectrometer counts in the thousands on typical high-throughput cryogenic CMB receivers, enabling the survey depths necessary for precision cosmology at high redshift.

\paragraph{Technology R\&D roadmap} Current approaches to mm-wave spectroscopy (diffraction gratings, Fourier Transform or Fabry-Perot spectroscopy, heterodyne detection) are difficult to scale to large spectrometer counts. \textit{On-chip} spectroscopy, in which the detector and spectrometer are integrated on a silicon wafer, offers a natural path to maximizing the sensitivity of mm-wave detectors. However, while prototype on-chip spectrometers are now being demonstrated \cite{endo2019, karkare2020}, their spatial packing density is still significantly lower than CMB focal planes (primarily due to the physical extent of the spectrometer on the wafer). Innovation in focal plane geometry and layout will be key to enabling high-density close-packed arrays. Similarly, while on-chip spectrometers have demonstrated spectral resolution of $R \sim 300 - 500$, improving to $R \sim 1000$ would move LIM experiments into the regime of spectroscopic surveys and significantly improve science return and systematics mitigation. This requires development of new low-loss dielectric materials. Finally, spectroscopic pixels require significantly more detectors than their broadband counterparts. Readout development based on state-of-the-art FPGA platforms, such as the RF system-on-chip, promise to dramatically reduce overall readout costs to \$1--2 per channel \cite{lowe2020}.

\subsubsection{R\&D for future high angular resolution CMB experiments} 

For CMB science, one proposed next-generation aim  is to pursue much finer resolution in a wide-area survey. For example, the CMB-HD telescope \cite{2022arXiv220305728T}, a concept under development by some members of the CMB community, targets $15''$ resolution at 150 GHz over 20,000 square degrees.

\paragraph{Science goals.}  The CMB-HD concept seeks to measure cosmic structure by mapping matter via gravitational lensing to $k \sim 10\,h$Mpc$^{-1}$ scales (thus probing dark matter physics) and mapping gas via the thermal and kinematic Sunyaev-Zeldovich effects.  The survey's sensitivity would rule out or detect any light thermal relic in the early Universe and rule out or detect inflationary magnetic fields.  Other science goals include constraints on axion-like particles, cosmic birefringence, the summed neutrino mass, dark energy, and a variety of astrophysical objects.  

\paragraph{Instrument description.}  The envisioned CMB-HD design calls for a pair of off-axis Dragone telescopes with 30-m primary and 26-m secondary mirrors and image-correcting cold optics.  The telescope focal planes will host 1.6 million detectors ($>3 \times$CMB-S4) in seven frequency bands from 30 GHz to 350 GHz.  Each detector pixel will measure two frequencies and two linear polarizations.

\paragraph{Technology R\&D roadmap.} For CMB-HD, research and development efforts are required for the telescope, cryostat, detectors, and detector readout.  
The crossed Dragone optical configuration requires four $\sim30$~m mirrors. Though this design is scaled-up from similar architectures employed by the Simons Observatory large-aperture telescope and CCAT-prime, the nearly 25x more massive mirrors along with more stringent pointing requirements present new challenges for telescope design. 
The mount for CMB-HD must bear much more weight.  In order to achieve sufficient optical stability on a timescale of tens of seconds under thermal, gravitational, and wind forces while scanning, the mirror surface will likely require active shape correction such as a laser metrology system currently being explored by the GBT 100-m telescope.
The conceptual design assumes horn-fed detectors, which at higher frequencies, will likely require new mulitplexing capabilities to realize the required pixel density. Potential technologies include new microwave-SQUID TES multiplexers currently being fielded by experiments like Simons Observatory or Microwave Kinetic Inductance Detectors (MKIDs) similar to technologies being developed for the MUSTANG-2 receiver on the GBT and the TolTEC reciever on the LMT at frequencies from 90 GHz to 270 GHz.  Significant engineering effort is needed to develop the design and mature the project plan.

\subsubsection{R\&D for future Gravitational Wave observatories} 

The next generation of gravitational-wave observatories can explore a wide range of fundamental physics phenomena throughout the history of the universe. These phenomena include access to the universe’s binary black hole population throughout cosmic time, to the universe’s expansion history independent of the cosmic distance ladders, to stochastic gravitational waves from early-universe phase transitions, to warped space-time in the strong-field and high-velocity limit, to the equation of state of nuclear matter at neutron star and post-merger densities, and to dark matter candidates through their interaction in extreme astrophysical environments or their interaction with the detector itself~\cite{Ballmer:2022uxx}.

Scaling the current gravitational-wave detector technology to the needs of the next generation of observatories Cosmic Explorer and Einstein Telescope requires research targeting improvements in squeezing and quantum metrology techniques, the production of large (320~kg) low-loss fused silica optics for test masses, optical coatings with reduced mechanical dissipation, and a low-cost ultra-high vacuum system. Accessing the scientifically interesting low-frequency band also requires improved active seismic isolation, including systems to subtract the direct Newtonian coupling of the seismic motion.

One possible upgrade for the Cosmic Explorer facilities is the technology currently being developed for the LIGO Voyager concept, consisting of a 2 micron laser and cryogenic silicon test masses. This approach would also require the production of large (320~kg) single crystal silicon test masses, a cryogenic cooling system with low vibrational coupling, and improved 2~um wavelength laser technology, particularly low-noise lasers and high quantum-efficiency photo diodes.

\section{Theory and Analysis} 
\label{sec:analysis}

Theoretical astrophysics, cosmological model building and data analysis are increasingly important to advancing the cosmic frontier.  As data becomes larger and more complex, theory and data analysis become critical parts of an experimental mission.  In fact, all the facilities this report covers have similar challenges associated with foreground signal subtraction: foreground signature from dust for CMB, foreground contamination to the 21cm line, and Newtonian noise subtraction for GWO.

To illustrate just one example of the importance of coupling the development of theory and analysis to that of experiment and observations, consider the measurement of primordial gravitational waves via CMB B modes.  Future surveys of these modes will be limited  by galactic dust and gravitational lensing. Without the continued development of the theory and analytical techniques to model and/or remove these additional signals, it will be impossible to reach the potential of these surveys.  

Sensitivity to fundamental physics in cosmological probes, as described above,  are increasingly limited by astrophysical foregrounds rather then experimental noise.  In many cases, there is no clear line distinguishing the theoretical contributions needed to achieve the goals of a specific survey and the broader activities of the theory community.  

A special feature of cosmic surveys is the dual role of astrophysical effects as a signal and a source of noise. The apparent foregrounds may themselves encode important information about the fundamental laws. For example, gravitational lensing decreases our sensitivity to parameters in the primary CMB, such as $r$ and $N_{\rm eff}$, but is itself used to measure the sum of the neutrino masses, $\sum m_\nu$. These kinds of secondary signals have already been proven to be powerful cosmological probes in their own right.

In this section, we will highlight a few concrete examples where theoretical and analysis techniques are essential and require future investment.  See~\cite{Snowmass2021:Inflation,Snowmass2021:TheoryCosmo,alvarez_snowmass,Dvorkin:2022jyg,Ferraro:2022cmj,Flauger:2022hie} and reference therein for a more thorough discussion of the theoretical implications described below.

\subsection{From Theory to Observations}

\subsubsection*{Fundamental Physics in Cosmic Surveys}

The cosmological information from the next generation of surveys will increasingly arise from nonlinear structures at low redshift, whether it comes in the form of CMB secondaries (lensing, SZ) or directly mapping galaxies or other tracers of nonlinear structure. Isolating the physics of inflation, dark matter, neutrinos, etc. from the physics of structure formation itself is essential to maximizing the scientific return from these surveys.

Both past, current and future analyses of these surveys have relied on theoretical insights to make this split possible. Famously, the use of the BAO as a probe of the expansion~\cite{Beutler:2019ojk} of the universe arose from the understanding that the signal was robust to nonlinear corrections~\cite{Eisenstein:2006nj}. Fundamentally, the BAO signal is associated with the acoustic horizon at recombination that is vastly larger than the scale associated with nonlinearity.  Related ideas have inspired the measurement of $\Neff$~\cite{Baumann:2017lmt,Baumann:2019keh} and primordial features~\cite{Beutler:2019ojk} in LSS survey and play a critical role in these analyses that are needed to achieve the sensitivity outlined in this report.  
Less targeted analyses do not reproduce the same sensitivities because the usual split between the linear and nonlinear regimes would exclude modes that contain these signals.

The search for primordial non-Gaussianity in LSS surveys has also been driven by theoretical developments. Future constraints on local non-Gaussianity will be driven by LSS surveys (with the assistance of CMB lensing~\cite{Schmittfull:2017ffw}) because of the discovery of scale-dependent bias~\cite{Dalal:2007cu}.  Nonlinear structures, like galaxies, form differently in the presence of local non-Gaussianity and give rise to a signal at large distances that cannot arise from Newtonian physics. Further theoretical work showed how this same effect enables a measurement of $\fnlloc$ without cosmic variance~\cite{Seljak:2008xr}. From understanding the space of inflationary models, it has also been understood that this is a unique signal of multi-field inflation~\cite{Creminelli:2004yq} and the above analysis can be generalized to extract the mass and spin of these additional fields~\cite{Gleyzes:2016tdh}.

In contrast, the search for equilateral-type non-Gaussianity will depend crucially on continued theoretical developments in the next decade~\cite{Alvarez:2014vva}. The equilateral bispectrum generated during inflation is highly degenerate with the bispectrum that arises from nonlinear effects and presents a serious challenge to these analyses.  Analyses are further challenged by the presence of redshift space distortions and bulk flows.  A number of theoretical techniques~\cite{McDonald:2009dh,Baumann:2010tm,Carrasco:2012cv} have been developed to confront these challenges that use our fundamental understanding of both the inflationary physics and the nonlinear structure formation to try to disentangle these effects. In addition, there is reason to believe that the initial concerns about the correlation of bispectra overstates the degeneracy in the maps that could be accessible with a number of new analysis techniques~\cite{2022JCAP...08..061B}.  However, these ideas are still far from producing competitive results~\cite{Cabass:2022wjy,DAmico:2022gki} and continued investments in these theoretical tools is necessary to reach the observational aspirations of the community.

More generally, one broad appeal of mapping the universe on large scales is that the history of the universe and any forces that shaped its evolution will be encoded in these maps. As a result, the value of these surveys is expected to grow as new uses for these maps arise from theoretical progress. 

\subsubsection*{Theoretical studies of FOPTs}
Theoretical and numerical modelling of phase transitions is a very active area. Four parameters are critical for determining the GW production: the nucleation temperature $T_*$, the bubble wall velocity $v_w$, the FOPT's strength $\alpha$ and its inverse duration $\beta$. Multiple open questions are under investigation focused on estimating these parameters. First, a perturbative treatment of the finite temperature potential is known to breakdown. The central problem is that the expansion parameter at finite temperature involves a mode occupation which diverges when the mass vanishes \cite{Linde:1980ts}. Currently, only the technique of dimensional reduction \cite{Kajantie:1995dw,Farakos:1994xh} performed at NLO using an $\hbar$ expansion provides a prescription to calculate thermodynamic parameters at $O(g^4)$ in a gauge independent way \cite{Croon:2020cgk,Gould:2021oba}. This method is challenging to use and has been applied to benchmarks in very few models. Proposed alternatives to dimensional reduction \cite{Curtin:2016urg,Croon:2021vtc} are in need of further development and testing. Second, an accurate evaluation of the nucleation rate $\Gamma_{\mathrm{nuc}}$ and its evolution with temperature is critical for defining the characteristic time scales of the transition. 
For sufficiently fast transitions, $T_{\ast}$ and $\beta$ can be obtained by linearizing the rate  near $T_{\ast}$. This breaks down for slow transitions, which can be of great phenomenological interest, where the next order corrections must be accounted for~\cite{Ellis:2018mja}. A number of other issues that affect the nucleation rate also require further study \cite{GW_EarlyUniv_WP}. 
Third, the bubble wall speed can be calculated via different formalisms whose applicability depends on the relative strengths of the transition that determine
whether the terminal speed will be only mildly relativistic or ultrarelativistic. This is also investigated by many authors~\cite{Konstandin:2010dm,BarrosoMancha:2020fay,Balaji:2020yrx,Ai:2021kak,Cline:2021iff,Bodeker:2009qy,Bodeker:2017cim,Hoeche:2020rsg,Gouttenoire:2021kjv,GW_EarlyUniv_WP}.

In non-thermal transitions, the vacuum energy released in phase transitions can far exceed the surrounding radiation energy ~\cite{Randall:2006py,Espinosa:2008kw}.
Here the bubble expansion mode has two possibilities~\cite{Espinosa:2010hh}: (i) strong detonation, where the wall reaches a terminal velocity due to balancing between the outward pressure and the friction, and GW production comes from a highly relativistic and concentrated fluid around the bubbles, and (ii) a runaway, where the wall continues to accelerate until it collides producing GWs~\cite{Ellis:2019oqb}. Both scenarios are being investigated, e.g. through the bulk flow runaway model~\cite{Jinno:2017fby,Konstandin:2017sat,Megevand:2021juo}, and the sound shell detonation model~\cite{Hindmarsh:2016lnk,Hindmarsh:2019phv}.

In addition, both purely hydrodynamic and magneto-hydrodynamic (MHD) turbulence are expected to source GWs \cite{Kamionkowski:1993fg}.
Past analyses have evaluated the GW production using semi-analytical modelling \cite{Kosowsky:2001xp,Dolgov:2002ra,Caprini:2006jb,Gogoberidze:2007an,Kahniashvili:2008pe,Caprini:2009yp,Niksa:2018ofa}. Simulation-based approaches are also being pursued \cite{Cutting:2019zws,RoperPol:2019wvy,Kahniashvili:2020jgm,RoperPol:2021xnd,RoperPol:2022iel}.

\subsubsection*{GW-EM Correlations}

Many SGWB models, astrophysical and cosmological, also yield predictions for other observables, such as the CMB, the distribution of galaxies across the sky and redshift, and the distribution of dark matter throughout the universe. It is therefore expected that cross-correlating the SGWB spatial structure with spatial structures in electromagnetic (EM) observables would enable new probes of the underlying physical models and of the earliest phases of the evolution of the universe~\cite{contaldi,jenkins_cosmstr,Jenkins:2018a,Jenkins:2018b,Jenkins:2019a,Jenkins:2019b,Jenkins:2019c,Cusin:2017a,Cusin:2017b,Cusin:2018,Cusin:2018_2, Cusin:2019,Cusin:2019b,Alonso:2020,Cusin:2019c,Cusin:2018avf,CanasHerrera,ghostpaper,bartolo,bartolo2020,dallarmi,Bellomo:2021mer}. Such spatial correlations can be studied in terms of the angular power spectrum:
\begin{equation}
    D(\theta) = \langle \delta\Omega_{\rm GW} (\hat{e}_1,f), \delta X(\hat{e}_2)\rangle = \sum_{lm} \frac{2l+1}{4\pi} D_{l}(f) P_l(\cos\theta)
\end{equation}
where $X(\hat{e}_2)$ describes an EM observation such as the CMB or galaxy count distribution in sky direction $\hat{e}_2$, $\Omega_{\rm GW} (\hat{e}_1,f)$ is the SGWB energy density normalized to the critical energy density in the universe at frequency $f$ and direction $\hat{e}_1$, $\theta$ is the angle between the two sky directions $\hat{e}_1$ and $\hat{e}_2$, and $P_l$ are Legendre polynomials. Predictions for the angular power spectrum can already be found in the literature. 
In the case of phase transitions, the PT would have nucleated at slightly different redshifts in different causally disconnected regions of the universe, giving rise to anisotropy in the SGWB. The SGWB angular structure would not be affected by interactions with the plasma (i.e. effects such as Silk damping and baryon acoustic oscillations are not relevant for GWs), resulting in a simple angular spectrum: $C_l^{\rm GW} \sim [l (l+1)]^{-1}$ \cite{ghostpaper}. Assuming the PT happened after inflation, the primordial density fluctuations that led to the CMB angular spectrum would also have been present during the PT,  imprinting a SGWB anisotropy at least as large as
the CMB anisotropy~\cite{ghostpaper}.  

In the case of cosmic strings, the angular spectrum would depend on fundamental parameters of cosmic strings (string tension, reconnection probability), and on the network dynamics model. While the isotropic (monopole) component of this SGWB may be within reach of the advanced or 3G detectors~\cite{O1cosmstr}, the anisotropy amplitudes are found to be $10^{4} - 10^{6}$ times smaller than the isotropic component, depending on the string tension and network dynamics~\cite{olmez2,jenkins_cosmstr}. This level of anisotropy may be within reach of the 3G detectors. Correlating the anisotropy of this SGWB with anisotropy in the CMB or large scale structure may reveal details about the formation and dynamics of the cosmic string network.

In the case of primordial black holes (PBH), cross correlating the sky-map of the SGWB due to binary black hole (BBH) signals with the sky-maps of galaxy distribution or dark matter distribution could provide additional insights on the origin of black holes \cite{raccanelli1,raccanelli2,nishikawa,scelfo}. In particular, PBH binaries are more likely to form in low-mass halos where typical velocities are smaller and binary formation through GW emission is more likely. On the other hand, stellar black holes are more likely to form binaries in more luminous galaxies (with more massive halos). Studying the correlations of the BBH SGWB anisotropy with distributuon of visible and dark matter on the sky can therefore be used to probe the origin of the BBH systems and their potential dark matter component.

Techniques for conducting GW-EM cross correlation studies are currently under development~\cite{yang,smiththrane,banagiri}. However, much more remains to be done in order to fully explore the potential of this approach: develop appropriate statistical formalisms for parameter estimation; systematic studies to understand the angular resolution of GW detector networks; development of theoretical models of SGWB-EM anisotropy correlation; studies delineating the astrophysical and cosmological components of the SGWB, and others.

\subsection{Simulations and Analysis}

\subsubsection*{Astrophysical modelling} 
Large-scale cosmological simulations are central to modelling astrophysical foregrounds and systematic effects. These simulations provide the testing suite to understand the interplay of these effects with the signals from gravitational waves. The challenges facing the simulation community include simulating large enough volumes to capture both large scale effects while still having resolution to reach low mass objects \cite{alvarez_snowmass}. Relying too closely on sub-grid modelling can lead to biases in the foreground simulations, and an inability to distinguish new physics from foreground modelling \cite{2021arXiv210804215B,2017A&A...603A..62V}. Similarly, accurate modelling of neutrino physics in terms of their impact on the clustering of matter (which in turn impacts the lensing power spectrum of the CMB, and is therefore degenerate with other parameters) may require a hybrid approach between hydrodynamical and N-body simulations, perturbative methods and improved sampling of the full six-dimensional neutrino phase space \cite{2017arXiv170704698S,2019arXiv190713167M,2022arXiv220202840F}. Another need for progress in simulations comes from $21\,\textrm{cm}$ experiments. While the Dark Ages is relatively straightforward to simulate (often involving just linear physics), the later phases of the reionization and post-reionization universe are non-trivial. Reionization simulations require large dynamic ranges in scale, since the reionization is driven by the uncertain small-scale astrophysics of the first luminous sources, modulated by large scale cosmology. The post-reionization universe is arguably slightly simpler, but uncertainties still exist (such as the uncertainty in the halo mass to HI mass relation), which must be modeled or at least parametrized with enough flexibility. These large-volume and high-resolution simulations are not only required to model a cosmological effect itself. Estimates of the covariance matrix typically require many realizations of a given model to supplement any analytical estimates of the covariance. Here approximate methods and machine learning techniques prove useful. Finally, with cross correlations between multiple probes (e.g., $21\,\textrm{cm}$ and galaxy surveys) being important both from a standpoint of science and from a standpoint of systematics, self-consistent simulations will be of crucial importance.

\subsubsection*{Data analysis pipeline development}
As models become more and more complex, efficient sampling of parameter space for statistical inference becomes prohibitively slow with standard Monte Carlo sampling techniques. In some cases, a smaller number of high-quality simulations are produced and statistical `emulators' are used to interpolate between nodes when sampling. Development of emulators that connect theories across a range of observables, and that include the foreground modelling described above are needed to integrate this theoretical framework into an experimental context.

Similarly, many of the current emulators have been built around models close to the $\Lambda$CDM paradigm, and on gravity-only simulations. Extending the model space and coupling the advances in hydrodynamical modelling with those in fundamental theory model-building will be key to ensuring that future observations are able to make contact with theory \cite{alvarez_snowmass}.

High performance computing requirements form an integral part of planning and costing for future cosmic experiments like CMB-S4 (which will have an estimated 70 TB/day or 800 MB/s data rate) or any $21\,\textrm{cm}$ cosmology experiment (which already have such a data rate). Efficient pipeline development for signal processing and map-making of large volumes of data, foreground cleaning and parameter sampling will be needed to ensure processing of data in a timely manner for release to the community. Much of the algorithmic development of analysis pipelines and tools starts with individual researchers undergoing university-funded research. Supporting this research is critical to developing the tools and techniques needed to maximize scientific return on investment in experiment.

\subsubsection*{Shared tools and frameworks}
Using bespoke software for individual experiments can lead to replication of analysis pipelines on common elements. While redundancy and independence of different groups is essential to reducing systematic experimental bias, development of common and shared tools for some parts of the analysis pipeline will lead to improvement in delivery of results. An example of efforts in this area is the Core Cosmology Library \cite{Chisari_2019}, a common library of cosmological observables given an input cosmological model. In particular this propagates the modelling of astrophysical uncertainties across probes, removing inconsistencies in approach linking fundamental physics to astrophysical uncertainties, which is of particular interest when cross-correlating different probes. 

Coordination between groups and experiments in software development and theory support is critical to make the most of the observations planned for the coming decade. Several white papers submitted to this group report on the need for dedicated theory and simulation efforts e.g. to model the distribution of dark matter on astrophysical scales, to distinguish between dark matter physics and early universe models \cite{bechtol_snowmass}, 21-cm modelling for upcoming experiments \cite{puma_snowmass} and to develop the tools and techniques for data mining of the large data sets that will be generated by upcoming facilities \cite{nord_snowmass}.

\subsection{Astrophysical Foregrounds}

\subsubsection*{Cosmic Microwave Background}
The cosmic microwave background to date has provided our most precise measurement of cosmological parameters. As the next generation of surveys produces maps of increasing depth~\cite{Chang:2022tzj,2022arXiv220308024A,2022arXiv220305728T}, the use of CMB data will require theoretical techniques that separate the primary CMB from a variety of secondary effects.  CMB photons are gravitationally lensed by the intervening matter~\cite{Lewis:2006fu} and scattered~\cite{Carlstrom:2002na} by ionized electrons.  Ongoing theoretical work into the nature of these secondary effects has led to techniques to separate these various contributions to the CMB maps based on their statistical properties and frequency dependence. CMB secondaries thus give rise to new maps for the distribution of matter in the universe and the locations of high redshift galaxy clusters.  From these secondaries, we gain new insights into the history of the universe and fundamental physics.  CMB lensing with the next generation of cosmic surveys is a central tool in constraints on (and eventually, detection of) a non-zero sum of the neutrino masses~\cite{2022arXiv220307377A, Dvorkin:2019jgs} or the energy density in other hot relics~\cite{Dvorkin:2022jyg}.

Galactic foregrounds present an additional challenge to CMB measurements, particularly the constraints on $r$ from polarization B-modes~\cite{BICEP2:2015nss,BICEP:2021xfz}.  Polarized dust emission in our galaxy contaminates our measurement of both E-modes and B-modes~\cite{Planck:2018gnk}.  The amplitude of the dust B-mode at CMB frequencies is larger than the gravitational wave signals consistent with current limit.  As a result, dust foregrounds must be removed from the maps in order to reach future $r$ targets~\cite{CMB-S4:2020lpa}.  Our understanding of dust emission from first principle is insufficient for these purposes but has been bolstered by simulations~\cite{Kim:2019xov,Kritsuk:2017aab} and data-driven techniques~\cite{Thorne:2021nux,Krachmalnicoff:2020rln}.  Continued research into the dust foregrounds and techniques to remove them will be essential for the success of surveys like the Simons Observatory~\cite{2019JCAP...02..056A}, CMB-S4~\cite{2022arXiv220308024A} and CMB-HD~\cite{2022arXiv220305728T}.

\subsubsection*{21 cm Foreground Removal}

Foregrounds---and their interaction with the non-idealities of one's instrument---are arguably the chief obstacle in $21\,\textrm{cm}$ cosmology. The challenge is in some ways greater than with the CMB, given that even the coolest parts of our Galaxy contain foregrounds that dwarf the cosmological signal by orders of magnitude. Compounding this problem is foreground modeling \emph{uncertainty}, which is considerable given the lack of high-quality, all-sky maps at low frequencies. Further observations at relevant low-frequency bands will reduce this uncertainty, but this must be coupled with detailed studies of algorithms that incorporate knowledge of how foregrounds appear in the data when processed through instrumental systematics, as discussed in Section~\ref{sec:21cmRD}.

\subsubsection*{SGWB Foreground Removal} 

Since the detection of the first binary black hole merger in 2014, GW detectors have observed nearly 100 such events. With the increased sensitivity, the upcoming observation runs of Advanced LIGO, Advanced Virgo, and KAGRA are expected to yield one detection per day. The next (3rd) generation of GW detectors, Einstein Telescope and Cosmic Explorer, are expected to observe $10^5-10^6$ binary merger events per year. These signals will form astrophysical foreground masking the cosmological contributions to the SGWB. 

Removal of this foreground is a significant challenge and the necessary technology currently does not exist. The first challenge is to enable parameter estimation of multiple merger signals overlapping in time (and in frequency) domain. At the moment, parameter estimation techniques rely on the assumption of a given time segment containing at most one binary merger. The second challenge is to remove the individually observed binary signals. This could be done by notching out the individually detected binary signals in time-frequency space, by subtracting the binary signals from the GW detector strain data~\cite{PhysRevD.73.042001,PhysRevD.102.063009,PhysRevLett.118.151105}, and by simultaneously fitting all binary signals along with the SGWB present in a given data segment~\cite{smiththrane,PhysRevLett.125.241101}. At the moment, none of these approaches have been developed to the point that their application to 3G GW detector data is possible.

\section{Conclusion}

Measurements of the early Universe have the potential to access physics at GUT scales and provide evidence for light relic particles, axions, phase transitions, and neutrinos. Physics at these scales can be measured through a few complementary probes: measurements of the Cosmic Microwave Background at large and small scales, surveys of structure through cosmic time, and gravitational wave observatories. 

CMB-S4 and GWO seek to detect primordial gravitational waves. CMB-S4 would allow a measurement or constraint on the amplitude of inflationary gravitational waves down to $r<0.001$. If measured by CMB-S4, inflation would have occurred near the energy scale associated with grand unified theories. GWOs can search for deviations of the primordial gravitational wave spectrum from the predicted scale invariant shape and constrain more exotic and complex inflationary scenarios. And, in combination, CMB and GWOs may also be used to identify or rule out alternatives to inflation.

The primordial spectrum of quantum fluctuations generated during inflation can measure the interactions of the inflaton, discussed in this white paper through a constraint or measurement of primordial Non-Gaussianity (PNG). `2D' measurements from the CMB have nearly reached their statistical limit, and further improvements must come from `3D' measurements of Large Scale Structure. Future experiments, such as a 21\,cm interferometric array, Stage-V spectroscopic survey, and mm-LIM survey aim to measure wide enough sky area and redshift depth to target this science case.

The envisioned research activities for 2025--35 fall into two directions: major facilities and enabling capabilities. Major facilities include:
\begin{itemize}
    \item constructing and operating the CMB-S4 experiment, 
    \item operating and upgrading existing gravitational wave observatories (LIGO), and 
    \item developing, constructing and operating a Stage-V spectroscopic facility (Spec-S5)
\end{itemize}
Research into enabling capabilities includes:
\begin{itemize}
    \item Research into theory with a program that is aligned with the major facilities described above and encompasses a continuum of research including: theoretical model building, predicting and calculating new observable phenomena, modeling and simulating astrophysical and cosmological signals, and building analysis pipelines.
    \item Investing in new technologies to provide the needed technical foundation to execute the next major facilities in 2035+. These technologies include developing new CMB detectors and instrumentation; developing new technologies for future gravitational wave observatories (e.g.\ CBE); and developing technologies for long-wave intensity mapping surveys including 21-cm and mm-wave. This technology development will include fielding smaller-scale instruments to provide a staged approach to developing the needed technical maturity for executing a major survey in the next decade.
\end{itemize}

%We strongly recommend that we take advantage of the unique physics possible in CF5 to build a program of experiments and R\&D:
%\begin{itemize}
%    \item Support upcoming and ongoing efforts in the CMB, GWO, and optical surveys of galaxies
%    \item Support theory and analysis required to take advantage of combined surveys, large data sets, and model-building required to exploit the full physics available in these measurements (including accessing non-linear scales and foreground removal). This also includes research in theory required to explore where these measurements lie in fundamental particle physics. 
%    \item Explore R\&D for future experiments that have capabilities beyond the 2030-era experiments, including 21cm and mm-waveLIM, CMB measurements, and GWOs. 

%\end{itemize}

% References
\bibliographystyle{unsrt_mod}
\bibliography{main.bib}

\begin{thebibliography}{100}

\bibitem{Snowmass2021:Inflation}
Ana {Ach{\'u}carro}, Matteo {Biagetti}, Matteo {Braglia}, et~al.
\newblock {Inflation: Theory and Observations}.
\newblock {\em arXiv e-prints}, page arXiv:2203.08128, March 2022.

\bibitem{Snowmass2021:TheoryCosmo}
Mustafa~A. {Amin}, Francis-Yan {Cyr-Racine}, Tim {Eifler}, et~al.
\newblock {Snowmass2021 Theory Frontier White Paper: Data-Driven Cosmology}.
\newblock {\em arXiv e-prints}, page arXiv:2203.07946, March 2022.

\bibitem{Inflation-NonG}
Pieter~Daniel {Meerburg}, Daniel {Green}, Raphael {Flauger}, et~al.
\newblock {Primordial Non-Gaussianity}.
\newblock {\em BAAS}, 51(3):107, May 2019.

\bibitem{Inflation-PowerS}
Anze {Slosar}, Xingang {Chen}, Cora {Dvorkin}, et~al.
\newblock {Scratches from the Past: Inflationary Archaeology through Features
  in the Power Spectrum of Primordial Fluctuations}.
\newblock {\em BAAS}, 51(3):98, May 2019.

\bibitem{2019arXiv190704473A}
Kevork {Abazajian}, Graeme {Addison}, Peter {Adshead}, et~al.
\newblock {CMB-S4 Science Case, Reference Design, and Project Plan}.
\newblock {\em arXiv e-prints}, page arXiv:1907.04473, July 2019.

\bibitem{1997PhRvL..78.2058K}
Marc {Kamionkowski}, Arthur {Kosowsky}, and Albert {Stebbins}.
\newblock {A Probe of Primordial Gravity Waves and Vorticity}.
\newblock {\em \prl}, 78(11):2058--2061, March 1997.

\bibitem{1997PhRvL..78.2054S}
Uro{\v{s}} {Seljak} and Matias {Zaldarriaga}.
\newblock {Signature of Gravity Waves in the Polarization of the Microwave
  Background}.
\newblock {\em \prl}, 78(11):2054--2057, March 1997.

\bibitem{2021PhRvL.127o1301A}
P.~A.~R. {Ade}, Z.~{Ahmed}, M.~{Amiri}, et~al.
\newblock {Improved Constraints on Primordial Gravitational Waves using Planck,
  WMAP, and BICEP/Keck Observations through the 2018 Observing Season}.
\newblock {\em \prl}, 127(15):151301, October 2021.

\bibitem{2020PhRvD.101l2003S}
J.~T. {Sayre}, C.~L. {Reichardt}, J.~W. {Henning}, et~al.
\newblock {Measurements of B -mode polarization of the cosmic microwave
  background from 500 square degrees of SPTpol data}.
\newblock {\em \prd}, 101(12):122003, June 2020.

\bibitem{2011ApJS..192...18K}
E.~{Komatsu}, K.~M. {Smith}, J.~{Dunkley}, et~al.
\newblock {Seven-year Wilkinson Microwave Anisotropy Probe (WMAP) Observations:
  Cosmological Interpretation}.
\newblock {\em \apjs}, 192(2):18, February 2011.

\bibitem{2012ApJ...760..145Q}
{QUIET Collaboration}, D.~{Araujo}, C.~{Bischoff}, et~al.
\newblock {Second Season QUIET Observations: Measurements of the Cosmic
  Microwave Background Polarization Power Spectrum at 95 GHz}.
\newblock {\em \apj}, 760(2):145, December 2012.

\bibitem{2016SPIE.9914E..1KH}
Kathleen {Harrington}, Tobias {Marriage}, Aamir {Ali}, et~al.
\newblock {The Cosmology Large Angular Scale Surveyor}.
\newblock In Wayne~S. {Holland} and Jonas {Zmuidzinas}, editors, {\em
  Millimeter, Submillimeter, and Far-Infrared Detectors and Instrumentation for
  Astronomy VIII}, volume 9914 of {\em Society of Photo-Optical Instrumentation
  Engineers (SPIE) Conference Series}, page 99141K, July 2016.

\bibitem{2018JCAP...09..005K}
Akito {Kusaka}, John {Appel}, Thomas {Essinger-Hileman}, et~al.
\newblock {Results from the Atacama B-mode Search (ABS) experiment}.
\newblock {\em \jcap}, 2018(9):005, September 2018.

\bibitem{2020A&A...641A...6P}
{Planck Collaboration}, N.~{Aghanim}, Y.~{Akrami}, et~al.
\newblock {Planck 2018 results. VI. Cosmological parameters}.
\newblock {\em \aap}, 641:A6, September 2020.

\bibitem{2020ApJ...897...55P}
{Polarbear Collaboration}, S.~{Adachi}, M.~A.~O. {Aguilar Fa{\'u}ndez}, et~al.
\newblock {A Measurement of the Degree-scale CMB B-mode Angular Power Spectrum
  with POLARBEAR}.
\newblock {\em \apj}, 897(1):55, July 2020.

\bibitem{2021arXiv210313334S}
{SPIDER Collaboration}, P.~A.~R. {Ade}, M.~{Amiri}, et~al.
\newblock {A Constraint on Primordial $B$-Modes from the First Flight of the
  SPIDER Balloon-Borne Telescope}.
\newblock {\em arXiv e-prints}, page arXiv:2103.13334, March 2021.

\bibitem{Chang:2022tzj}
Clarence~L. {Chang}, Kevin~M. {Huffenberger}, Bradford~A. {Benson}, et~al.
\newblock {Snowmass2021 Cosmic Frontier: Cosmic Microwave Background
  Measurements White Paper}.
\newblock {\em arXiv e-prints}, page arXiv:2203.07638, March 2022.

\bibitem{2019JCAP...02..056A}
Peter {Ade}, James {Aguirre}, Zeeshan {Ahmed}, et~al.
\newblock {The Simons Observatory: science goals and forecasts}.
\newblock {\em \jcap}, 2019(2):056, February 2019.

\bibitem{2018SPIE10708E..07H}
Howard {Hui}, P.~A.~R. {Ade}, Z.~{Ahmed}, et~al.
\newblock {BICEP Array: a multi-frequency degree-scale CMB polarimeter}.
\newblock In Jonas {Zmuidzinas} and Jian-Rong {Gao}, editors, {\em Millimeter,
  Submillimeter, and Far-Infrared Detectors and Instrumentation for Astronomy
  IX}, volume 10708 of {\em Society of Photo-Optical Instrumentation Engineers
  (SPIE) Conference Series}, page 1070807, July 2018.

\bibitem{2022ApJS..258...42S}
J.~A. {Sobrin}, A.~J. {Anderson}, A.~N. {Bender}, et~al.
\newblock {The Design and Integrated Performance of SPT-3G}.
\newblock {\em \apjs}, 258(2):42, February 2022.

\bibitem{PhysRevD.104.022004}
R.~Abbott et~al.
\newblock Upper limits on the isotropic gravitational-wave background from
  advanced ligo and advanced virgo's third observing run.
\newblock {\em Phys. Rev. D}, 104:022004, 2021.

\bibitem{PhysRevX.6.011035}
Paul~D. Lasky et~al.
\newblock Gravitational-wave cosmology across 29 decades in frequency.
\newblock {\em Phys. Rev. X}, 6:011035, 2016.

\bibitem{EinsteinTelescope}
{Einstein gravitational wave Telescope (ET) conceptual design study,
  ET-0106C-10}.
\newblock https://tds.ego-gw.it/ql/?c=7954, 2011.

\bibitem{CosmicExplorer}
B.P. Abbott et~al.
\newblock Exploring the sensitivity of next generation gravitational wave
  detectors.
\newblock {\em Class. Quant. Grav.}, 34:044001, 2017.

\bibitem{LISA}
K.~Danzmann et~al.
\newblock {LISA, Laser Interferometer Space Antenna}.
\newblock
  https://www.elisascience.org/files/publications/LISA\_L3\_20170120.pdf, 2017.

\bibitem{turner}
M.S. Turner.
\newblock Detectability of inflation-produced gravitational waves.
\newblock {\em Phys. Rev. D}, 55:R435, 1997.

\bibitem{peloso_parviol}
N.~Barnaby, E.~Pajer, and M.~Peloso.
\newblock Gauge field production in axion inflation: Consequences for
  monodromy, non-gaussianity in the cmb, and gravitational waves at
  interferometers.
\newblock {\em Phys. Rev. D}, 85:023525, 2012.

\bibitem{boylebuonanno}
L.A. Boyle and A.~Buonanno.
\newblock Relating gravitational wave constraints from primordial
  nucleosynthesis, pulsar timing, laser interferometers, and the cmb:
  implications for the early universe.
\newblock {\em Phys. Rev. D}, 78:043531, 2008.

\bibitem{Gasperini:1992em}
M.~Gasperini and G.~Veneziano.
\newblock {Pre - big bang in string cosmology}.
\newblock {\em Astropart. Phys.}, 1:317--339, 1993.

\bibitem{Gasperini:2016gre}
M.~Gasperini.
\newblock {Observable gravitational waves in pre-big bang cosmology: an
  update}.
\newblock {\em JCAP}, 12:010, 2016.

\bibitem{Khoury:2001wf}
Justin Khoury, Burt~A. Ovrut, Paul~J. Steinhardt, and Neil Turok.
\newblock {The Ekpyrotic universe: Colliding branes and the origin of the hot
  big bang}.
\newblock {\em Phys. Rev. D}, 64:123522, 2001.

\bibitem{Brandenberger:1988aj}
Robert~H. Brandenberger and C.~Vafa.
\newblock {Superstrings in the Early Universe}.
\newblock {\em Nucl. Phys. B}, 316:391--410, 1989.

\bibitem{Battefeld:2014uga}
D.~Battefeld and Patrick Peter.
\newblock {A Critical Review of Classical Bouncing Cosmologies}.
\newblock {\em Phys. Rept.}, 571:1--66, 2015.

\bibitem{Finelli:2001sr}
Fabio Finelli and Robert Brandenberger.
\newblock {On the generation of a scale invariant spectrum of adiabatic
  fluctuations in cosmological models with a contracting phase}.
\newblock {\em Phys. Rev. D}, 65:103522, 2002.

\bibitem{Brandenberger:2012zb}
Robert~H. Brandenberger.
\newblock {The Matter Bounce Alternative to Inflationary Cosmology}.
\newblock 6 2012.

\bibitem{Dvorkin:2022jyg}
Cora {Dvorkin}, Joel {Meyers}, Peter {Adshead}, et~al.
\newblock {The Physics of Light Relics}.
\newblock {\em arXiv e-prints}, page arXiv:2203.07943, March 2022.

\bibitem{Ellis:2019oqb}
John Ellis, Marek Lewicki, Jos\'e~Miguel No, and Ville Vaskonen.
\newblock {Gravitational wave energy budget in strongly supercooled phase
  transitions}.
\newblock {\em JCAP}, 06:024, 2019.

\bibitem{DelleRose:2019pgi}
Luigi Delle~Rose, Giuliano Panico, Michele Redi, and Andrea Tesi.
\newblock {Gravitational Waves from Supercool Axions}.
\newblock {\em JHEP}, 04:025, 2020.

\bibitem{An:2022}
H.~An, K.-F. Lyu, L.-T. Wang, and S~Zhou.
\newblock {Gravitational Waves from an Inflation Triggered First-Order Phase
  Transition}.
\newblock {\em arXiv:2201.05171}, 2022.

\bibitem{Siemens:2006yp}
Xavier Siemens, Vuk Mandic, and Jolien Creighton.
\newblock {Gravitational wave stochastic background from cosmic
  (super)strings}.
\newblock {\em Phys. Rev. Lett.}, 98:111101, 2007.

\bibitem{Hindmarsh:2017gnf}
Mark Hindmarsh, Stephan~J. Huber, Kari Rummukainen, and David~J. Weir.
\newblock {Shape of the acoustic gravitational wave power spectrum from a first
  order phase transition}.
\newblock {\em Phys. Rev. D}, 96(10):103520, 2017.
\newblock [Erratum: Phys.Rev.D 101, 089902 (2020)].

\bibitem{Hindmarsh:2019phv}
Mark Hindmarsh and Mulham Hijazi.
\newblock {Gravitational waves from first order cosmological phase transitions
  in the Sound Shell Model}.
\newblock {\em JCAP}, 12:062, 2019.

\bibitem{Niksa:2018ofa}
Peter Niksa, Martin Schlederer, and G\"unter Sigl.
\newblock {Gravitational Waves produced by Compressible MHD Turbulence from
  Cosmological Phase Transitions}.
\newblock {\em Class. Quant. Grav.}, 35(14):144001, 2018.

\bibitem{Caprini:2009yp}
Chiara Caprini, Ruth Durrer, and Geraldine Servant.
\newblock {The stochastic gravitational wave background from turbulence and
  magnetic fields generated by a first-order phase transition}.
\newblock {\em JCAP}, 12:024, 2009.

\bibitem{RoperPol:2019wvy}
Alberto Roper~Pol, Sayan Mandal, Axel Brandenburg, et~al.
\newblock {Numerical simulations of gravitational waves from early-universe
  turbulence}.
\newblock {\em Phys. Rev. D}, 102(8):083512, 2020.

\bibitem{Jinno:2020eqg}
Ryusuke Jinno, Thomas Konstandin, and Henrique Rubira.
\newblock {A hybrid simulation of gravitational wave production in first-order
  phase transitions}.
\newblock {\em JCAP}, 04:014, 2021.

\bibitem{Ramsey-Musolf:2019lsf}
Michael~J. Ramsey-Musolf.
\newblock {The electroweak phase transition: a collider target}.
\newblock {\em JHEP}, 09:179, 2020.

\bibitem{Profumo:2007wc}
Stefano Profumo, Michael~J. Ramsey-Musolf, and Gabe Shaughnessy.
\newblock {Singlet Higgs phenomenology and the electroweak phase transition}.
\newblock {\em JHEP}, 08:010, 2007.

\bibitem{Delaunay:2007wb}
Cedric Delaunay, Christophe Grojean, and James~D. Wells.
\newblock {Dynamics of Non-renormalizable Electroweak Symmetry Breaking}.
\newblock {\em JHEP}, 04:029, 2008.

\bibitem{Huang:2016cjm}
Peisi Huang, Andrew~J. Long, and Lian-Tao Wang.
\newblock {Probing the Electroweak Phase Transition with Higgs Factories and
  Gravitational Waves}.
\newblock {\em Phys. Rev. D}, 94(7):075008, 2016.

\bibitem{Chala:2018ari}
Mikael Chala, Claudius Krause, and Germano Nardini.
\newblock {Signals of the electroweak phase transition at colliders and
  gravitational wave observatories}.
\newblock {\em JHEP}, 07:062, 2018.

\bibitem{Croon:2020cgk}
Djuna Croon, Oliver Gould, Philipp Schicho, et~al.
\newblock {Theoretical uncertainties for cosmological first-order phase
  transitions}.
\newblock {\em JHEP}, 04:055, 2021.

\bibitem{Grojean:2006bp}
Christophe Grojean and Geraldine Servant.
\newblock {Gravitational Waves from Phase Transitions at the Electroweak Scale
  and Beyond}.
\newblock {\em Phys. Rev. D}, 75:043507, 2007.

\bibitem{Alves:2018jsw}
Alexandre Alves, Tathagata Ghosh, Huai-Ke Guo, et~al.
\newblock {Collider and Gravitational Wave Complementarity in Exploring the
  Singlet Extension of the Standard Model}.
\newblock {\em JHEP}, 04:052, 2019.

\bibitem{Alves:2020bpi}
Alexandre Alves, Dorival Gon\c{c}alves, Tathagata Ghosh, et~al.
\newblock {Di-Higgs Blind Spots in Gravitational Wave Signals}.
\newblock {\em Phys. Lett. B}, 818:136377, 2021.

\bibitem{Vaskonen:2016yiu}
Ville Vaskonen.
\newblock {Electroweak baryogenesis and gravitational waves from a real scalar
  singlet}.
\newblock {\em Phys. Rev. D}, 95(12):123515, 2017.

\bibitem{Dorsch:2016nrg}
G.~C. Dorsch, S.~J. Huber, T.~Konstandin, and J.~M. No.
\newblock {A Second Higgs Doublet in the Early Universe: Baryogenesis and
  Gravitational Waves}.
\newblock {\em JCAP}, 05:052, 2017.

\bibitem{Chao:2017vrq}
Wei Chao, Huai-Ke Guo, and Jing Shu.
\newblock {Gravitational Wave Signals of Electroweak Phase Transition Triggered
  by Dark Matter}.
\newblock {\em JCAP}, 09:009, 2017.

\bibitem{Wang:2019pet}
Xiao Wang, Fa~Peng Huang, and Xinmin Zhang.
\newblock {Gravitational wave and collider signals in complex two-Higgs doublet
  model with dynamical CP-violation at finite temperature}.
\newblock {\em Phys. Rev. D}, 101(1):015015, 2020.

\bibitem{Demidov:2017lzf}
S.~V. Demidov, D.~S. Gorbunov, and D.~V. Kirpichnikov.
\newblock {Gravitational waves from phase transition in split NMSSM}.
\newblock {\em Phys. Lett. B}, 779:191--194, 2018.

\bibitem{Ahriche:2018rao}
Amine Ahriche, Katsuya Hashino, Shinya Kanemura, and Salah Nasri.
\newblock {Gravitational Waves from Phase Transitions in Models with Charged
  Singlets}.
\newblock {\em Phys. Lett. B}, 789:119--126, 2019.

\bibitem{Huang:2017rzf}
Fa~Peng Huang and Jiang-Hao Yu.
\newblock {Exploring inert dark matter blind spots with gravitational wave
  signatures}.
\newblock {\em Phys. Rev. D}, 98(9):095022, 2018.

\bibitem{Mohamadnejad:2019vzg}
Ahmad Mohamadnejad.
\newblock {Gravitational waves from scale-invariant vector dark matter model:
  Probing below the neutrino-floor}.
\newblock {\em Eur. Phys. J. C}, 80(3):197, 2020.

\bibitem{Baldes:2018nel}
Iason Baldes and G\'eraldine Servant.
\newblock {High scale electroweak phase transition: baryogenesis
  \textbackslash{}\& symmetry non-restoration}.
\newblock {\em JHEP}, 10:053, 2018.

\bibitem{Huang:2018aja}
Fa~Peng Huang, Zhuoni Qian, and Mengchao Zhang.
\newblock {Exploring dynamical CP violation induced baryogenesis by
  gravitational waves and colliders}.
\newblock {\em Phys. Rev. D}, 98(1):015014, 2018.

\bibitem{Ellis:2019flb}
Sebastian A.~R. Ellis, Seyda Ipek, and Graham White.
\newblock {Electroweak Baryogenesis from Temperature-Varying Couplings}.
\newblock {\em JHEP}, 08:002, 2019.

\bibitem{Alves:2018oct}
Alexandre Alves, Tathagata Ghosh, Huai-Ke Guo, and Kuver Sinha.
\newblock {Resonant Di-Higgs Production at Gravitational Wave Benchmarks: A
  Collider Study using Machine Learning}.
\newblock {\em JHEP}, 12:070, 2018.

\bibitem{Alves:2019igs}
Alexandre Alves, Dorival Gon\c{c}alves, Tathagata Ghosh, et~al.
\newblock {Di-Higgs Production in the $4b$ Channel and Gravitational Wave
  Complementarity}.
\newblock {\em JHEP}, 03:053, 2020.

\bibitem{Cline:2021iff}
James~M. Cline, Avi Friedlander, Dong-Ming He, et~al.
\newblock {Baryogenesis and gravity waves from a UV-completed electroweak phase
  transition}.
\newblock {\em Phys. Rev. D}, 103(12):123529, 2021.

\bibitem{Chao:2021xqv}
Wei Chao, Huai-Ke Guo, and Xiu-Fei Li.
\newblock {First Order Color Symmetry Breaking and Restoration Triggered by
  Electroweak Symmetry Non-restoration}.
\newblock 12 2021.

\bibitem{Liu:2021mhn}
Jia Liu, Xiao-Ping Wang, and Ke-Pan Xie.
\newblock {Searching for lepton portal dark matter with colliders and
  gravitational waves}.
\newblock {\em JHEP}, 06:149, 2021.

\bibitem{Zhang:2021alu}
Zhao Zhang, Chengfeng Cai, Xue-Min Jiang, et~al.
\newblock {Phase transition gravitational waves from pseudo-Nambu-Goldstone
  dark matter and two Higgs doublets}.
\newblock {\em JHEP}, 05:160, 2021.

\bibitem{Cai:2022bcf}
Rong-Gen Cai, Katsuya Hashino, Shao-Jiang Wang, and Jiang-Hao Yu.
\newblock {Gravitational waves from patterns of electroweak symmetry breaking:
  an effective perspective}.
\newblock 2 2022.

\bibitem{Schwarz:2009ii}
Dominik~J. Schwarz and Maik Stuke.
\newblock {Lepton asymmetry and the cosmic QCD transition}.
\newblock {\em JCAP}, 11:025, 2009.
\newblock [Erratum: JCAP 10, E01 (2010)].

\bibitem{Middeldorf-Wygas:2020glx}
Mandy~M. Middeldorf-Wygas, Isabel~M. Oldengott, Dietrich B\"odeker, and
  Dominik~J. Schwarz.
\newblock {The cosmic QCD transition for large lepton flavour asymmetries}.
\newblock 8 2020.

\bibitem{Caprini:2010xv}
Chiara Caprini, Ruth Durrer, and Xavier Siemens.
\newblock {Detection of gravitational waves from the QCD phase transition with
  pulsar timing arrays}.
\newblock {\em Phys. Rev. D}, 82:063511, 2010.

\bibitem{vonHarling:2017yew}
Benedict von Harling and Geraldine Servant.
\newblock {QCD-induced Electroweak Phase Transition}.
\newblock {\em JHEP}, 01:159, 2018.

\bibitem{Weinberg:1974hy}
Steven Weinberg.
\newblock {Gauge and Global Symmetries at High Temperature}.
\newblock {\em Phys. Rev. D}, 9:3357--3378, 1974.

\bibitem{Land:1992sm}
David Land and Eric~D. Carlson.
\newblock {Two stage phase transition in two Higgs models}.
\newblock {\em Phys. Lett. B}, 292:107--112, 1992.

\bibitem{Patel:2012pi}
Hiren~H. Patel and Michael~J. Ramsey-Musolf.
\newblock {Stepping Into Electroweak Symmetry Breaking: Phase Transitions and
  Higgs Phenomenology}.
\newblock {\em Phys. Rev. D}, 88:035013, 2013.

\bibitem{Patel:2013zla}
Hiren~H. Patel, Michael~J. Ramsey-Musolf, and Mark~B. Wise.
\newblock {Color Breaking in the Early Universe}.
\newblock {\em Phys. Rev. D}, 88(1):015003, 2013.

\bibitem{Blinov:2015sna}
Nikita Blinov, Jonathan Kozaczuk, David~E. Morrissey, and Carlos Tamarit.
\newblock {Electroweak Baryogenesis from Exotic Electroweak Symmetry Breaking}.
\newblock {\em Phys. Rev. D}, 92(3):035012, 2015.

\bibitem{Niemi:2018asa}
Lauri Niemi, Hiren~H. Patel, Michael~J. Ramsey-Musolf, et~al.
\newblock {Electroweak phase transition in the real triplet extension of the
  SM: Dimensional reduction}.
\newblock {\em Phys. Rev. D}, 100(3):035002, 2019.

\bibitem{Croon:2018new}
Djuna Croon and Graham White.
\newblock {Exotic Gravitational Wave Signatures from Simultaneous Phase
  Transitions}.
\newblock {\em JHEP}, 05:210, 2018.

\bibitem{Morais:2018uou}
Ant\'onio~P. Morais, Roman Pasechnik, and Thibault Vieu.
\newblock {Multi-peaked signatures of primordial gravitational waves from
  multi-step electroweak phase transition}.
\newblock {\em PoS}, EPS-HEP2019:054, 2020.

\bibitem{Morais:2019fnm}
Ant\'onio~P. Morais and Roman Pasechnik.
\newblock {Probing multi-step electroweak phase transition with multi-peaked
  primordial gravitational waves spectra}.
\newblock {\em JCAP}, 04:036, 2020.

\bibitem{Angelescu:2018dkk}
Andrei Angelescu and Peisi Huang.
\newblock {Multistep Strongly First Order Phase Transitions from New Fermions
  at the TeV Scale}.
\newblock {\em Phys. Rev. D}, 99(5):055023, 2019.

\bibitem{TripletGW2022}
Leon Friedrich, Michael~J. Ramsey-Musolf, Tuomas V.~I. Tenkanen, and Van~Que
  Tran.
\newblock {Addressing the Gravitational Wave-Collider Inverse Problem}.
\newblock {\em Phys. Rev. D}, 103(11):115035, 2021.

\bibitem{Jinno:2016knw}
Ryusuke Jinno and Masahiro Takimoto.
\newblock {Probing a classically conformal B-L model with gravitational waves}.
\newblock {\em Phys. Rev. D}, 95(1):015020, 2017.

\bibitem{Chao:2017ilw}
Wei Chao, Wen-Feng Cui, Huai-Ke Guo, and Jing Shu.
\newblock {Gravitational wave imprint of new symmetry breaking}.
\newblock {\em Chin. Phys. C}, 44(12):123102, 2020.

\bibitem{Brdar:2018num}
Vedran Brdar, Alexander~J. Helmboldt, and Jisuke Kubo.
\newblock {Gravitational Waves from First-Order Phase Transitions: LIGO as a
  Window to Unexplored Seesaw Scales}.
\newblock {\em JCAP}, 02:021, 2019.

\bibitem{Okada:2018xdh}
Nobuchika Okada and Osamu Seto.
\newblock {Probing the seesaw scale with gravitational waves}.
\newblock {\em Phys. Rev. D}, 98(6):063532, 2018.

\bibitem{Marzo:2018nov}
Carlo Marzo, Luca Marzola, and Ville Vaskonen.
\newblock {Phase transition and vacuum stability in the classically conformal
  B\textendash{}L model}.
\newblock {\em Eur. Phys. J. C}, 79(7):601, 2019.

\bibitem{Bian:2019szo}
Ligong Bian, Wei Cheng, Huai-Ke Guo, and Yongchao Zhang.
\newblock {Cosmological implications of a B \ensuremath{-} L charged hidden
  scalar: leptogenesis and gravitational waves}.
\newblock {\em Chin. Phys. C}, 45(11):113104, 2021.

\bibitem{Hasegawa:2019amx}
Taiki Hasegawa, Nobuchika Okada, and Osamu Seto.
\newblock {Gravitational waves from the minimal gauged $U(1)_{B-L}$ model}.
\newblock {\em Phys. Rev. D}, 99(9):095039, 2019.

\bibitem{Okada:2020vvb}
Nobuchika Okada, Osamu Seto, and Hikaru Uchida.
\newblock {Gravitational waves from breaking of an extra $U(1)$ in $SO(10)$
  grand unification}.
\newblock {\em PTEP}, 2021(3):033B01, 2021.

\bibitem{Fornal:2020esl}
Bartosz Fornal and Barmak Shams Es~Haghi.
\newblock {Baryon and Lepton Number Violation from Gravitational Waves}.
\newblock {\em Phys. Rev. D}, 102(11):115037, 2020.

\bibitem{Greljo:2019xan}
Admir Greljo, Toby Opferkuch, and Ben~A. Stefanek.
\newblock {Gravitational Imprints of Flavor Hierarchies}.
\newblock {\em Phys. Rev. Lett.}, 124(17):171802, 2020.

\bibitem{Fornal:2020ngq}
Bartosz Fornal.
\newblock {Gravitational Wave Signatures of Lepton Universality Violation}.
\newblock {\em Phys. Rev. D}, 103(1):015018, 2021.

\bibitem{Dev:2019njv}
P.~S.~Bhupal Dev, Francesc Ferrer, Yiyang Zhang, and Yongchao Zhang.
\newblock {Gravitational Waves from First-Order Phase Transition in a Simple
  Axion-Like Particle Model}.
\newblock {\em JCAP}, 11:006, 2019.

\bibitem{VonHarling:2019rgb}
Benedict Von~Harling, Alex Pomarol, Oriol Pujol\`as, and Fabrizio Rompineve.
\newblock {Peccei-Quinn Phase Transition at LIGO}.
\newblock {\em JHEP}, 04:195, 2020.

\bibitem{Hashino:2018zsi}
Katsuya Hashino, Mitsuru Kakizaki, Shinya Kanemura, et~al.
\newblock {Gravitational waves from first order electroweak phase transition in
  models with the U(1)$_{X}$ gauge symmetry}.
\newblock {\em JHEP}, 06:088, 2018.

\bibitem{Huang:2017laj}
Fa~Peng Huang and Xinmin Zhang.
\newblock {Probing the gauge symmetry breaking of the early universe in 3-3-1
  models and beyond by gravitational waves}.
\newblock {\em Phys. Lett. B}, 788:288--294, 2019.

\bibitem{Croon:2018kqn}
Djuna Croon, Tom\'as~E. Gonzalo, and Graham White.
\newblock {Gravitational Waves from a Pati-Salam Phase Transition}.
\newblock {\em JHEP}, 02:083, 2019.

\bibitem{Brdar:2019fur}
Vedran Brdar, Lukas Graf, Alexander~J. Helmboldt, and Xun-Jie Xu.
\newblock {Gravitational Waves as a Probe of Left-Right Symmetry Breaking}.
\newblock {\em JCAP}, 12:027, 2019.

\bibitem{Huang:2020bbe}
Wei-Chih Huang, Francesco Sannino, and Zhi-Wei Wang.
\newblock {Gravitational Waves from Pati-Salam Dynamics}.
\newblock {\em Phys. Rev. D}, 102(9):095025, 2020.

\bibitem{Fornal:2021ovz}
Bartosz Fornal, Barmak Shams Es~Haghi, Jiang-Hao Yu, and Yue Zhao.
\newblock {Gravitational Waves from Mini-Split SUSY}.
\newblock {\em Phys. Rev. D}, 104:115005, 2021.

\bibitem{Craig:2020jfv}
Nathaniel Craig, Noam Levi, Alberto Mariotti, and Diego Redigolo.
\newblock {Ripples in Spacetime from Broken Supersymmetry}.
\newblock {\em JHEP}, 21:184, 2020.

\bibitem{Apreda:2001us}
Riccardo Apreda, Michele Maggiore, Alberto Nicolis, and Antonio Riotto.
\newblock {Gravitational waves from electroweak phase transitions}.
\newblock {\em Nucl. Phys. B}, 631:342--368, 2002.

\bibitem{Bian:2017wfv}
Ligong Bian, Huai-Ke Guo, and Jing Shu.
\newblock {Gravitational Waves, baryon asymmetry of the universe and electric
  dipole moment in the CP-violating NMSSM}.
\newblock {\em Chin. Phys. C}, 42(9):093106, 2018.
\newblock [Erratum: Chin.Phys.C 43, 129101 (2019)].

\bibitem{Schwaller:2015tja}
Pedro Schwaller.
\newblock {Gravitational Waves from a Dark Phase Transition}.
\newblock {\em Phys. Rev. Lett.}, 115(18):181101, 2015.

\bibitem{Baldes:2018emh}
Iason Baldes and Camilo Garcia-Cely.
\newblock {Strong gravitational radiation from a simple dark matter model}.
\newblock {\em JHEP}, 05:190, 2019.

\bibitem{Breitbach:2018ddu}
Moritz Breitbach, Joachim Kopp, Eric Madge, et~al.
\newblock {Dark, Cold, and Noisy: Constraining Secluded Hidden Sectors with
  Gravitational Waves}.
\newblock {\em JCAP}, 07:007, 2019.

\bibitem{Croon:2018erz}
Djuna Croon, Ver\'onica Sanz, and Graham White.
\newblock {Model Discrimination in Gravitational Wave spectra from Dark Phase
  Transitions}.
\newblock {\em JHEP}, 08:203, 2018.

\bibitem{Hall:2019ank}
Eleanor Hall, Thomas Konstandin, Robert McGehee, et~al.
\newblock {Baryogenesis From a Dark First-Order Phase Transition}.
\newblock {\em JHEP}, 04:042, 2020.

\bibitem{Baldes:2017rcu}
Iason Baldes.
\newblock {Gravitational waves from the asymmetric-dark-matter generating phase
  transition}.
\newblock {\em JCAP}, 05:028, 2017.

\bibitem{Croon:2019rqu}
Djuna Croon, Alexander Kusenko, Anupam Mazumdar, and Graham White.
\newblock {Solitosynthesis and Gravitational Waves}.
\newblock {\em Phys. Rev. D}, 101(8):085010, 2020.

\bibitem{Hall:2019rld}
Eleanor Hall, Thomas Konstandin, Robert McGehee, and Hitoshi Murayama.
\newblock {Asymmetric Matters from a Dark First-Order Phase Transition}.
\newblock 11 2019.

\bibitem{Chao:2020adk}
Wei Chao, Xiu-Fei Li, and Lei Wang.
\newblock {Filtered pseudo-scalar dark matter and gravitational waves from
  first order phase transition}.
\newblock {\em JCAP}, 06:038, 2021.

\bibitem{Dent:2022bcd}
James~B. Dent, Bhaskar Dutta, Sumit Ghosh, et~al.
\newblock {Sensitivity to Dark Sector Scales from Gravitational Wave
  Signatures}.
\newblock 3 2022.

\bibitem{Li:2020eun}
Mingqiu Li, Qi-Shu Yan, Yongchao Zhang, and Zhijie Zhao.
\newblock {Prospects of gravitational waves in the minimal left-right symmetric
  model}.
\newblock {\em JHEP}, 03:267, 2021.

\bibitem{DiBari:2021dri}
Pasquale Di~Bari, Danny Marfatia, and Ye-Ling Zhou.
\newblock {Gravitational waves from first-order phase transitions in Majoron
  models of neutrino mass}.
\newblock {\em JHEP}, 10:193, 2021.

\bibitem{Zhou:2022mlz}
Ruiyu Zhou, Ligong Bian, and Yong Du.
\newblock {Electroweak Phase Transition and Gravitational Waves in the Type-II
  Seesaw Model}.
\newblock 3 2022.

\bibitem{Helmboldt:2019pan}
Alexander~J. Helmboldt, Jisuke Kubo, and Susan van~der Woude.
\newblock {Observational prospects for gravitational waves from hidden or dark
  chiral phase transitions}.
\newblock {\em Phys. Rev. D}, 100(5):055025, 2019.

\bibitem{Aoki:2019mlt}
Mayumi Aoki and Jisuke Kubo.
\newblock {Gravitational waves from chiral phase transition in a conformally
  extended standard model}.
\newblock {\em JCAP}, 04:001, 2020.

\bibitem{Croon:2019ugf}
Djuna Croon, Jessica~N. Howard, Seyda Ipek, and Timothy M.~P. Tait.
\newblock {QCD baryogenesis}.
\newblock {\em Phys. Rev. D}, 101(5):055042, 2020.

\bibitem{Croon:2019iuh}
Djuna Croon, Rachel Houtz, and Ver\'onica Sanz.
\newblock {Dynamical Axions and Gravitational Waves}.
\newblock {\em JHEP}, 07:146, 2019.

\bibitem{Garcia-Bellido:2021zgu}
Juan Garcia-Bellido, Hitoshi Murayama, and Graham White.
\newblock {Exploring the early Universe with Gaia and Theia}.
\newblock {\em JCAP}, 12(12):023, 2021.

\bibitem{Huang:2020crf}
Wei-Chih Huang, Manuel Reichert, Francesco Sannino, and Zhi-Wei Wang.
\newblock {Testing the dark SU(N) Yang-Mills theory confined landscape: From
  the lattice to gravitational waves}.
\newblock {\em Phys. Rev. D}, 104(3):035005, 2021.

\bibitem{Halverson:2020xpg}
James Halverson, Cody Long, Anindita Maiti, et~al.
\newblock {Gravitational waves from dark Yang-Mills sectors}.
\newblock {\em JHEP}, 05:154, 2021.

\bibitem{Kang:2021epo}
Zhaofeng Kang, Shinya Matsuzaki, and Jiang Zhu.
\newblock {Dark confinement-deconfinement phase transition: a roadmap from
  Polyakov loop models to gravitational waves}.
\newblock {\em JHEP}, 09:060, 2021.

\bibitem{Kajantie:1996mn}
K.~Kajantie, M.~Laine, K.~Rummukainen, and Mikhail~E. Shaposhnikov.
\newblock {Is there a~ hot electroweak phase transition at $m_H \gtrsim m_W$?}
\newblock {\em Phys. Rev. Lett.}, 77:2887--2890, 1996.

\bibitem{Morrissey:2012db}
David~E. Morrissey and Michael~J. Ramsey-Musolf.
\newblock {Electroweak baryogenesis}.
\newblock {\em New J. Phys.}, 14:125003, 2012.

\bibitem{Barrow:2022gsu}
J.~L. Barrow et~al.
\newblock {Theories and Experiments for Testable Baryogenesis Mechanisms: A
  Snowmass White Paper}.
\newblock 3 2022.

\bibitem{Asadi:2022njl}
Pouya Asadi et~al.
\newblock {Early-Universe Model Building}.
\newblock 3 2022.

\bibitem{Nelson:1993nf}
Ann~E. Nelson and Nathan Seiberg.
\newblock {R symmetry breaking versus supersymmetry breaking}.
\newblock {\em Nucl. Phys. B}, 416:46--62, 1994.

\bibitem{Jeannerot:2003qv}
Rachel Jeannerot, Jonathan Rocher, and Mairi Sakellariadou.
\newblock {How generic is cosmic string formation in SUSY GUTs}.
\newblock {\em Phys. Rev. D}, 68:103514, 2003.

\bibitem{Kibble:1976sj}
T.~W.~B. Kibble.
\newblock {Topology of Cosmic Domains and Strings}.
\newblock {\em J. Phys. A}, 9:1387--1398, 1976.

\bibitem{Vilenkin:2000jqa}
A.~Vilenkin and E.~P.~S. Shellard.
\newblock {\em {Cosmic Strings and Other Topological Defects}}.
\newblock Cambridge University Press, 7 2000.

\bibitem{Vilenkin:1981zs}
A.~Vilenkin.
\newblock {Gravitational Field of Vacuum Domain Walls and Strings}.
\newblock {\em Phys. Rev. D}, 23:852--857, 1981.

\bibitem{Sakellariadou:1990ne}
M.~Sakellariadou.
\newblock {Gravitational waves emitted from infinite strings}.
\newblock {\em Phys. Rev. D}, 42:354--360, 1990.
\newblock [Erratum: Phys.Rev.D 43, 4150 (1991)].

\bibitem{Sakellariadou:1991sd}
M.~Sakellariadou.
\newblock {Radiation of Nambu-Goldstone bosons from infinitely long cosmic
  strings}.
\newblock {\em Phys. Rev. D}, 44:3767--3773, 1991.

\bibitem{Gleiser:1998na}
Marcelo Gleiser and Ronald Roberts.
\newblock {Gravitational waves from collapsing vacuum domains}.
\newblock {\em Phys. Rev. Lett.}, 81:5497--5500, 1998.

\bibitem{Hiramatsu:2010yz}
Takashi Hiramatsu, Masahiro Kawasaki, and Ken'ichi Saikawa.
\newblock {Gravitational Waves from Collapsing Domain Walls}.
\newblock {\em JCAP}, 05:032, 2010.

\bibitem{Dunsky:2021tih}
David~I. Dunsky, Anish Ghoshal, Hitoshi Murayama, et~al.
\newblock {Gravitational Wave Gastronomy}.
\newblock 11 2021.

\bibitem{Fenu:2009qf}
Elisa Fenu, Daniel~G. Figueroa, Ruth Durrer, and Juan Garcia-Bellido.
\newblock {Gravitational waves from self-ordering scalar fields}.
\newblock {\em JCAP}, 10:005, 2009.

\bibitem{Vachaspati:1984gt}
Tanmay Vachaspati and Alexander Vilenkin.
\newblock {Gravitational Radiation from Cosmic Strings}.
\newblock {\em Phys. Rev. D}, 31:3052, 1985.

\bibitem{Blanco-Pillado:2017oxo}
Jose~J. Blanco-Pillado and Ken~D. Olum.
\newblock {Stochastic gravitational wave background from smoothed cosmic string
  loops}.
\newblock {\em Phys. Rev. D}, 96(10):104046, 2017.

\bibitem{Blanco-Pillado:2017rnf}
Jose~J. Blanco-Pillado, Ken~D. Olum, and Xavier Siemens.
\newblock {New limits on cosmic strings from gravitational wave observation}.
\newblock {\em Phys. Lett. B}, 778:392--396, 2018.

\bibitem{Ringeval:2017eww}
Christophe Ringeval and Teruaki Suyama.
\newblock {Stochastic gravitational waves from cosmic string loops in scaling}.
\newblock {\em JCAP}, 12:027, 2017.

\bibitem{Vilenkin:1981bx}
A.~Vilenkin.
\newblock {Gravitational radiation from cosmic strings}.
\newblock {\em Phys. Lett. B}, 107:47--50, 1981.

\bibitem{Hogan:1984is}
C.~J. Hogan and M.~J. Rees.
\newblock {Gravitational interactions of cosmic strings}.
\newblock {\em Nature}, 311:109--113, 1984.

\bibitem{DePies:2007bm}
Matthew~R. DePies and Craig~J. Hogan.
\newblock {Stochastic Gravitational Wave Background from Light Cosmic Strings}.
\newblock {\em Phys. Rev. D}, 75:125006, 2007.

\bibitem{Olmez:2010bi}
S.~Olmez, V.~Mandic, and X.~Siemens.
\newblock {Gravitational-Wave Stochastic Background from Kinks and Cusps on
  Cosmic Strings}.
\newblock {\em Phys. Rev. D}, 81:104028, 2010.

\bibitem{Vachaspati:2015cma}
Tanmay Vachaspati, Levon Pogosian, and Daniele Steer.
\newblock {Cosmic Strings}.
\newblock {\em Scholarpedia}, 10(2):31682, 2015.

\bibitem{Sarangi:2002yt}
Saswat Sarangi and S.~H.~Henry Tye.
\newblock {Cosmic string production towards the end of brane inflation}.
\newblock {\em Phys. Lett. B}, 536:185--192, 2002.

\bibitem{Jones:2002cv}
Nicholas~T. Jones, Horace Stoica, and S.~H.~Henry Tye.
\newblock {Brane interaction as the origin of inflation}.
\newblock {\em JHEP}, 07:051, 2002.

\bibitem{polchinski}
E.J. Copeland, R.C. Myers, and J.~Polchinski.
\newblock Cosmic f- and d-strings.
\newblock {\em JHEP}, 0406:013, 2004.

\bibitem{2006AAS...209.7413H}
Craig~J. {Hogan}.
\newblock {Gravitational Waves from Cosmic Superstrings}.
\newblock In {\em American Astronomical Society Meeting Abstracts}, volume 209
  of {\em American Astronomical Society Meeting Abstracts}, page 74.13,
  December 2006.

\bibitem{LIGOScientific:2017ikf}
B.~P. Abbott et~al.
\newblock {Constraints on cosmic strings using data from the first Advanced
  LIGO observing run}.
\newblock {\em Phys. Rev. D}, 97(10):102002, 2018.

\bibitem{Chang:2019mza}
Chia-Feng Chang and Yanou Cui.
\newblock {Stochastic Gravitational Wave Background from Global Cosmic
  Strings}.
\newblock {\em Phys. Dark Univ.}, 29:100604, 2020.

\bibitem{GW_EarlyUniv_WP}
R.~Caldwell et~al.
\newblock {Detection of Early-Universe Gravitational Wave Signatures and
  Fundamental Physics}.
\newblock {\em arXiv:2203.07972}, 2022.

\bibitem{DV2}
T.~Damour and A.~Vilenkin.
\newblock Gravitational radiation from cosmic (super)strings: Bursts,
  stochastic background, and observational windows.
\newblock {\em Phys. Rev. D}, 71:063510, 2005.

\bibitem{Martins:1996jp}
C.~J. A.~P. Martins and E.~P.~S. Shellard.
\newblock {Quantitative string evolution}.
\newblock {\em Phys. Rev. D}, 54:2535--2556, 1996.

\bibitem{Martins:2000cs}
C.~J. A.~P. Martins and E.~P.~S. Shellard.
\newblock {Extending the velocity dependent one scale string evolution model}.
\newblock {\em Phys. Rev. D}, 65:043514, 2002.

\bibitem{Blanco-Pillado:2013qja}
Jose~J. {Blanco-Pillado}, Ken~D. Olum, and Benjamin Shlaer.
\newblock The number of cosmic string loops.
\newblock {\em Phys. Rev.}, D89(2):023512, 2014.

\bibitem{Blanco-Pillado:2011egf}
Jose~J. Blanco-Pillado, Ken~D. Olum, and Benjamin Shlaer.
\newblock {Large parallel cosmic string simulations: New results on loop
  production}.
\newblock {\em Phys. Rev. D}, 83:083514, 2011.

\bibitem{Ringeval:2005kr}
Christophe Ringeval, Mairi Sakellariadou, and Francois Bouchet.
\newblock {Cosmological evolution of cosmic string loops}.
\newblock {\em JCAP}, 02:023, 2007.

\bibitem{Lorenz:2010sm}
Larissa Lorenz, Christophe Ringeval, and Mairi Sakellariadou.
\newblock {Cosmic string loop distribution on all length scales and at any
  redshift}.
\newblock {\em JCAP}, 10:003, 2010.

\bibitem{Auclair:2019zoz}
Pierre Auclair, Christophe Ringeval, Mairi Sakellariadou, and Daniele Steer.
\newblock {Cosmic string loop production functions}.
\newblock {\em JCAP}, 06:015, 2019.

\bibitem{Auclair:2019wcv}
Pierre Auclair et~al.
\newblock {Probing the gravitational wave background from cosmic strings with
  LISA}.
\newblock {\em JCAP}, 04:034, 2020.

\bibitem{Cui:2017ufi}
Yanou Cui, Marek Lewicki, David~E. Morrissey, and James~D. Wells.
\newblock {Cosmic Archaeology with Gravitational Waves from Cosmic Strings}.
\newblock {\em Phys. Rev. D}, 97(12):123505, 2018.

\bibitem{Cui:2018rwi}
Yanou Cui, Marek Lewicki, David~E. Morrissey, and James~D. Wells.
\newblock {Probing the pre-BBN universe with gravitational waves from cosmic
  strings}.
\newblock {\em JHEP}, 01:081, 2019.

\bibitem{Gouttenoire:2019kij}
Yann Gouttenoire, G\'eraldine Servant, and Peera Simakachorn.
\newblock {Beyond the Standard Models with Cosmic Strings}.
\newblock {\em JCAP}, 07:032, 2020.

\bibitem{LIGOScientific:2021nrg}
R.~Abbott et~al.
\newblock {Constraints on Cosmic Strings Using Data from the Third Advanced
  LIGO\textendash{}Virgo Observing Run}.
\newblock {\em Phys. Rev. Lett.}, 126(24):241102, 2021.

\bibitem{Ellis:2020ena}
John Ellis and Marek Lewicki.
\newblock {Cosmic String Interpretation of NANOGrav Pulsar Timing Data}.
\newblock {\em Phys. Rev. Lett.}, 126(4):041304, 2021.

\bibitem{Blanco-Pillado:2021ygr}
Jose~J. Blanco-Pillado, Ken~D. Olum, and Jeremy~M. Wachter.
\newblock {Comparison of cosmic string and superstring models to NANOGrav
  12.5-year results}.
\newblock {\em Phys. Rev. D}, 103(10):103512, 2021.

\bibitem{NANOGrav:2020bcs}
Zaven Arzoumanian et~al.
\newblock {The NANOGrav 12.5 yr Data Set: Search for an Isotropic Stochastic
  Gravitational-wave Background}.
\newblock {\em Astrophys. J. Lett.}, 905(2):L34, 2020.

\bibitem{Punturo:2010zz}
M.~Punturo et~al.
\newblock {The Einstein Telescope: A third-generation gravitational wave
  observatory}.
\newblock {\em Class. Quant. Grav.}, 27:194002, 2010.

\bibitem{Yagi:2011wg}
Kent Yagi and Naoki Seto.
\newblock {Detector configuration of DECIGO/BBO and identification of
  cosmological neutron-star binaries}.
\newblock {\em Phys. Rev.}, D83:044011, 2011.

\bibitem{AEDGE:2019nxb}
Yousef~Abou El-Neaj et~al.
\newblock {AEDGE: Atomic Experiment for Dark Matter and Gravity Exploration in
  Space}.
\newblock {\em EPJ Quant. Technol.}, 7:6, 2020.

\bibitem{Hild:2010id}
S.~Hild et~al.
\newblock {Sensitivity Studies for Third-Generation Gravitational Wave
  Observatories}.
\newblock {\em Class. Quant. Grav.}, 28:094013, 2011.

\bibitem{Sesana:2019vho}
Alberto Sesana et~al.
\newblock {Unveiling the gravitational universe at $\mu$-Hz frequencies}.
\newblock {\em Exper. Astron.}, 51(3):1333--1383, 2021.

\bibitem{Theia:2017xtk}
C\'eline Boehm et~al.
\newblock {Theia: Faint objects in motion or the new astrometry frontier}.
\newblock 7 2017.

\bibitem{Boileau:2021gbr}
Guillaume Boileau, Alexander~C. Jenkins, Mairi Sakellariadou, et~al.
\newblock {Ability of LISA to detect a gravitational-wave background of
  cosmological origin: the cosmic string case}.
\newblock 9 2021.

\bibitem{2016PhR...643....1M}
David J.~E. {Marsh}.
\newblock {Axion cosmology}.
\newblock {\em \physrep}, 643:1--79, July 2016.

\bibitem{1992PhLB..289...67H}
Diego {Harari} and Pierre {Sikivie}.
\newblock {Effects of a Nambu-Goldstone boson on the polarization of radio
  galaxies and the cosmic microwave background}.
\newblock {\em Physics Letters B}, 289(1-2):67--72, September 1992.

\bibitem{2006PhLB..642..192A}
Luca {Amendola} and Riccardo {Barbieri}.
\newblock {Dark matter from an ultra-light pseudo-Goldsone-boson}.
\newblock {\em Physics Letters B}, 642(3):192--196, November 2006.

\bibitem{Hlo_ek_2018}
Ren{\'{e} }e Hlo{\v{z}}ek, David J~E Marsh, and Daniel Grin.
\newblock Using the full power of the cosmic microwave background to probe
  axion dark matter.
\newblock {\em Monthly Notices of the Royal Astronomical Society},
  476(3):3063--3085, feb 2018.

\bibitem{cmbs4_sciencebook}
Kevork~N. {Abazajian}, Peter {Adshead}, Zeeshan {Ahmed}, et~al.
\newblock {CMB-S4 Science Book, First Edition}.
\newblock {\em arXiv e-prints}, page arXiv:1610.02743, October 2016.

\bibitem{2022arXiv220314923J}
J.~{Jaeckel}, G.~{Rybka}, and L.~{Winslow}.
\newblock {Axion Dark Matter}.
\newblock {\em arXiv e-prints}, page arXiv:2203.14923, March 2022.

\bibitem{2017PhRvD..96f1301C}
Francisco X.~Linares {Cede{\~n}o}, Alma~X. {Gonz{\'a}lez-Morales}, and
  L.~Arturo {Ure{\~n}a-L{\'o}pez}.
\newblock {Cosmological signatures of ultralight dark matter with an axionlike
  potential}.
\newblock {\em \prd}, 96(6):061301, September 2017.

\bibitem{2013MNRAS.434.1619C}
J.~{Chluba} and D.~{Grin}.
\newblock {CMB spectral distortions from small-scale isocurvature
  fluctuations}.
\newblock {\em \mnras}, 434(2):1619--1635, September 2013.

\bibitem{1998PhRvL..81.3067C}
Sean~M. {Carroll}.
\newblock {Quintessence and the Rest of the World: Suppressing Long-Range
  Interactions}.
\newblock {\em \prl}, 81(15):3067--3070, October 1998.

\bibitem{2008PhRvD..78j3516L}
Mingzhe {Li} and Xinmin {Zhang}.
\newblock {Cosmological CPT violating effect on CMB polarization}.
\newblock {\em \prd}, 78(10):103516, November 2008.

\bibitem{2009PhRvL.103e1302P}
Maxim {Pospelov}, Adam {Ritz}, and Constantinos {Skordis}.
\newblock {Pseudoscalar Perturbations and Polarization of the Cosmic Microwave
  Background}.
\newblock {\em \prl}, 103(5):051302, July 2009.

\bibitem{2009PhRvD..79f3002F}
Fabio {Finelli} and Matteo {Galaverni}.
\newblock {Rotation of linear polarization plane and circular polarization from
  cosmological pseudoscalar fields}.
\newblock {\em \prd}, 79(6):063002, March 2009.

\bibitem{carosi_2013}
G.~Carosi, A.~Friedland, M.~Giannotti, et~al.
\newblock Probing the axion-photon coupling: phenomenological and experimental
  perspectives. a snowmass white paper, 2013.

\bibitem{2018MNRAS.476.3063H}
Ren{\'e}e {Hlo{\v{z}}ek}, David J.~E. {Marsh}, and Daniel {Grin}.
\newblock {Using the full power of the cosmic microwave background to probe
  axion dark matter}.
\newblock {\em \mnras}, 476(3):3063--3085, May 2018.

\bibitem{Fedderke_2019}
Michael~A. Fedderke, Peter~W. Graham, and Surjeet Rajendran.
\newblock Axion dark matter detection with {CMB} polarization.
\newblock {\em Physical Review D}, 100(1), jul 2019.

\bibitem{Namikawa_2020}
Toshiya Namikawa, Yilun Guan, Omar Darwish, et~al.
\newblock Atacama cosmology telescope: Constraints on cosmic birefringence.
\newblock {\em Physical Review D}, 101(8), apr 2020.

\bibitem{2013PhRvD..87d7303G}
Vera {Gluscevic}, Marc {Kamionkowski}, and Duncan {Hanson}.
\newblock {Patchy screening of the cosmic microwave background by inhomogeneous
  reionization}.
\newblock {\em \prd}, 87(4):047303, February 2013.

\bibitem{2022arXiv220305728T}
{The CMB-HD Collaboration}, {:}, Simone {Aiola}, et~al.
\newblock {Snowmass2021 CMB-HD White Paper}.
\newblock {\em arXiv e-prints}, page arXiv:2203.05728, March 2022.

\bibitem{2022arXiv220308024A}
Kevork {Abazajian}, Arwa {Abdulghafour}, Graeme~E. {Addison}, et~al.
\newblock {Snowmass 2021 CMB-S4 White Paper}.
\newblock {\em arXiv e-prints}, page arXiv:2203.08024, March 2022.

\bibitem{PUMAAPC}
Anze {Slosar} et~al.
\newblock {Packed Ultra-wideband Mapping Array (PUMA): A Radio Telescope for
  Cosmology and Transients}.
\newblock In {\em Bulletin of the American Astronomical Society}, volume~51,
  page~53, September 2019.

\bibitem{puma_snowmass}
Emanuele {Castorina}, Simon {Foreman}, Dionysios {Karagiannis}, et~al.
\newblock {Packed Ultra-wideband Mapping Array (PUMA): Astro2020 RFI Response}.
\newblock {\em arXiv e-prints}, page arXiv:2002.05072, February 2020.

\bibitem{Knox2019HubbleHunter}
L.~{Knox} and M.~{Millea}.
\newblock {Hubble constant hunter's guide}.
\newblock {\em \prd}, 101(4):043533, February 2020.

\bibitem{CHIMEresults}
{The CHIME Collaboration}.
\newblock {Detection of Cosmological 21 cm Emission with the Canadian Hydrogen
  Intensity Mapping Experiment}.
\newblock {\em arXiv e-prints}, page arXiv:2202.01242, February 2022.

\bibitem{2022MNRAS.510.3495W}
Laura {Wolz}, Alkistis {Pourtsidou}, Kiyoshi~W. {Masui}, et~al.
\newblock {H I constraints from the cross-correlation of eBOSS galaxies and
  Green Bank Telescope intensity maps}.
\newblock {\em MNRAS}, 510(3):3495--3511, March 2022.

\bibitem{2021RAA....21...30L}
Lin-Cheng {Li}, Lister {Staveley-Smith}, and Jonghwan {Rhee}.
\newblock {An HI intensity mapping survey with a Phased Array Feed}.
\newblock {\em Research in Astronomy and Astrophysics}, 21(2):030, March 2021.

\bibitem{2020MNRAS.498.5916T}
Denis {Tramonte} and Yin-Zhe {Ma}.
\newblock {The neutral hydrogen distribution in large-scale haloes from 21-cm
  intensity maps}.
\newblock {\em MNRAS}, 498(4):5916--5935, November 2020.

\bibitem{2010Natur.466..463C}
{T.-C.} {Chang}, {U.-L.} {Pen}, K.~{Bandura}, and J.~B. {Peterson}.
\newblock {An intensity map of hydrogen 21-cm emission at redshift z\~{}0.8}.
\newblock {\em Nature}, 466:463--465, July 2010.

\bibitem{2013ApJ...763L..20M}
K~W Masui, E.~R. Switzer, N~Banavar, et~al.
\newblock {Measurement of 21 cm Brightness Fluctuations at z ~ 0.8 in
  Cross-correlation}.
\newblock {\em ApJ Letters}, 763(1):L20, January 2013.

\bibitem{2010PhRvD..81j3527M}
K.~W. {Masui}, P.~{McDonald}, and U.-L. {Pen}.
\newblock {Near-term measurements with 21 cm intensity mapping: Neutral
  hydrogen fraction and BAO at z < 2}.
\newblock {\em Phys. Rev. D}, 81(10), May 2010.

\bibitem{2013MNRAS.tmpL.125S}
E.~R. Switzer, K~W Masui, K~Bandura, et~al.
\newblock {Determination of z ˜ 0.8 neutral hydrogen fluctuations using the 21
  cm intensity mapping autocorrelation}.
\newblock {\em MNRAS Letters}, page L125, June 2013.

\bibitem{2015ApJ...809...62P}
Jonathan~C. {Pober}, Zaki~S. {Ali}, Aaron~R. {Parsons}, et~al.
\newblock {PAPER-64 Constraints On Reionization. II. The Temperature of the z
  =8.4 Intergalactic Medium}.
\newblock {\em ApJ}, 809(1):62, August 2015.

\bibitem{2019ApJ...883..133K}
Matthew {Kolopanis}, Daniel~C. {Jacobs}, Carina {Cheng}, et~al.
\newblock {A Simplified, Lossless Reanalysis of PAPER-64}.
\newblock {\em ApJ}, 883(2):133, October 2019.

\bibitem{2019ApJ...887..141L}
W.~{Li}, J.~C. {Pober}, N.~{Barry}, et~al.
\newblock {First Season MWA Phase II Epoch of Reionization Power Spectrum
  Results at Redshift 7}.
\newblock {\em ApJ}, 887(2):141, December 2019.

\bibitem{2016ApJ...833..102B}
A.~P. {Beardsley}, B.~J. {Hazelton}, I.~S. {Sullivan}, et~al.
\newblock {First Season MWA EoR Power spectrum Results at Redshift 7}.
\newblock {\em ApJ}, 833(1):102, December 2016.

\bibitem{2016MNRAS.460.4320E}
A.~{Ewall-Wice}, Joshua~S. {Dillon}, J.~N. {Hewitt}, et~al.
\newblock {First limits on the 21 cm power spectrum during the Epoch of X-ray
  heating}.
\newblock {\em MNRAS}, 460(4):4320--4347, August 2016.

\bibitem{EdgesDetection}
J.~D. {Bowman}, A.~E.~E {Rogers}, R.~A. {Monsalve}, et~al.
\newblock {An absorption profile centred at 78 megahertz in the sky-averaged
  spectrum}.
\newblock {\em Nature}, 555:67--70, 2018.

\bibitem{SARAS}
Saurabh {Singh}, Jishnu {Nambissan T.}, Ravi {Subrahmanyan}, et~al.
\newblock {On the detection of a cosmic dawn signal in the radio background}.
\newblock {\em Nature Astronomy}, February 2022.

\bibitem{aguirre2019roadmap}
Aaron {Parsons}, James~E. {Aguirre}, Adam~P. {Beardsley}, et~al.
\newblock {A Roadmap for Astrophysics and Cosmology with High-Redshift 21 cm
  Intensity Mapping}.
\newblock In {\em Bulletin of the American Astronomical Society}, volume~51,
  page 241, September 2019.

\bibitem{2017PASA...34...33P}
P.~{Procopio} et~al.
\newblock {A High-Resolution Foreground Model for the MWA EoR1 Field: Model and
  Implications for EoR Power Spectrum Analysis}.
\newblock {\em \pasa}, 34:e033, August 2017.

\bibitem{pober_et_al2013b}
J.~C. {Pober} et~al.
\newblock {Opening the 21 cm Epoch of Reionization Window: Measurements of
  Foreground Isolation with PAPER}.
\newblock {\em \apjl}, 768:L36, May 2013.

\bibitem{seo_and_hirata2016}
H.-J. {Seo} and C.~M. {Hirata}.
\newblock {The foreground wedge and 21-cm BAO surveys}.
\newblock {\em \mnras}, 456:3142, March 2016.

\bibitem{2012MNRAS.419.3491L}
A.~{Liu} and M.~{Tegmark}.
\newblock {How well can we measure and understand foregrounds with 21-cm
  experiments?}
\newblock {\em \mnras}, 419:3491, February 2012.

\bibitem{Shaw:2014vy}
J.~Richard {Shaw}, Kris {Sigurdson}, Michael {Sitwell}, et~al.
\newblock {Coaxing cosmic 21 cm fluctuations from the polarized sky using m
  -mode analysis}.
\newblock {\em \prd}, 91(8):083514, April 2015.

\bibitem{Shaw:2013tb}
J.~Richard {Shaw}, Kris {Sigurdson}, Ue-Li {Pen}, et~al.
\newblock {All-sky Interferometry with Spherical Harmonic Transit Telescopes}.
\newblock {\em \apj}, 781(2):57, February 2014.

\bibitem{2022arXiv220111806B}
Kalyani {Bhopi}, Will {Tyndall}, Pranav {Sanghavi}, et~al.
\newblock {A Digital Calibration Source for 21cm Cosmology Telescopes}.
\newblock {\em arXiv e-prints}, page arXiv:2201.11806, January 2022.

\bibitem{karkare2022}
Kirit~S. {Karkare}, Azadeh {Moradinezhad Dizgah}, Garrett~K. {Keating}, et~al.
\newblock {Snowmass 2021 Cosmic Frontier White Paper: Cosmology with
  Millimeter-Wave Line Intensity Mapping}.
\newblock {\em arXiv e-prints}, page arXiv:2203.07258, March 2022.

\bibitem{kovetz2019}
Ely {Kovetz}, Patrick~C. {Breysse}, Adam {Lidz}, et~al.
\newblock {Astrophysics and Cosmology with Line-Intensity Mapping}.
\newblock {\em \baas}, 51(3):101, May 2019.

\bibitem{moradinezhaddizgah2019}
Azadeh {Moradinezhad Dizgah} and Garrett~K. {Keating}.
\newblock {Line Intensity Mapping with [C II] and CO(1-0) as Probes of
  Primordial Non-Gaussianity}.
\newblock {\em \apj}, 872(2):126, February 2019.

\bibitem{karkare2018}
Kirit~S. {Karkare} and Simeon {Bird}.
\newblock {Constraining the expansion history and early dark energy with line
  intensity mapping}.
\newblock {\em \prd}, 98(4):043529, August 2018.

\bibitem{moradinezhaddizgah2022}
Azadeh {Moradinezhad Dizgah}, Garrett~K. {Keating}, Kirit~S. {Karkare}, et~al.
\newblock {Neutrino Properties with Ground-based Millimeter-wavelength Line
  Intensity Mapping}.
\newblock {\em \apj}, 926(2):137, February 2022.

\bibitem{endo2019}
Akira {Endo}, Kenichi {Karatsu}, Yoichi {Tamura}, et~al.
\newblock {First light demonstration of the integrated superconducting
  spectrometer}.
\newblock {\em Nature Astronomy}, 3:989--996, August 2019.

\bibitem{karkare2020}
K.~S. {Karkare}, P.~S. {Barry}, C.~M. {Bradford}, et~al.
\newblock {Full-Array Noise Performance of Deployment-Grade SuperSpec mm-Wave
  On-Chip Spectrometers}.
\newblock {\em Journal of Low Temperature Physics}, 199(3-4):849--857, February
  2020.

\bibitem{lowe2020}
Ian {Lowe}, Gabriele {Coppi}, Peter A.~R. {Ade}, et~al.
\newblock {The Balloon-borne Large Aperture Submillimeter Telescope
  Observatory}.
\newblock In {\em Society of Photo-Optical Instrumentation Engineers (SPIE)
  Conference Series}, volume 11445 of {\em Society of Photo-Optical
  Instrumentation Engineers (SPIE) Conference Series}, page 114457A, December
  2020.

\bibitem{Ballmer:2022uxx}
Stefan~W. {Ballmer}, Rana {Adhikari}, Leonardo {Badurina}, et~al.
\newblock {Snowmass2021 Cosmic Frontier White Paper: Future Gravitational-Wave
  Detector Facilities}.
\newblock {\em arXiv e-prints}, page arXiv:2203.08228, March 2022.

\bibitem{alvarez_snowmass}
Marcelo~A. {Alvarez}, Arka {Banerjee}, Simon {Birrer}, et~al.
\newblock {Snowmass2021 Computational Frontier White Paper: Cosmological
  Simulations and Modeling}.
\newblock {\em arXiv e-prints}, page arXiv:2203.07347, March 2022.

\bibitem{Ferraro:2022cmj}
Simone {Ferraro}, Noah {Sailer}, Anze {Slosar}, and Martin {White}.
\newblock {Snowmass2021 Cosmic Frontier White Paper: Cosmology and Fundamental
  Physics from the three-dimensional Large Scale Structure}.
\newblock {\em arXiv e-prints}, page arXiv:2203.07506, March 2022.

\bibitem{Flauger:2022hie}
Raphael {Flauger}, Victor {Gorbenko}, Austin {Joyce}, et~al.
\newblock {Snowmass White Paper: Cosmology at the Theory Frontier}.
\newblock {\em arXiv e-prints}, page arXiv:2203.07629, March 2022.

\bibitem{Beutler:2019ojk}
Florian Beutler, Matteo Biagetti, Daniel Green, et~al.
\newblock {Primordial Features from Linear to Nonlinear Scales}.
\newblock {\em Phys. Rev. Res.}, 1(3):033209, 2019.

\bibitem{Eisenstein:2006nj}
Daniel~J. Eisenstein, Hee-jong Seo, and Martin~J. White.
\newblock {On the Robustness of the Acoustic Scale in the Low-Redshift
  Clustering of Matter}.
\newblock {\em Astrophys. J.}, 664:660--674, 2007.

\bibitem{Baumann:2017lmt}
Daniel Baumann, Daniel Green, and Matias Zaldarriaga.
\newblock {Phases of New Physics in the BAO Spectrum}.
\newblock {\em JCAP}, 11:007, 2017.

\bibitem{Baumann:2019keh}
Daniel~D Baumann, Florian Beutler, Raphael Flauger, et~al.
\newblock {First constraint on the neutrino-induced phase shift in the spectrum
  of baryon acoustic oscillations}.
\newblock {\em Nature Phys.}, 15:465--469, 2019.

\bibitem{Schmittfull:2017ffw}
Marcel Schmittfull and Uros Seljak.
\newblock {Parameter constraints from cross-correlation of CMB lensing with
  galaxy clustering}.
\newblock {\em Phys. Rev. D}, 97(12):123540, 2018.

\bibitem{Dalal:2007cu}
Neal Dalal, Olivier Dore, Dragan Huterer, and Alexander Shirokov.
\newblock {The imprints of primordial non-gaussianities on large-scale
  structure: scale dependent bias and abundance of virialized objects}.
\newblock {\em Phys. Rev. D}, 77:123514, 2008.

\bibitem{Seljak:2008xr}
Uros Seljak.
\newblock {Extracting primordial non-gaussianity without cosmic variance}.
\newblock {\em Phys. Rev. Lett.}, 102:021302, 2009.

\bibitem{Creminelli:2004yq}
Paolo Creminelli and Matias Zaldarriaga.
\newblock {Single field consistency relation for the 3-point function}.
\newblock {\em JCAP}, 10:006, 2004.

\bibitem{Gleyzes:2016tdh}
J\'er\^ome Gleyzes, Roland de~Putter, Daniel Green, and Olivier Dor\'e.
\newblock {Biasing and the search for primordial non-Gaussianity beyond the
  local type}.
\newblock {\em JCAP}, 04:002, 2017.

\bibitem{Alvarez:2014vva}
Marcelo Alvarez et~al.
\newblock {Testing Inflation with Large Scale Structure: Connecting Hopes with
  Reality}.
\newblock 12 2014.

\bibitem{McDonald:2009dh}
Patrick McDonald and Arabindo Roy.
\newblock {Clustering of dark matter tracers: generalizing bias for the coming
  era of precision LSS}.
\newblock {\em JCAP}, 08:020, 2009.

\bibitem{Baumann:2010tm}
Daniel Baumann, Alberto Nicolis, Leonardo Senatore, and Matias Zaldarriaga.
\newblock {Cosmological Non-Linearities as an Effective Fluid}.
\newblock {\em JCAP}, 07:051, 2012.

\bibitem{Carrasco:2012cv}
John Joseph~M. Carrasco, Mark~P. Hertzberg, and Leonardo Senatore.
\newblock {The Effective Field Theory of Cosmological Large Scale Structures}.
\newblock {\em JHEP}, 09:082, 2012.

\bibitem{2022JCAP...08..061B}
Daniel {Baumann} and Daniel {Green}.
\newblock {The power of locality: primordial non-Gaussianity at the map level}.
\newblock {\em \jcap}, 2022(8):061, August 2022.

\bibitem{Cabass:2022wjy}
Giovanni Cabass, Mikhail~M. Ivanov, Oliver H.~E. Philcox, et~al.
\newblock {Constraints on Single-Field Inflation from the BOSS Galaxy Survey}.
\newblock 1 2022.

\bibitem{DAmico:2022gki}
Guido D'Amico, Matthew Lewandowski, Leonardo Senatore, and Pierre Zhang.
\newblock {Limits on primordial non-Gaussianities from BOSS galaxy-clustering
  data}.
\newblock 1 2022.

\bibitem{Linde:1980ts}
Andrei~D. Linde.
\newblock {Infrared Problem in Thermodynamics of the Yang-Mills Gas}.
\newblock {\em Phys. Lett. B}, 96:289--292, 1980.

\bibitem{Kajantie:1995dw}
K.~Kajantie, M.~Laine, K.~Rummukainen, and Mikhail~E. Shaposhnikov.
\newblock {Generic rules for high temperature dimensional reduction and their
  application to the standard model}.
\newblock {\em Nucl. Phys. B}, 458:90--136, 1996.

\bibitem{Farakos:1994xh}
K.~Farakos, K.~Kajantie, K.~Rummukainen, and Mikhail~E. Shaposhnikov.
\newblock {3-d physics and the electroweak phase transition: A Framework for
  lattice Monte Carlo analysis}.
\newblock {\em Nucl. Phys. B}, 442:317--363, 1995.

\bibitem{Gould:2021oba}
Oliver Gould and Tuomas V.~I. Tenkanen.
\newblock {On the perturbative expansion at high temperature and implications
  for cosmological phase transitions}.
\newblock {\em JHEP}, 06:069, 2021.

\bibitem{Curtin:2016urg}
David Curtin, Patrick Meade, and Harikrishnan Ramani.
\newblock {Thermal Resummation and Phase Transitions}.
\newblock {\em Eur. Phys. J. C}, 78(9):787, 2018.

\bibitem{Croon:2021vtc}
Djuna Croon, Eleanor Hall, and Hitoshi Murayama.
\newblock {Non-perturbative methods for false vacuum decay}.
\newblock 4 2021.

\bibitem{Ellis:2018mja}
John Ellis, Marek Lewicki, and Jos\'e~Miguel No.
\newblock {On the Maximal Strength of a First-Order Electroweak Phase
  Transition and its Gravitational Wave Signal}.
\newblock {\em JCAP}, 04:003, 2019.

\bibitem{Konstandin:2010dm}
Thomas Konstandin and Jose~M. No.
\newblock {Hydrodynamic obstruction to bubble expansion}.
\newblock {\em JCAP}, 02:008, 2011.

\bibitem{BarrosoMancha:2020fay}
Marc Barroso~Mancha, Tomislav Prokopec, and Bogumila Swiezewska.
\newblock {Field-theoretic derivation of bubble-wall force}.
\newblock {\em JHEP}, 01:070, 2021.

\bibitem{Balaji:2020yrx}
Shyam Balaji, Michael Spannowsky, and Carlos Tamarit.
\newblock {Cosmological bubble friction in local equilibrium}.
\newblock {\em JCAP}, 03:051, 2021.

\bibitem{Ai:2021kak}
Wen-Yuan Ai, Bjorn Garbrecht, and Carlos Tamarit.
\newblock {Bubble wall velocities in local equilibrium}.
\newblock 9 2021.

\bibitem{Bodeker:2009qy}
Dietrich Bodeker and Guy~D. Moore.
\newblock {Can electroweak bubble walls run away?}
\newblock {\em JCAP}, 05:009, 2009.

\bibitem{Bodeker:2017cim}
Dietrich Bodeker and Guy~D. Moore.
\newblock {Electroweak Bubble Wall Speed Limit}.
\newblock {\em JCAP}, 05:025, 2017.

\bibitem{Hoeche:2020rsg}
Stefan H\"oche, Jonathan Kozaczuk, Andrew~J. Long, et~al.
\newblock {Towards an all-orders calculation of the electroweak bubble wall
  velocity}.
\newblock {\em JCAP}, 03:009, 2021.

\bibitem{Gouttenoire:2021kjv}
Yann Gouttenoire, Ryusuke Jinno, and Filippo Sala.
\newblock {Friction pressure on relativistic bubble walls}.
\newblock 12 2021.

\bibitem{Randall:2006py}
Lisa Randall and Geraldine Servant.
\newblock {Gravitational waves from warped spacetime}.
\newblock {\em JHEP}, 05:054, 2007.

\bibitem{Espinosa:2008kw}
J.~R. Espinosa, T.~Konstandin, J.~M. No, and M.~Quiros.
\newblock {Some Cosmological Implications of Hidden Sectors}.
\newblock {\em Phys. Rev. D}, 78:123528, 2008.

\bibitem{Espinosa:2010hh}
Jose~R. Espinosa, Thomas Konstandin, Jose~M. No, and Geraldine Servant.
\newblock {Energy Budget of Cosmological First-order Phase Transitions}.
\newblock {\em JCAP}, 06:028, 2010.

\bibitem{Jinno:2017fby}
Ryusuke Jinno and Masahiro Takimoto.
\newblock {Gravitational waves from bubble dynamics: Beyond the Envelope}.
\newblock {\em JCAP}, 01:060, 2019.

\bibitem{Konstandin:2017sat}
Thomas Konstandin.
\newblock {Gravitational radiation from a bulk flow model}.
\newblock {\em JCAP}, 03:047, 2018.

\bibitem{Megevand:2021juo}
Ariel Megevand and Federico~Agustin Membiela.
\newblock {Gravitational waves from bubble walls}.
\newblock {\em JCAP}, 10:073, 2021.

\bibitem{Hindmarsh:2016lnk}
Mark Hindmarsh.
\newblock {Sound shell model for acoustic gravitational wave production at a
  first-order phase transition in the early Universe}.
\newblock {\em Phys. Rev. Lett.}, 120(7):071301, 2018.

\bibitem{Kamionkowski:1993fg}
Marc Kamionkowski, Arthur Kosowsky, and Michael~S. Turner.
\newblock {Gravitational radiation from first order phase transitions}.
\newblock {\em Phys. Rev. D}, 49:2837--2851, 1994.

\bibitem{Kosowsky:2001xp}
Arthur Kosowsky, Andrew Mack, and Tinatin Kahniashvili.
\newblock {Gravitational radiation from cosmological turbulence}.
\newblock {\em Phys. Rev. D}, 66:024030, 2002.

\bibitem{Dolgov:2002ra}
Alexander~D. Dolgov, Dario Grasso, and Alberto Nicolis.
\newblock {Relic backgrounds of gravitational waves from cosmic turbulence}.
\newblock {\em Phys. Rev. D}, 66:103505, 2002.

\bibitem{Caprini:2006jb}
Chiara Caprini and Ruth Durrer.
\newblock {Gravitational waves from stochastic relativistic sources: Primordial
  turbulence and magnetic fields}.
\newblock {\em Phys. Rev. D}, 74:063521, 2006.

\bibitem{Gogoberidze:2007an}
Grigol Gogoberidze, Tina Kahniashvili, and Arthur Kosowsky.
\newblock {The Spectrum of Gravitational Radiation from Primordial Turbulence}.
\newblock {\em Phys. Rev. D}, 76:083002, 2007.

\bibitem{Kahniashvili:2008pe}
Tina Kahniashvili, Leonardo Campanelli, Grigol Gogoberidze, et~al.
\newblock {Gravitational Radiation from Primordial Helical Inverse Cascade MHD
  Turbulence}.
\newblock {\em Phys. Rev. D}, 78:123006, 2008.
\newblock [Erratum: Phys.Rev.D 79, 109901 (2009)].

\bibitem{Cutting:2019zws}
Daniel Cutting, Mark Hindmarsh, and David~J. Weir.
\newblock {Vorticity, kinetic energy, and suppressed gravitational wave
  production in strong first order phase transitions}.
\newblock {\em Phys. Rev. Lett.}, 125(2):021302, 2020.

\bibitem{Kahniashvili:2020jgm}
Tina Kahniashvili, Axel Brandenburg, Grigol Gogoberidze, et~al.
\newblock {Circular polarization of gravitational waves from early-Universe
  helical turbulence}.
\newblock {\em Phys. Rev. Res.}, 3(1):013193, 2021.

\bibitem{RoperPol:2021xnd}
Alberto Roper~Pol, Sayan Mandal, Axel Brandenburg, and Tina Kahniashvili.
\newblock {Polarization of gravitational waves from helical MHD turbulent
  sources}.
\newblock 7 2021.

\bibitem{RoperPol:2022iel}
Alberto Roper~Pol, Chiara Caprini, Andrii Neronov, and Dmitri Semikoz.
\newblock {The gravitational wave signal from primordial magnetic fields in the
  Pulsar Timing Array frequency band}.
\newblock 1 2022.

\bibitem{contaldi}
C.~Contaldi.
\newblock Anisotropies of gravitational wave backgrounds: A line of sight
  approach.
\newblock {\em Physics Letters B}, 771:9, 2017.

\bibitem{jenkins_cosmstr}
A.C. Jenkins and M.~Sakellariadou.
\newblock Anisotropies in the stochastic gravitational-wave background:
  Formalism and the cosmic string case.
\newblock {\em Phys. Rev. D}, 98:063509, 2018.

\bibitem{Jenkins:2018a}
A.C. Jenkins, M.~M.~Sakellariadou, T.~Regimbau, and E.~Slezak.
\newblock Anisotropies in the astrophysical gravitational-wave background:
  Predictions for the detection of compact binaries by ligo and virgo.
\newblock {\em Phys. Rev. D}, 98:063501, 2018.

\bibitem{Jenkins:2018b}
A.C. Jenkins, R.~O'Shaughnessy, M.~Sakellariadou, and D.~Wysocki.
\newblock Anisotropies in the astrophysical gravitational-wave background: The
  impact of black hole distributions.
\newblock {\em Phys. Rev. Lett.}, 122:111101, 2019.

\bibitem{Jenkins:2019a}
A.C. Jenkins and M.~Sakellariadou.
\newblock Shot noise in the astrophysical gravitational-wave background.
\newblock {\em Phys. Rev. D}, 100:063508, 2019.

\bibitem{Jenkins:2019b}
A.C. Jenkins, J.D. Romano, and M.~Sakellariadou.
\newblock Estimating the angular power spectrum of the gravitational-wave
  background in the presence of shot noise.
\newblock {\em Phys. Rev. D}, 100:083501, 2019.

\bibitem{Jenkins:2019c}
D.~Bertacca et~al.
\newblock Projection effects on the observed angular spectrum of the
  astrophysical stochastic gravitational wave background.
\newblock {\em Phys. Rev. D}, 101:103513, 2020.

\bibitem{Cusin:2017a}
G.~Cusin, C.~Pitrou, and J.-Ph. Uzan.
\newblock Anisotropy of the astrophysical gravitational wave background:
  Analytic expression of the angular power spectrum and correlation with
  cosmological observations.
\newblock {\em Phys. Rev. Lett.}, 96:103019, 2017.

\bibitem{Cusin:2017b}
G.~Cusin, C.~Pitrou, and J.-Ph. Uzan.
\newblock The signal of the stochastic gravitational wave background and the
  angular correlation of its energy density.
\newblock {\em Phys. Rev. D}, 97:123527, 2018.

\bibitem{Cusin:2018}
G.~Cusin, I.~Dvorkin, C.~Pitrou, and J.-Ph. Uzan.
\newblock First predictions of the angular power spectrum of the astrophysical
  gravitational wave background.
\newblock {\em Phys. Rev. Lett.}, 120:231101, 2018.

\bibitem{Cusin:2018_2}
G.~Cusin, I.~Dvorkin, C.~Pitrou, and J.-Ph. Uzan.
\newblock Comment on the article ``anisotropies in the astrophysical
  gravitational-wave background: The impact of black hole distributions'' by
  a.c. jenkins et al. [arxiv:1810.13435].
\newblock {\em arXiv:1811.03582}, 2018.

\bibitem{Cusin:2019}
G.~Cusin, I.~Dvorkin, C.~Pitrou, and J.-Ph. Uzan.
\newblock Properties of the stochastic astrophysical gravitational wave
  background: astrophysical sources dependencies.
\newblock {\em Phys. Rev. D}, 100:063004, 2019.

\bibitem{Cusin:2019b}
C.~Pitrou, G.~Cusin, and J.-Ph. Uzan.
\newblock A unified view of anisotropies in the astrophysical gravitational
  wave background.
\newblock {\em Phys. Rev. D}, 101:081301, 2020.

\bibitem{Alonso:2020}
D.~Alonso, G.~Cusin, P.G. Ferreira, and C.~Pitrou.
\newblock Detecting the anisotropic astrophysical gravitational wave background
  in the presence of shot noise through cross-correlations.
\newblock {\em Phys. Rev. D}, 102:023002, 2020.

\bibitem{Cusin:2019c}
G.~Cusin, I.~Dvorkin, C.~Pitrou, and J.-P. Uzan.
\newblock Stochastic gravitational wave background anisotropies in the mhz
  band: astrophysical dependencies.
\newblock {\em Mon. Not. Roy. Astron. Soc.}, 493:L1, 2019.

\bibitem{Cusin:2018avf}
Giulia Cusin, Ruth Durrer, and Pedro~G. Ferreira.
\newblock {Polarization of a stochastic gravitational wave background through
  diffusion by massive structures}.
\newblock {\em Phys. Rev. D}, 99(2):023534, 2019.

\bibitem{CanasHerrera}
G.~Canas-Herrera, O.~Contigiani, and V.~Vardanyan.
\newblock Cross-correlation of the astrophysical gravitational-wave background
  with galaxy clustering.
\newblock {\em Phys. Rev. D}, 102:043513, 2020.

\bibitem{ghostpaper}
M.~Geller, A.~Hook, R.~Sundrum, and Y.~Tsai.
\newblock Primordial anisotropies in the gravitational wave background from
  cosmological phase transitions.
\newblock {\em Phys. Rev. Lett.}, 121:201303, 2018.

\bibitem{bartolo}
N.~Bartolo et~al.
\newblock Anisotropies and non-gaussianity of the cosmological gravitational
  wave background.
\newblock {\em Phys. Rev. D}, 100:121501, 2019.

\bibitem{bartolo2020}
N.~Bartolo et~al.
\newblock Characterizing the cosmological gravitational wave background:
  Anisotropies and non-gaussianity.
\newblock {\em Phys. Rev. D}, 102:023527, 2020.

\bibitem{dallarmi}
L.V. Dall'Armi, A.~Ricciardone, N.~Bartolo, et~al.
\newblock The imprint of relativistic particles on the anisotropies of the
  stochastic gravitational-wave background.
\newblock {\em Phys. Rev. D}, 103:023522, 2021.

\bibitem{Bellomo:2021mer}
Nicola Bellomo, Daniele Bertacca, Alexander~C. Jenkins, et~al.
\newblock {CLASS\_GWB: robust modeling of the astrophysical gravitational wave
  background anisotropies}.
\newblock 10 2021.

\bibitem{O1cosmstr}
B.P. Abbott et~al.
\newblock Constraints on cosmic strings using data from the first advanced ligo
  observing run.
\newblock {\em Phys. Rev. D}, 97:102002, 2018.

\bibitem{olmez2}
S.~Olmez, V.~Mandic, and X.~Siemens.
\newblock Anisotropies in the gravitational-wave stochastic background.
\newblock {\em J. Cosm. Astrop. Phys.}, 07:009, 2012.

\bibitem{raccanelli1}
A.~Raccanelli, E.D. Kovetz, S.~Bird, et~al.
\newblock Determining the progenitors of merging black-hole binaries.
\newblock {\em Phys. Rev. D}, 94:023516, 2016.

\bibitem{raccanelli2}
A.~Raccanelli.
\newblock Gravitational wave astronomy with radio galaxy surveys.
\newblock {\em Mon. Not. Royal Astro. S.}, 469:656, 2017.

\bibitem{nishikawa}
H.~Nishikawa, E.D. Kovetz, M.~Kamionkowski, and J.~Silk.
\newblock Primordial-black-hole mergers in dark-matter spikes.
\newblock {\em Phys. Rev. D}, 99:043533, 2019.

\bibitem{scelfo}
G.~Scelfo, N.~Bellomo, A.~Raccanelli, et~al.
\newblock Gw$\times$lss: chasing the progenitors of merging binary black holes.
\newblock {\em JCAP}, 2018:039, 2018.

\bibitem{yang}
K.Z. Yang, V.~Mandic, C.~Scarlata, and S.~Banagiri.
\newblock Searching for cross-correlation between stochastic gravitational wave
  background and galaxy number counts.
\newblock {\em Mon. Notices Royal Astron. Soc.}, 500:1666, 2021.

\bibitem{smiththrane}
R.~Smith and E.~Thrane.
\newblock {\em Phys. Rev. X}, 8:021019, 2018.

\bibitem{banagiri}
S.~Banagiri, V.~Mandic, C.~Scarlata, and K.Z. Yang.
\newblock Measuring angular n-point correlations of binary black-hole merger
  gravitational-wave events with hierarchical bayesian inference.
\newblock {\em Phys. Rev. D}, 102:063007, 2020.

\bibitem{2021arXiv210804215B}
Adrian~E. {Bayer}, Arka {Banerjee}, and Uros {Seljak}.
\newblock {Beware of Fake $\nu$s: The Effect of Massive Neutrinos on the
  Non-Linear Evolution of Cosmic Structure}.
\newblock {\em arXiv e-prints}, page arXiv:2108.04215, August 2021.

\bibitem{2017A&A...603A..62V}
F.~{Vansyngel}, F.~{Boulanger}, T.~{Ghosh}, et~al.
\newblock {Statistical simulations of the dust foreground to cosmic microwave
  background polarization}.
\newblock {\em \aap}, 603:A62, July 2017.

\bibitem{2017arXiv170704698S}
Leonardo {Senatore} and Matias {Zaldarriaga}.
\newblock {The Effective Field Theory of Large-Scale Structure in the presence
  of Massive Neutrinos}.
\newblock {\em arXiv e-prints}, page arXiv:1707.04698, July 2017.

\bibitem{2019arXiv190713167M}
Thomas {McClintock}, Eduardo {Rozo}, Arka {Banerjee}, et~al.
\newblock {The Aemulus Project IV: Emulating Halo Bias}.
\newblock {\em arXiv e-prints}, page arXiv:1907.13167, July 2019.

\bibitem{2022arXiv220202840F}
Nicholas {Frontiere}, J.~D. {Emberson}, Michael {Buehlmann}, et~al.
\newblock {Simulating Hydrodynamics in Cosmology with CRK-HACC}.
\newblock {\em arXiv e-prints}, page arXiv:2202.02840, February 2022.

\bibitem{Chisari_2019}
Nora~Elisa Chisari, David Alonso, Elisabeth Krause, et~al.
\newblock Core cosmology library: Precision cosmological predictions for
  {LSST}.
\newblock {\em The Astrophysical Journal Supplement Series}, 242(1):2, may
  2019.

\bibitem{bechtol_snowmass}
Keith {Bechtol}, Simon {Birrer}, Francis-Yan {Cyr-Racine}, et~al.
\newblock {Snowmass2021 Cosmic Frontier White Paper: Dark Matter Physics from
  Halo Measurements}.
\newblock {\em arXiv e-prints}, page arXiv:2203.07354, March 2022.

\bibitem{nord_snowmass}
Brian {Nord}, Andrew~J. {Connolly}, Jamie {Kinney}, et~al.
\newblock {Algorithms and Statistical Models for Scientific Discovery in the
  Petabyte Era}.
\newblock In {\em Bulletin of the American Astronomical Society}, volume~51,
  page 224, September 2019.

\bibitem{Lewis:2006fu}
Antony Lewis and Anthony Challinor.
\newblock {Weak gravitational lensing of the CMB}.
\newblock {\em Phys. Rept.}, 429:1--65, 2006.

\bibitem{Carlstrom:2002na}
John~E. Carlstrom, Gilbert~P. Holder, and Erik~D. Reese.
\newblock {Cosmology with the Sunyaev-Zel'dovich effect}.
\newblock {\em Ann. Rev. Astron. Astrophys.}, 40:643--680, 2002.

\bibitem{2022arXiv220307377A}
Kevork~N. {Abazajian}, Nikita {Blinov}, Thejs {Brinckmann}, et~al.
\newblock {Synergy between cosmological and laboratory searches in neutrino
  physics: a white paper}.
\newblock {\em arXiv e-prints}, page arXiv:2203.07377, March 2022.

\bibitem{Dvorkin:2019jgs}
Cora Dvorkin et~al.
\newblock {Neutrino Mass from Cosmology: Probing Physics Beyond the Standard
  Model}.
\newblock 3 2019.

\bibitem{BICEP2:2015nss}
P.~A.~R. Ade et~al.
\newblock {Joint Analysis of BICEP2/$Keck Array$ and $Planck$ Data}.
\newblock {\em Phys. Rev. Lett.}, 114:101301, 2015.

\bibitem{BICEP:2021xfz}
P.~A.~R. Ade et~al.
\newblock {Improved Constraints on Primordial Gravitational Waves using Planck,
  WMAP, and BICEP/Keck Observations through the 2018 Observing Season}.
\newblock {\em Phys. Rev. Lett.}, 127(15):151301, 2021.

\bibitem{Planck:2018gnk}
Y.~Akrami et~al.
\newblock {Planck 2018 results. XI. Polarized dust foregrounds}.
\newblock {\em Astron. Astrophys.}, 641:A11, 2020.

\bibitem{CMB-S4:2020lpa}
Kevork Abazajian et~al.
\newblock {CMB-S4: Forecasting Constraints on Primordial Gravitational Waves}.
\newblock {\em Astrophys. J.}, 926(1):54, 2022.

\bibitem{Kim:2019xov}
Chang-Goo Kim, Steve~K. Choi, and Raphael Flauger.
\newblock {Dust Polarization Maps from TIGRESS: E/B power asymmetry and TE
  correlation}.
\newblock 1 2019.

\bibitem{Kritsuk:2017aab}
Alexei~G. Kritsuk, Raphael Flauger, and Sergey~D. Ustyugov.
\newblock {Dust-polarization maps for local interstellar turbulence}.
\newblock {\em Phys. Rev. Lett.}, 121(2):021104, 2018.

\bibitem{Thorne:2021nux}
Ben Thorne, Lloyd Knox, and Karthik Prabhu.
\newblock {A generative model of galactic dust emission using variational
  autoencoders}.
\newblock {\em Mon. Not. Roy. Astron. Soc.}, 504(2):2603--2613, 2021.

\bibitem{Krachmalnicoff:2020rln}
Nicoletta Krachmalnicoff and Giuseppe Puglisi.
\newblock {ForSE: A GAN-based Algorithm for Extending CMB Foreground Models to
  Subdegree Angular Scales}.
\newblock {\em Astrophys. J.}, 911(1):42, 2021.

\bibitem{PhysRevD.73.042001}
Curt Cutler and Jan Harms.
\newblock Big bang observer and the neutron-star-binary subtraction problem.
\newblock {\em Phys. Rev. D}, 73:042001, Feb 2006.

\bibitem{PhysRevD.102.063009}
Ashish Sharma and Jan Harms.
\newblock Searching for cosmological gravitational-wave backgrounds with
  third-generation detectors in the presence of an astrophysical foreground.
\newblock {\em Phys. Rev. D}, 102:063009, Sep 2020.

\bibitem{PhysRevLett.118.151105}
T.~Regimbau, M.~Evans, N.~Christensen, et~al.
\newblock Digging deeper: Observing primordial gravitational waves below the
  binary-black-hole-produced stochastic background.
\newblock {\em Phys. Rev. Lett.}, 118:151105, Apr 2017.

\bibitem{PhysRevLett.125.241101}
Sylvia Biscoveanu, Colm Talbot, Eric Thrane, and Rory Smith.
\newblock Measuring the primordial gravitational-wave background in the
  presence of astrophysical foregrounds.
\newblock {\em Phys. Rev. Lett.}, 125:241101, Dec 2020.

\end{thebibliography}

\end{document}